\documentclass[twocolumn]{aastex6}

\newcommand{\sqd}{\ensuremath{{\rm deg}^2}}

\fullcollaborationName{The CZTI Collaboration and the GROWTH collaboration}
\shorttitle{A tale of two transients: GW170104 and GRB~170105A}
\shortauthors{Bhalerao et al}

\begin{document}

\title{A tale of two transients: GW170104 and GRB170105A}
\slugcomment{Accepted for publication in the Astrophysical Journal}

\author{V. Bhalerao\altaffilmark{1,2}, 
M. M. Kasliwal\altaffilmark{3},
D. Bhattacharya\altaffilmark{4},
A. Corsi\altaffilmark{5},
E. Aarthy\altaffilmark{6},
S. M. Adams\altaffilmark{3},
N. Blagorodnova\altaffilmark{3},
T. Cantwell\altaffilmark{7},
S. B. Cenko\altaffilmark{8,9},
R. Fender\altaffilmark{10},
D. Frail\altaffilmark{11}
R. Itoh\altaffilmark{12},
J. Jencson\altaffilmark{3},
N. Kawai\altaffilmark{12},
A.~K.~H. Kong\altaffilmark{13,14},
T. Kupfer\altaffilmark{3},
A. Kutyrev\altaffilmark{15}, 
J. Mao\altaffilmark{16,21,22},
S. Mate\altaffilmark{2},
N.~P.~S. Mithun\altaffilmark{6},
K. Mooley\altaffilmark{10,24},
D.~A. Perley\altaffilmark{17,23},
Y.~C. Perrott\altaffilmark{20},
R. M. Quimby\altaffilmark{18},
A.~R. Rao\altaffilmark{19},
L.~P. Singer\altaffilmark{8},
V. Sharma\altaffilmark{4},
D.~J. Titterington\altaffilmark{20},
E. Troja\altaffilmark{8,15}, 
S.~V. Vadawale\altaffilmark{6},
A. Vibhute\altaffilmark{4},
H. Vedantham\altaffilmark{3},
S. Veilleux\altaffilmark{15}
}
\altaffiltext{1}{Email: \url{varunb@iitb.ac.in}}
\altaffiltext{2}{Department of Physics, Indian Institute of Technology Bombay, Mumbai 400076, India}
\altaffiltext{3}{Cahill Center for Astronomy and Astrophysics, California Institute of Technology, Pasadena, CA 91125, USA}
\altaffiltext{4}{Inter-University Centre for Astronomy and Astrophysics, P. O. Bag 4, Ganeshkhind, Pune 411007, India}
\altaffiltext{5}{Department of Physics and Astronomy, Texas Tech University, Box 1051, Lubbock, TX 79409-1051, USA}
\altaffiltext{6}{Physical Research Laboratory, Ahmedabad, India}
\altaffiltext{7}{University of Manchester, Alan Turing Building, Oxford Road, Manchester M13 9PL, UK}
\altaffiltext{8}{Astrophysics Science Division, NASA Goddard Space Flight Center, 8800 Greenbelt Road, Greenbelt, MD 20771, USA}
\altaffiltext{9}{Joint  Space-Science  Institute,  University  of  Maryland, College Park, MD 20742}
\altaffiltext{10}{Centre for Astrophysical Surveys, University of Oxford, Denys Wilkinson Building, Keble Road, Oxford OX1 3RH, UK}
\altaffiltext{11}{National Radio Astronomy Observatory, P.O. Box O, Socorro, NM 87801, USA}
\altaffiltext{12}{Department of Physics, Tokyo Institute of Technology, 2-12-1 Ookayama, Meguro-ku, Tokyo 152-8551, Japan}
\altaffiltext{13}{Institute of Astronomy and Department of Physics, National Tsing Hua University, Hsinchu 30013, Taiwan}
\altaffiltext{14}{Astrophysics, Department of Physics, University of Oxford, Keble Road, Oxford OX1 3RH, UK}
\altaffiltext{15}{Department of Astronomy, University of Maryland, College Park, MD 20742, USA}
\altaffiltext{16}{Yunnan Observatories, Chinese Academy of Sciences, 650011 Kunming, Yunnan Province, China}
\altaffiltext{17}{Astrophysics Research Institute, Liverpool John Moores University, IC2, Liverpool Science Park, 146 Brownlow Hill, Liverpool L3 5RF, UK}
\altaffiltext{23}{Dark Cosmology Centre, Niels Bohr Institute, University of Copenhagen, Juliane Maries Vej 30, 2100 K{\o}benhavn {\O}, Denmark}
\altaffiltext{18}{Department of Astronomy, San Diego State University}
\altaffiltext{19}{Tata Institute of Fundamental Research, Homi Bhabha Road, Mumbai, India}
\altaffiltext{20}{Astrophysics Group, Cavendish Laboratory, 19 J. J. Thomson Avenue, Cambridge CB3 0HE, UK}
\altaffiltext{21}{Center for Astronomical Mega-Science, Chinese Academy of Sciences, 20A Datun Road, Chaoyang District, 100012 Beijing, China}
\altaffiltext{22}{Key Laboratory for the Structure and Evolution of Celestial Objects, Chinese Academy of Sciences, 650011 Kunming, China}
\altaffiltext{24}{Hintze Research Fellow}

\begin{abstract}
We present multi-wavelength follow-up campaigns by the AstroSat-CZTI and GROWTH collaborations to search for an electromagnetic counterpart to the gravitational wave event GW170104.  At the time of the GW170104 trigger, the AstroSat CZTI field-of-view covered 50.3\% of the sky localization. We do not detect any hard X-ray ($>$100\,keV) signal at this time, and place an upper limit of $\approx 4.5 \times 10^{-7}~{\rm erg~cm}^{-2}{\rm~s}^{-1}$ for a 1\,s timescale.  Separately, the ATLAS survey reported a rapidly fading optical source dubbed ATLAS17aeu in the error circle of GW170104. Our panchromatic investigation of ATLAS17aeu shows that it is the afterglow of an unrelated long, soft GRB~170105A, with only a fortuitous spatial coincidence with GW170104. We then discuss the properties of this transient in the context of standard long GRB afterglow models.
\end{abstract}


\section{Introduction}
The direct detection of gravitational waves (GW) by advanced detectors has started the era of GW astronomy~\citep{lsc16}. Astronomers from around the world teamed up with the LIGO and Virgo collaborations in the first observing run (O1) to search for electromagnetic (EM) counterparts to the GW candidates~\citep{aaa+16,aaa+16b}. 
Systematic searches for EM counterparts to GW150914, LVT151012, and  GW151226 did not find conclusive electromagnetic emission associated with them (see for instance \citealt{kcs+16,pck+16,bbv+16b,2016ApJ...820L..36S,2016ApJ...823L...2A,2016ApJ...823L..33S,2016ApJ...823L..34A,2016ApJ...825L...4T,2016MNRAS.460L..40E,2016MNRAS.462.4094S,2016PASJ...68L...9M,2016ApJ...826L..29C,2016ApJ...827L..40S,2016ApJ...829L..20A,2016ApJ...830L..11A,2016PhRvD..94l2007A,2017ApJ...835...82R,2017arXiv170306298A}, but also note a possible counterpart detected by Fermi GBM, \citealt{2016ApJ...826L...6C}).
This partnership continues in the ongoing second observing run (O2) of these advanced detectors, and EM partners have been sent several GW candidates for follow-up.

The scientific goals of an EM-GW search are to obtain precise source positions to break GW parameter degeneracies, measure source distance and redshift, study the host environment, characterize afterglow evolution, study ejecta composition and nucleosynthesis, and understand source energetics. Detection of EM counterparts can even extend GW detector reach by lowering false alarm rates.

On 2017-01-04 10:11:58.599~UTC, the LIGO Scientific Collaboration and Virgo (LVC) detected a candidate event G268556 and alerted partner astronomers~\citep{GW170104_detect}. The alert suggested that this was likely
the merger of two stellar mass black holes and provided an event localization with a 50\% (90\%) credible region spanning 400~\sqd\ (1600~\sqd). The false alarm probability was lower than one per six months. Offline analysis accounting for calibration uncertainties revised the localization area to a 50\% (90\%) credible region of about 500~\sqd\ (2000~\sqd). Further detailed analysis confirmed the astrophysical nature of this event --- now christened GW170104 --- with black hole masses of $\approx31~M_\odot$ and $\approx19~M_\odot$ and a redshift of $\approx 0.18$. The resultant $\approx49~M_\odot$ black hole is the second heaviest stellar-mass black hole known to date, exceeded only by GW150914~\citep{GW170104_main}.

Time-coincident searches for an X-ray counterpart to GW170104 yielded no significant detections (\S\ref{subsec:xraygw}). Searches for a spatially coincident optical counterpart yielded many candidates which is unsurprising given the dynamic nature of the optical sky \citep{GW_Global,GW170104_ATLAS,GW_Global2nd,GW_EWE,GW_iPTF,GW_CiPTF,GW_PANSTAR,GW_GLOBALOT}. While advances in wide-field optical imaging have overcome the challenge of mapping the coarse localizations of GW triggers, such efforts continue to be plagued with the challenge of false positives i.e.\ astrophysical events that appear to be both spatially and temporally coincident with the GW trigger but are unrelated \citep[and references therein]{aaa+16}. 

Most optical transients discovered in such large-area searches evolve slowly  on many-day to week timescales (e.g. supernovae, AGN). Thus, the report of ATLAS\,17aeu \citep{GW170104_ATLAS} fading by 0.7~mag~hr$^{-1}$ drove a ripple of excitement in the EM-GW community. 

The GROWTH collaboration\footnote{Global Relay of Observatories Watching Transients Happen; \url{http://growth.caltech.edu/}.} promptly imaged ATLAS\,17aeu with the Large Format Camera (LFC) mounted on the Palomar 200-inch Hale Telescope (P200),  the Large Monolithic Imager (LMI) on the Discovery Channel Telescope, the GMG telescope at Lijiang Observatory and the MITSuME telescope at Akeno Observatory (\S\ref{subsec:atlas}). We detected the transient and fit a power-law temporal decay of the form $F = F_0 (t - t_o)^{-\alpha}$. Intriguingly, the statistically robust power-law fit suggested an explosion time ($t_0$) that was offset from the GW trigger time by $21.5\pm1.0$~hours~\citep{GW170104_Palo}. The prospect of this event being an unrelated, untriggerred or off-axis Gamma Ray Burst (GRB) was rather small as there had only been two such optical reports to date: PTF\,11agg \citep{ckh+13} and iPTF\,14yb \citep{cup+15}. Nonetheless, we decided to trigger the {\em Swift} satellite (\S\ref{subsec:xrayprop}), the Karl G. Jansky Very Large Array (VLA) (\S\ref{subsec:radio}) and the Arcminute MicroKelvin Imager -- Large Array (\S\ref{subsec:radio}), and detected both an X-ray and radio counterpart. 

Motivated thus, upon checking data from the AstroSat Cadmium Zinc Telluride Imager (CZTI) and high energy archives, we found a GRB had actually been detected that would be consistent with the explosion time of ATLAS\,17aeu (\S\ref{sec:grb}). Furthermore,
AstroSat's localization confirmed that the spatial coincidence was also consistent with this hypothesis (\S\ref{subsec:loc}). In this paper, we report the efforts of the AstroSat CZTI and GROWTH collaborations to establish that the panchromatic properties of ATLAS17aeu are simply explained as the afterglow of GRB~170105A, unrelated to GW\,170104.

\section{GW170104: Search for electromagnetic counterparts}

\subsection{No X-ray counterpart}\label{subsec:xraygw}
We undertook an offline search for a hard X-ray counterpart to GW170104 in AstroSat CZTI data. CZTI is a hard X-ray coded aperture mask instrument that functions as an open detector above $\sim$100~keV~\citep{bbv+16}. CZTI has high sensitivity to hard X-ray transients and has detected over a hundred GRBs in 18 months of operation\footnote{CZTI GRB discoveries are distributed as GCN circulars and reported online at \url{http://astrosat.iucaa.in/czti/?q=grb}.}. Coincidence between the four identical, independent quadrants of CZTI serves as an excellent discriminant between astrophysical transients and instrumental noise.

\begin{figure}[htb]
\includegraphics[width=\columnwidth]{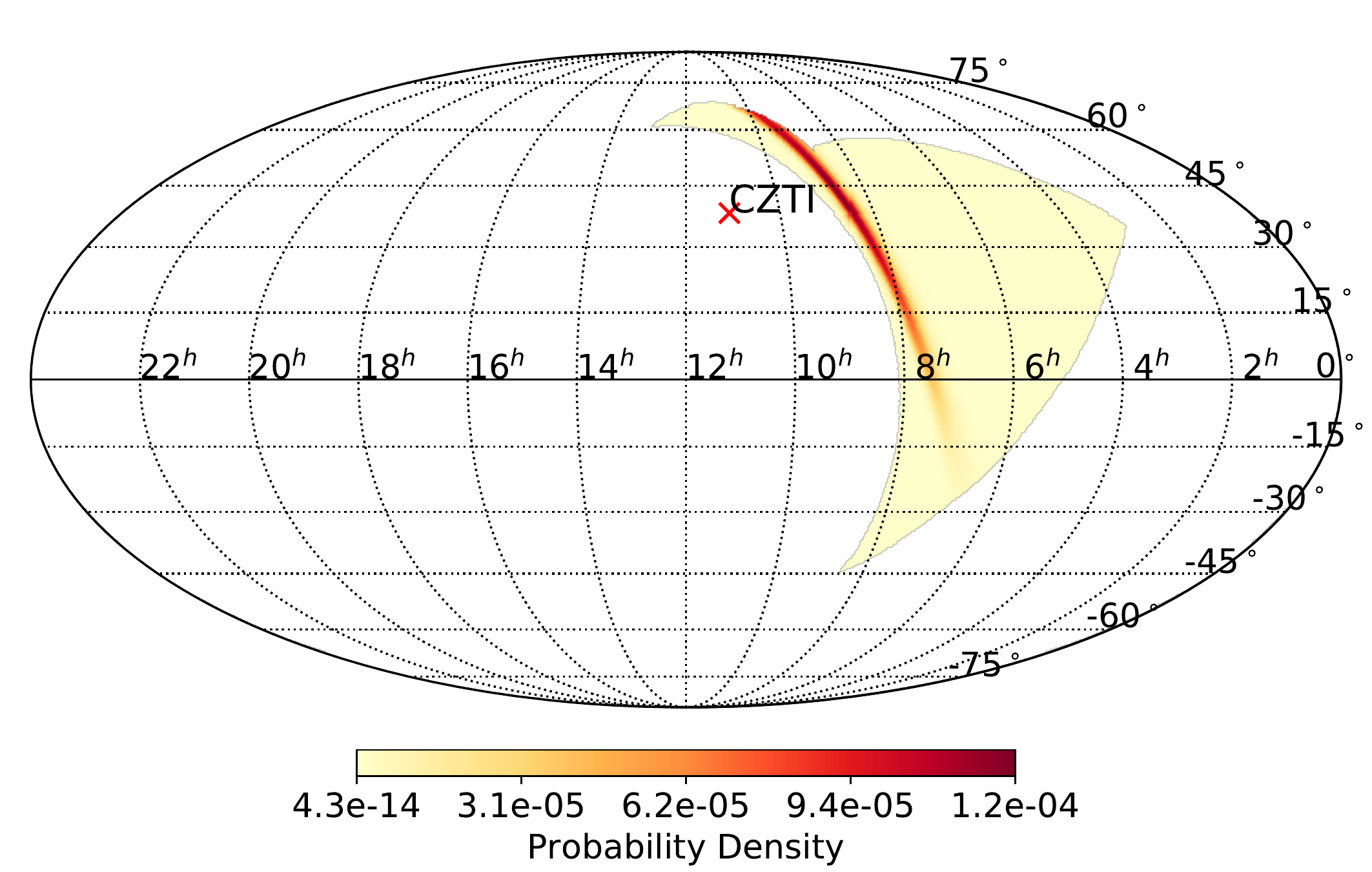}
\includegraphics[width=\columnwidth]{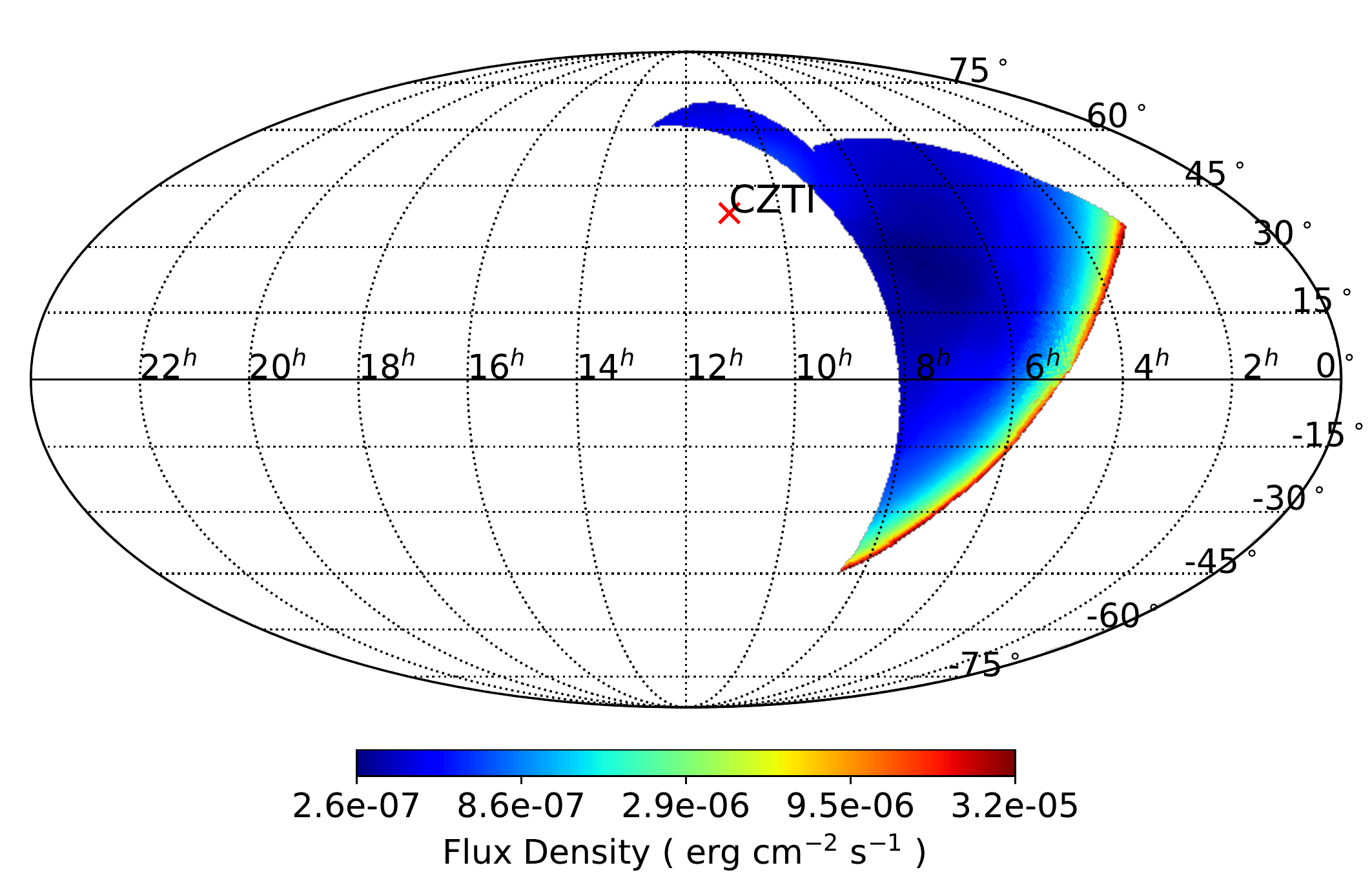}
\caption{\textit{Upper panel:} The LIGO sky position probability map for GW170104, masked to show only the sky visible to CZTI. The red cross marks the boresight of CZTI. Parts of the sky obscured by earth or by satellite elements are shown in white. The visible area encloses a 50.3\% probability of containing the GW source.
\textit{Lower panel:} The upper limits on hard X-ray emission from GW170104, from a search for 1~s transients. The variation of upper limits with position for other timescales is identical modulo an overall scaling factor.}
\label{fig:czti_coverage}
\end{figure}

Based on the refined localization map \citep{GW170104_finalskymap}, CZTI covered 50.3\% of the GW170104 probability region at the time of the trigger (Figure~\ref{fig:czti_coverage}, top panel). The rest of the localization was obscured by the earth or behind the focal plane. Following usual GRB search procedures for CZTI, data were first reduced with the CZTI pipeline to suppress noisy pixels and to generate event files. We then calculated ``dynamic spectra'' by binning data in 20~keV, 1~s time bins. The resultant two-dimensional distributions, effectively consisting of light curves in successive energy bins, were scrutinized for any transients. We normalized the light curve at each energy by subtracting the mean count rate and dividing by the standard deviation at that energy. The process was also repeated with 0.1~s and 10~s binning. In searches at all three timescales, no transient was detected in a 100~s window centered on the time of GW170104.

Next, we calculated upper limits on hard X-ray emission from GW170104. CZTI count rates show slow variations with longitude of the satellite. The detectors occasionally have flickering pixels, which can create false positives in a transient search. As a result, we used data from neighboring orbits to calculate the minimum counts required for a secure detection. For GW170104, we measured these noise properties using data from five orbits before and after the trigger (orbit 6867 to 6878). After default data reduction steps, we calculated light curves using events from 20 to 200~keV. These light curves were de-trended using a second order Savitzky Golay filter with a 100~s window.

We then calculated a cut-off rate for each quadrant such that the probability of getting counts above that rate in any 100~s window is 10\%.
Events where the count rates in all quadrants are above the respective cutoff rates in the same time bin, are considered as secure transient detections, with a false alarm probability of 0.01\%. We repeated this process for time scales of 0.1 and 10~s as well to calculate respective count rate upper limits. We assumed that the transient spectrum is described by a Band function with GRB-like parameters: $\alpha = -1$, $\beta = -2.5$, and $E_{\rm peak} = 300$~keV. With this spectral model, count rates were converted to direction-dependent upper limits on flux (Figure~\ref{fig:czti_coverage}) by using a ray tracing code~\citep{rch+16}. Weighting these flux upper limits with the probability of finding the GW counterpart in the respective directions, the effective flux limits over the sky visible to CZTI are $1.8 \times 10^{-7}~{\rm erg~cm}^{-2}{\rm~s}^{-1}$,  $4.5 \times 10^{-7}~{\rm erg~cm}^{-2}{\rm~s}^{-1}$, and $1.0 \times 10^{-7}~{\rm erg~cm}^{-2}{\rm~s}^{-1}$ for searches at 0.1~s, 1~s, and 10~s timescales respectively.
For reference, upper limits from other high energy instruments are given in Table~\ref{tab:obs_he}.

\floattable
\begin{deluxetable}{lRLcCCLl}
\tablewidth{\textwidth}
\tablecaption{X-ray and gamma ray searches for a counterpart to GW170104\label{tab:obs_he}}
\tablehead{\colhead{Instrument} & \colhead{Search} &\colhead{Search} &\colhead{Search} & \colhead{Energy range} & \colhead{Flux limit} & \colhead{Probability} & \colhead{Reference} \\
 & \colhead{Start} & \colhead{End} & \colhead{timescale} & \colhead{(keV)} & \colhead{${\rm erg~cm}^{-2}{\rm~s}^{-1}$}  & \colhead{coverage} & 
 }
\startdata
AstroSat CZTI    & -50 {\rm~s} & +50 {\rm~s} &  0.1~s   & 20-200   & 1.8 \times 10^{-7}  & 0.50 & This work\\
				                        &  & & 1~s      &   &  	$4.5 \times 10^{-7}$	&  & \\
				                        &  & & 10~s     &   &  	$1.0 \times 10^{-7}$	&  & \\
\textit{Fermi} GBM  & -30 {\rm~s} & +30 {\rm~s} & 0.265~s -- 8.192~s & \nodata & \nodata & 0.82 & \cite{GW170104_FermiGBM} \\
INTEGRAL (SPI/ACS)& -100 {\rm~s} & +100 {\rm~s} & 0.1~s  & $100-10^{5}$  & $5.0 \times 10^{-7}$ & 0.90 & \cite{GW170104_Integral}\\
				    &  &                       & 1~s    &               &  $1.6 \times 10^{-7}$	&  & \\
				    &  &                       & 10~s   &               &  $0.45 \times 10^{-7}$ &  &\\
\textit{Fermi} LAT           & 0 {\rm~s} & +10 {\rm~ks}  & \nodata	   & $>10^{5}$   & \nodata &   0.55  & \cite{GW170104_LAT}\\
AGILE-MCAL     & -11.2 {\rm~s} & +1.4 {\rm~s} & 1~s	 & $400-10^{5}$ & $5.45 - 6.18 \times 10^{-7}$  & 0.37 & \cite{GW170104_AGILE}\\
Super-AGILE    & -100 {\rm~s} & +100 {\rm~s} & 1~s	 & 20-60     & $1.5 - 6.6 \times 10^{-8}$ &   &\\
AGILE-GRID\tablenotemark{a}   & -500 {\rm~s} & +500 {\rm~s} & 2~s	& $3 \times 10^{4} - 10^{7}$ & 2.0 \times 10^{-6} &   0.4 &\cite{GW170104_AGILEGRID}\\
		                                 &  &  & 100~s	&            & $3.4 \times 10^{-8}$ &    & \\
CALET HXM       & -60 {\rm~s} & +60 {\rm~s} & \nodata	& 7-1000               &  \nodata   &   0.37 & \cite{GW170104_Calet}\\
CALET SGM       & -60 {\rm~s} & +60 {\rm~s} & \nodata	& $100-2 \times 10^{4}$ & \nodata   & 0.4  & \cite{GW170104_Calet}\\
CALET CAL       & -60 {\rm~s} & +60 {\rm~s} & \nodata   &  $>10^{7}$              & \nodata  & 0.3 & \cite{GW170104_Calet}\\
Lomonosov       & -10 {\rm~hr} & +10 {\rm~hr} & 1~s    & 20-800    &  	$1\times 10^{-7}$				 			& $\sim$0.5  & \cite{GW170104_lom}\\
\textit{Swift} BAT\tablenotemark{b}       & -100 {\rm~s} & +100 {\rm~s} & 1~s	   & 15-350    & 6.0 \times 10^{-8} &   0.48 & \cite{GW170104_bat}\\
MAXI GSC\tablenotemark{c}    & 0 {\rm~s} & + 92{\rm~min} & \nodata & 2-20	& 1.7 \times 10^{-9}		& 0.80 & \cite{GW170104_maxi}\\
               & 0 {\rm~s} & + 24{\rm~hr}  & \nodata &         		& 0.5 \times 10^{-9}		& 0.86	&\\
Konus-wind     & -100{\rm~s} & +100{\rm~s}  & 2.944   & $10-10^{4}$  & 3.3 \times 10^{-7} (5$\sigma$)   & \nodata  & \cite{GW170104_konus}\\      
\enddata
\tablecomments{The probability coverage is the total probability of the gravitational wave source being located in the sky region observed by any instrument. Note that for most instruments, the reported numbers are based on earlier versions of the gravitational wave localization sky map. The probability for CZTI is based on the revised \citet{GW170104_finalskymap} sky map.}
\tablenotetext{a}{\citet{GW170104_AGILEGRID} analyzed data in energy range from 30~MeV -- 10~GeV, in timescales from 2 to 1000 sec centered on the trigger time. We denote this search range as $-$500~s to +500~s.}
\tablenotetext{b}{\cite{GW170104_bat} report 4$\sigma$ upper limits for {\em Swift}-BAT.}
\tablenotetext{c}{We convert upper limits from \citet{GW170104_konus} to flux by assuming a spectrum with slope $-1.1$ and normalization $9.7{\rm~counts~s}^{-1}{\rm~cm}^{-2}{\rm~keV}^{-1}$ at 1~keV.}
\end{deluxetable}

\subsection{ATLAS17aeu: a candidate optical counterpart}\label{subsec:atlas}
ATLAS17aeu was discovered by the Asteroid Terrestrial-impact Last Alert System~\citep[ATLAS,][]{ATLAS,GW170104_ATLAS} as a fast-fading optical transient in the error region of GW170104. To determine the nature of the source and any possible association with GW170104, the GROWTH collaboration undertook the following observations with various telescopes worldwide. 

We imaged the position of ATLAS17aeu with the Large Format Camera \citep[LFC;][]{sms+00} on the Palomar 200-inch Hale telescope (P200).  
The LFC data were reduced with standard IRAF tasks and PSF photometry was performed using DAOPHOT.  Photometric calibration was done relative to Pan-STARRS DR1 \citep{cmm+16,fmc+16}.
We imaged the location of ATLAS17aeu with the Large Monolithic Imager (LMI) mounted on the 4.3\,m Discovery Channel Telescope (DCT) in Happy Jack, AZ.  The LMI images were processed using a custom IRAF pipeline for basic detrending (bias subtraction and flat fielding) and individual dithered images were combined using \texttt{SWarp} \citep{bmr+02}.  Transient photometry was measured using aperture photometry with the inclusion radius matched to the FWHM of the image point spread function.  Photometric calibration was performed relative to point sources from the Sloan Digital Sky Survey (SDSS; \citealt{sdss16}).

We observed the optical transient ATLAS17aeu with the 2.4-m GMG
telescope at the Lijiang Observatory in Yunnan, China. We obtained an
$R$-band image with the Yunnan Faint Object Spectrograph and Camera
(YFOSC) on 2017-01-07 14:55:35 UT. ATLAS17aeu was not detected with a
3-sigma limit of $m_{R} \gtrsim 22.3$~mag.

We undertook optical $g^\prime$, $Rc$ and $Ic$ band photometric observations of ATLAS17aeu on MJD~57760 with the 50~cm MITSuME telescope at Akeno Observatory, Japan~\citep{mitsume}. Data were reduced using standard CCD photometry procedures in PYRAF.

We also observed ATLAS17aeu on 2017 January 17.4 with the Wide Field Infrared Camera \citep[WIRC;][]{weh+03} on P200. We obtained a sequence of 52 well-dithered 45 s exposures to allow for accurate subtraction of the sky background, for a total integration time of 2340 s in the $J$-band. Imaging reductions, including flat-fielding, background subtraction, astrometric alignment, and stacking of individual frames were performed using a custom pipeline. The photometric zero point of the final image was measured using aperture photometry of 37 isolated 2MASS stars spread across the field, with the aperture radius set to match the typical seeing in the image. We convert the Vega system magnitudes to AB magnitudes following \citet{br07}. We detect nothing at the position of the transient to a 5$\sigma$ point source limiting magnitude of $m_J >$ 22.3\,mag. 

\begin{figure*}[!tbh]
\includegraphics[width=\textwidth]{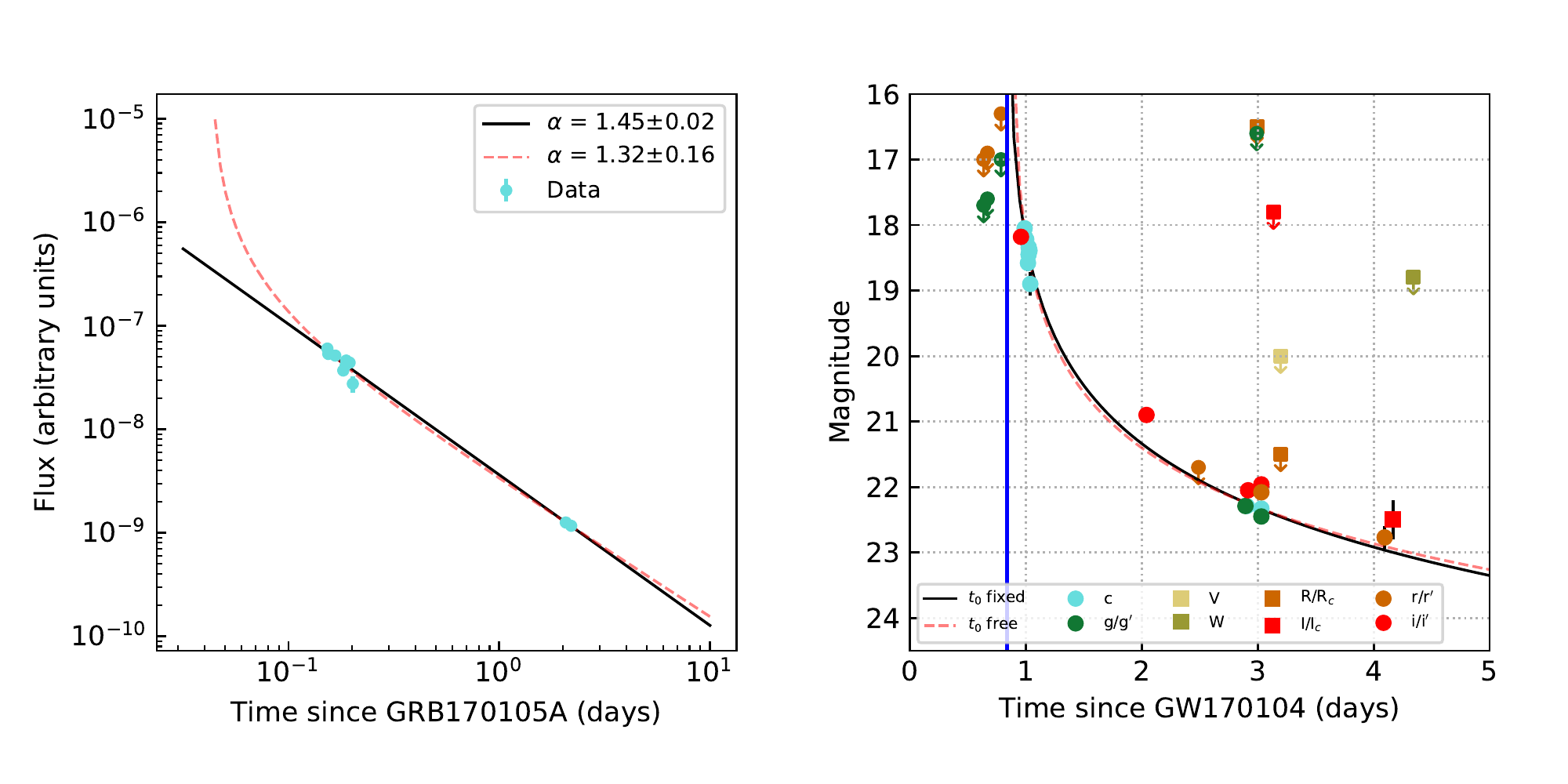}
\caption{\textit{Left panel:} The dashed red lines show a power-law fit of the form $F = F_0 (t - t_o)^{-\alpha}$, with $\alpha$ and $t_0$ as free parameters, to the cyan band data of ATLAS17aeu. We measure $\alpha = 1.32 \pm 0.16$ and $t_0 = 57758.303 \pm 0.045$ (MJD), $21.1 \pm 1.0$~hours after GW170104. This calculated explosion time is consistent with GRB~170105A, which occurred at MJD~57758.260. The solid black line shows a power law fit with $t_0$ fixed to the GRB time, and the X-axis shows days since MJD~57758.260. \textit{Right panel:} All optical and infrared photometry of ATLAS17aeu overplotted with the same power-law fit as the left panel. The X-axis is the time since GW170104. The solid blue vertical line marks the time of GRB~170105A. Points with downward pointing arrows are upper limits.}
\label{fig:powdecay}
\end{figure*}

We summarize all available optical and infrared photometry on the transient in Table~\ref{tab:obs_opt}. For non-detections, we report 5$\sigma$ upper limits. The early observations of ATLAS17aeu were in the ATLAS cyan band\footnote{ATLAS filter details are available at \url{http://www.fallingstar.com/specifications.php}.}. 
To account for the different bandpasses, we convert our P200/LFC and DCT/LMI multi-band photometry to the cyan filter assuming a power-law spectrum, $F_\nu \propto \nu^{-\beta}$ at each epoch. 
We then jointly fit these two data points along with the ATLAS photometry to a power law model of the form $F_\nu \propto (t - t_0)^{-\alpha}$. We refine the measurements of~\citet{GW170104_Palo} and obtain $t_{0, \rm MJD} = 57758.303 \pm 0.045$ and $\alpha = 1.32 \pm 0.16$ (Figure~\ref{fig:powdecay}). This is $21.1 \pm 1.1$ hours after the GW170104 trigger time (MJD~57757.425), which prompted us to search for any possible high energy event at this $t_0$.

\section{GRB~170105A: Observations}\label{sec:grb}
\subsection{X-ray detection}\label{subsec:xray}
We searched AstroSat CZTI data for any transients in the 3$\sigma$ window given by our preliminary power-law fits to ATLAS and LFC data, and found a burst peaking at 2017-01-05~06:14:06~UT \citep{czti_grb170105A}. This event was independently discovered and reported as GRB~170105A by the POLAR collaboration~\citep{mxh17}. This trigger time, MJD~57758.260, is consistent within a 1$\sigma$ range of the explosion time calculated for ATLAS17aeu in \S\ref{subsec:atlas}.
Inspection of CZTI data showed GRB~170105A had no photons above $\sim$100~keV --- making it much softer than typical GRBs detected by CZTI.
$T_{90}$ measured from quick CZTI analysis was 2.86~s\footnote{$T_{90}$ is defined as the interval during which 90\% of the counts from the GRB are received, starting from the instant when 5\% of the total counts are observed~\citep{kpk+95}.}, slightly longer than $T_{90} = 2.0 \pm 0.5$ reported by \citet{mxh17}. Careful reanalysis of the data allowed us to attain a lower noise floor, leading to detection of longer duration emission from the GRB. We measure $T_{90} = 15\pm1$~s, and detect 1070 photons in quadrants A and B. The soft spectrum and long $T_{90}$ confirm that GRB~170105A is a long soft burst \citep{kmf+93}.

\subsection{Localization}\label{subsec:loc}
GRB~170105A was outside the primary field of view of CZTI, and could not be localized using standard pipelines. A precise position was not available from other high energy missions either. This motivated us to undertake localization of the GRB from CZTI data by using various satellite elements as an effective coded aperture mask.

\begin{figure}[!tbh]
\includegraphics[width=\columnwidth]{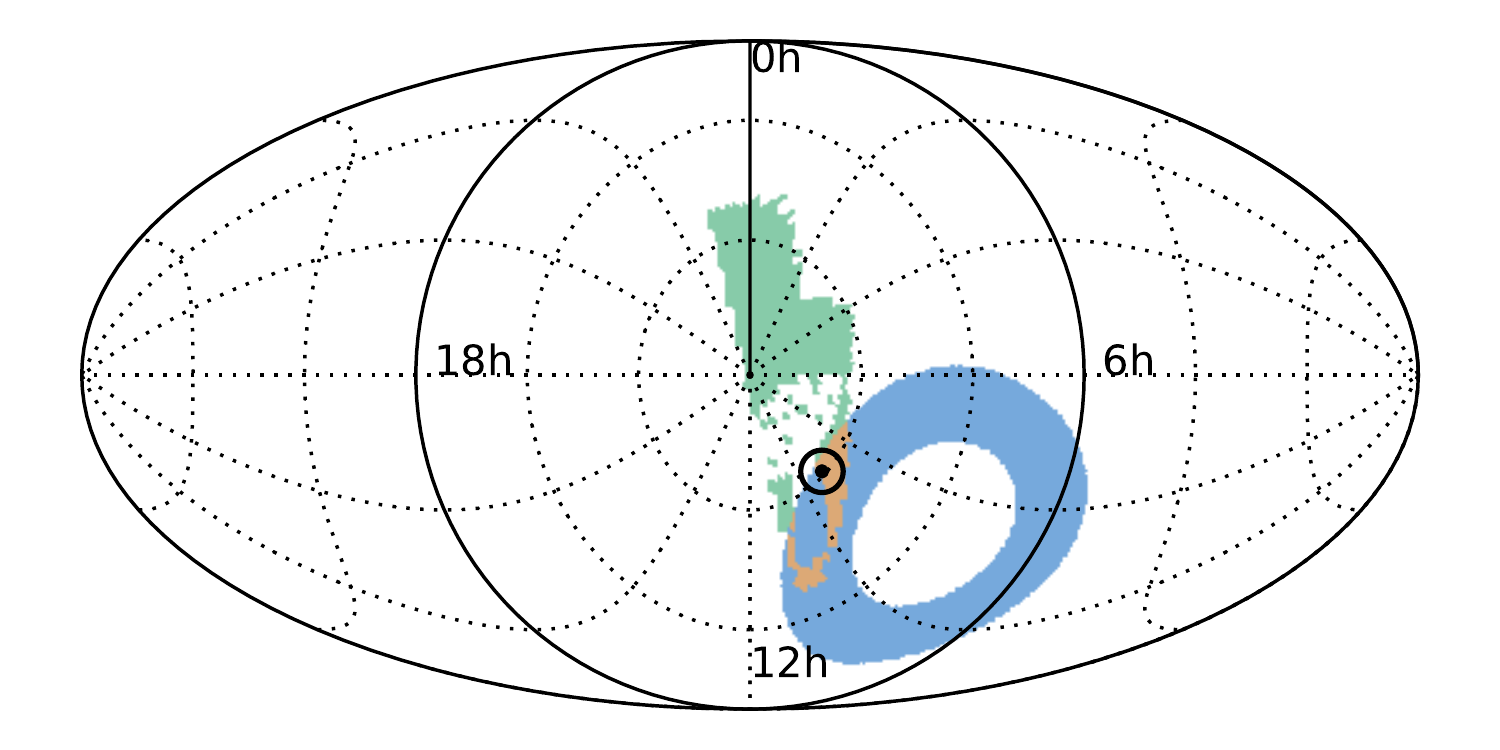}
\caption{Localization of GRB~170105A by CZTI and IPN. The 2$\sigma$ CZTI localization (green) spans 1148~\sqd, while the IPN annulus (blue) covers $\sim$2600~\sqd\ on the sky. The 192~\sqd\ common region (light brown) includes the location of ATLAS17aeu (black bulls-eye). 
}
\label{fig:loc}
\end{figure}

GRB~170105A was clearly detected in two of the four CZTI quadrants, with some scattered radiation seen in a third quadrant. The fact that we detected soft X-ray photons but no signal above $\sim$100~keV indicates that the photons had a relatively obscuration-free line of sight to quadrants A and B. However, lack of photons in quadrants C and D suggests obscuration by some satellite component in the line of sight, likely by the CZTI collimators themselves. Based on our experience with similar diagnostics for other GRBs, these criteria allow us to narrow down the GRB location to an octant of the sky. To further refine the localization, we used our raytrace code to calculate the ratio of count rates in quadrants A and B for photons incident at a representative energy of 50~keV. We selected the sky region where the counts ratio from these simulations is within $\pm2\sigma$ of the measured background-subtracted counts ratio. This constraint localizes the GRB to a 1148~\sqd\ area of the sky (Figure~\ref{fig:loc}). 

\citet{GW170104_IPN} used the Interplanetary Network (IPN) to localize GRB~170105A to a 2600~\sqd\ annulus on the sky. The CZTI and IPN localization regions have an overlap of 192~\sqd. These regions also have some area in common with the LIGO localization of GW170104. The probabilities that the GW source is contained in the CZTI, IPN and common regions are 6.0\%, 20.1\%, and 5.5\% respectively.

The position of ATLAS17aeu is consistent with the joint CZTI + IPN localization of GRB~170105A. Based on the temporal (\S\ref{subsec:atlas}) and spatial coincidence, we conclude that ATLAS17aeu is indeed the afterglow of GRB~170105A.

\floattable
\begin{deluxetable}{lCCCClCl}
\tablewidth{\textwidth}
\tablecaption{Optical observations of GRB~170105A\label{tab:obs_opt}}
\tablehead{\colhead{Time} & \colhead{$\Delta t_\mathrm{GW}$\tablenotemark{a}} & \colhead{$\Delta t_\mathrm{GRB}$\tablenotemark{b}} & \colhead{Filter} &\colhead{Mag\tablenotemark{c}} &\colhead{Telescope} &\colhead{Flux\tablenotemark{d}} &  \colhead{Reference} \\
\colhead{(MJD)} & \colhead{(days)} & \colhead{(days)} & & & & \colhead{($\mu$Jy)} &
}
\startdata
57758.0595 & 0.6345 & -0.2003 & g & $>$17.7 & SWASP/GOTO & $<$340 & \cite{GW170104_SWASP} \\
57758.0595 & 0.6345 & -0.2003 & r & $>$17.0 & SWASP/GOTO & $<$580 & ''\\
57758.0920 & 0.6670 & -0.1678 & g & $>$17.6 & SWASP/GOTO & $<$370 & ''\\
57758.0920 & 0.6670 & -0.1678 & r & $>$16.9 & SWASP/GOTO & $<$640 & ''\\
57758.2100 & 0.7850 & -0.0498 & g & $>$17.0 & SWASP/GOTO & $<$650 & '' \\ 
57758.2100 & 0.7850 & -0.0498 & r & $>$16.3 & SWASP/GOTO & $<$1120& ''\\ 
57758.3816 & 0.9566 & 0.1218 & i & 18.18 $\pm$ 0.04 & Pan-STARRS1 & 207 $\pm$6 & \cite{GW170104_Pan}\\
57758.4130 & 0.9880 & 0.1532 & cyan & 18.05 $\pm$ 0.09 & ATLAS & 242$\pm$17 & \cite{GW170104_ATLAS}\\
57758.4145 & 0.9895 & 0.1547 & cyan & 18.18 $\pm$ 0.1 & ATLAS & 214 $\pm$17 & ''\\
57758.4267 & 1.0017 & 0.1669 & cyan & 18.22 $\pm$ 0.1 & ATLAS & 207 $\pm$17 & ''\\
57758.4419 & 1.0169 & 0.1821 & cyan & 18.58 $\pm$ 0.13 & ATLAS & 148 $\pm$15 & ''\\
57758.4469 & 1.0219 & 0.1871 & cyan & 18.45 $\pm$ 0.11 & ATLAS & 167 $\pm$15 & ''\\
57758.4479 & 1.0229 & 0.1881 & cyan & 18.34 $\pm$ 0.11 & ATLAS & 185 $\pm$16 & ''\\
57758.4550 & 1.0300 & 0.1952 & cyan & 18.39 $\pm$ 0.11 & ATLAS & 177 $\pm$15 & ''\\
57758.4620 & 1.0370 & 0.2022 & cyan & 18.90 $\pm$ 0.18 & ATLAS & 148 $\pm$24 & ''\\
57759.4647 & 2.0397 & 1.2049 & i & 20.90 $\pm$ 0.12 & Pan-STARRS1 & 17 $\pm$2 & \cite{GW170104_Pan}\\
57759.9130 & 2.4880 & 1.6532 & r & $>$21.7 & Asiago & $<$8 & \cite{GW170104_asiago}\\
57760.3181 & 2.8931 & 2.0583 & g^\prime & 22.29 $\pm$ 0.03 & LFC/P200 & 4.9 $\pm$0.1 & This work, \cite{GW170104_Palo}\\
57760.3412 & 2.9162 & 2.0814 & i^\prime & 22.05 $\pm$ 0.06 & LFC/P200 & 5.8 $\pm$0.3 & ''\\
57760.4154 & 2.9904 & 2.1556 & g^\prime & $>$16.6 & Akeno/MITSuME & $<$590 & This work   \\
57760.4184 & 2.9934 & 2.1586 & R_c & $>$16.5 & Akeno/MITSuME & $<$530 & ''\\
57760.4556 & 3.0306 & 2.1958 & g^\prime & 22.47 $\pm$ 0.05 & DCT & 4.3 $\pm$0.2 & This work, \cite{GW170104_channel} \\
57760.4556 & 3.0306 & 2.1958 & r^\prime & 22.10 $\pm$ 0.04 & DCT & 5.4 $\pm$0.2 & ''\\
57760.4556 & 3.0306 & 2.1958 & i^\prime & 21.96 $\pm$ 0.04 & DCT & 6.4 $\pm$0.3 & ''\\
57760.5597 & 3.1347 & 2.2999 & I_c & $>$17.8 & Akeno/MITSuME & $<$200 & This work   \\
57760.6215 & 3.1965 & 2.3617 & V & $>$20.0 & Nanshan & $<$40 & \cite{GW170104_Xu} \\
57760.6219 & 3.1969 & 2.3621 & R & $>$21.5 & YFOSC & $<$4.3 & \cite{GW170104_GMG} \\
57761.5197 & 4.0947 & 3.2599 & r & 22.77 $\pm$ 0.17 & Gemini+GMOS & 2.9 $\pm$0.4 & \cite{GW170104_Pan} \\
57761.5917 & 4.1667 & 3.3319 & I & 22.5 $\pm$ 0.3 & TNG/DOLORES & 4 $\pm$1 & \cite{GW170104_TNG} \\
57761.7681 & 4.3431 & 3.5083 & White & $>$18.8 & 0.6/0.9m Schmidt & <110 & \cite{GW170104_Xu}\\
57770.3790 & 12.9540 & 12.1192 & J & $>$22.3 & WIRC/P200 & <10 & This work \\
57785.4027 & 27.9777 & 27.1429 & r^\prime & 23.99 $\pm$ 0.06 & DCT & 0.90\pm0.05  & This work \\
57828.1296 & 70.7046 & 69.8698 & r^\prime & $>$24.43 & DCT & <0.5 & This work \\ 
57828.1432 & 70.7182 & 69.8834 & i^\prime & $>$24.05 & DCT & <0.7 & This work \\ 
57828.1569 & 70.7319 & 69.8971 & z^\prime & $>$23.25 & DCT & <1.3 & This work \\ 
\enddata
\tablenotetext{a}{Difference between observation time and the GW170104 trigger (2017-01-04 10:11:58.599~UTC).}
\tablenotetext{b}{Difference between observation time and GRB170105A (2017-01-05~06:14:06~UT).}
\tablenotetext{c}{For non-detections, upper limits are 5$\sigma$, with the exception of a 2.5$\sigma$ limit for \cite{GW170104_asiago}.}
\tablenotetext{d}{Fluxes have been corrected for galactic extinction, $E(B-V) = 0.033$, using \citep{sfd98} values from the IRSA extinction calculator at \url{http://irsa.ipac.caltech.edu/applications/DUST/}.}
\tablecomments{LFC data around MJD~55760.32 are best fit with a power law, $f_\nu = 10^{-26} \times (h\nu)^{-1.2} {\rm~erg~cm}^{-2}{\rm~s}^{-1}{\rm~Hz}^{-1}$ where $h$ is the Planck constant and $\nu$ is frequency in Hz. This yields  $m_{\rm cyan} = 22.25\pm0.03$ and $F_\nu= 5.0 \pm 0.1~\mu$Jy. DCT data around MJD~55760.46 are best fit with a power law, $f_\nu = 10^{-30} \times (h\nu)^{-1.56} {\rm~erg~cm}^{-2}{\rm~s}^{-1}{\rm~Hz}^{-1}$, giving $m_{\rm cyan} = 22.33\pm0.05$ and $F_\nu = 4.5 \pm0.2~\mu$Jy.}
\end{deluxetable}

\subsection{X-ray properties}\label{subsec:xrayprop}
\begin{figure}[!tbh]
\includegraphics[width=\columnwidth]{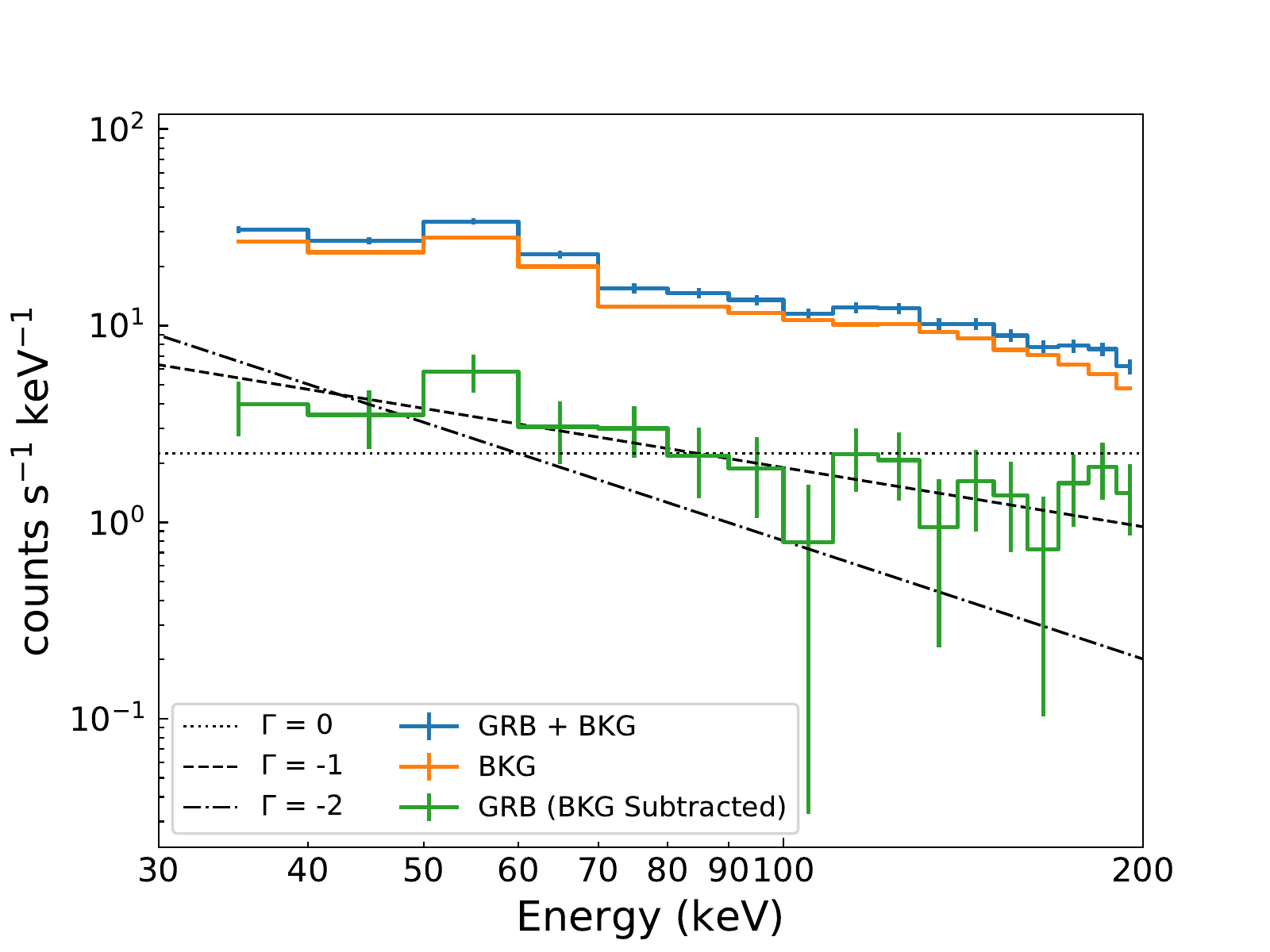}
\caption{CZTI spectrum of GRB~170105A. The total GRB + background spectrum (top, blue line) was extracted from a time window from $T_0$ - 1 s to $T_0$ + 20 s. The background spectrum (orange line) was extracted from a larger time window from $T_0$ - 596 s to $T_0$ - 96 s, where $T_0$ is the GRB~170105A trigger time (2017-01-05 06:14:06 UT). The difference (bottom, green line) shows the GRB spectrum, binned in 10 keV bins. Dotted, dashed and dot-dashed lines show power law spectra with photon index $\Gamma = $ 0, -1, and -2 respectively.}
\label{fig:xspec}
\end{figure}

GRB~170105A was outside the CZTI primary field of view, so we could not use the standard pipeline for spectral analysis. However, we can calculate some spectral properties of a GRB whose exact position is known, by estimating obscuration and scattering by various satellite elements along the line of sight. Based on arguments in \S\ref{subsec:loc}, we now use the position of ATLAS17aeu to calculate the GRB properties. 

We modeled the entire satellite in GEANT4 (Mate et al., in prep.), and simulated photons incident from the direction of ATLAS17aeu. GEANT4 accounts for absorption, fluorescence, and coherent as well as incoherent scattering to give the spatial and energy distribution of observed photons. We repeat these simulations for a range of energies from 20~keV to 2~MeV, taking note of photons in the final range of interest: 20--200~keV. Since the GEANT4 model does not include the intrinsic resolution of the detector, we broaden the derived spectra by a Gaussian kernel with a full width at half maximum of 6~keV. We note that in this method we currently overestimate the flux in various fluorescence lines in the 50--70~keV range, so we ignore this region in further analysis.

\begin{figure}[bht]
\includegraphics[width=0.9\columnwidth]{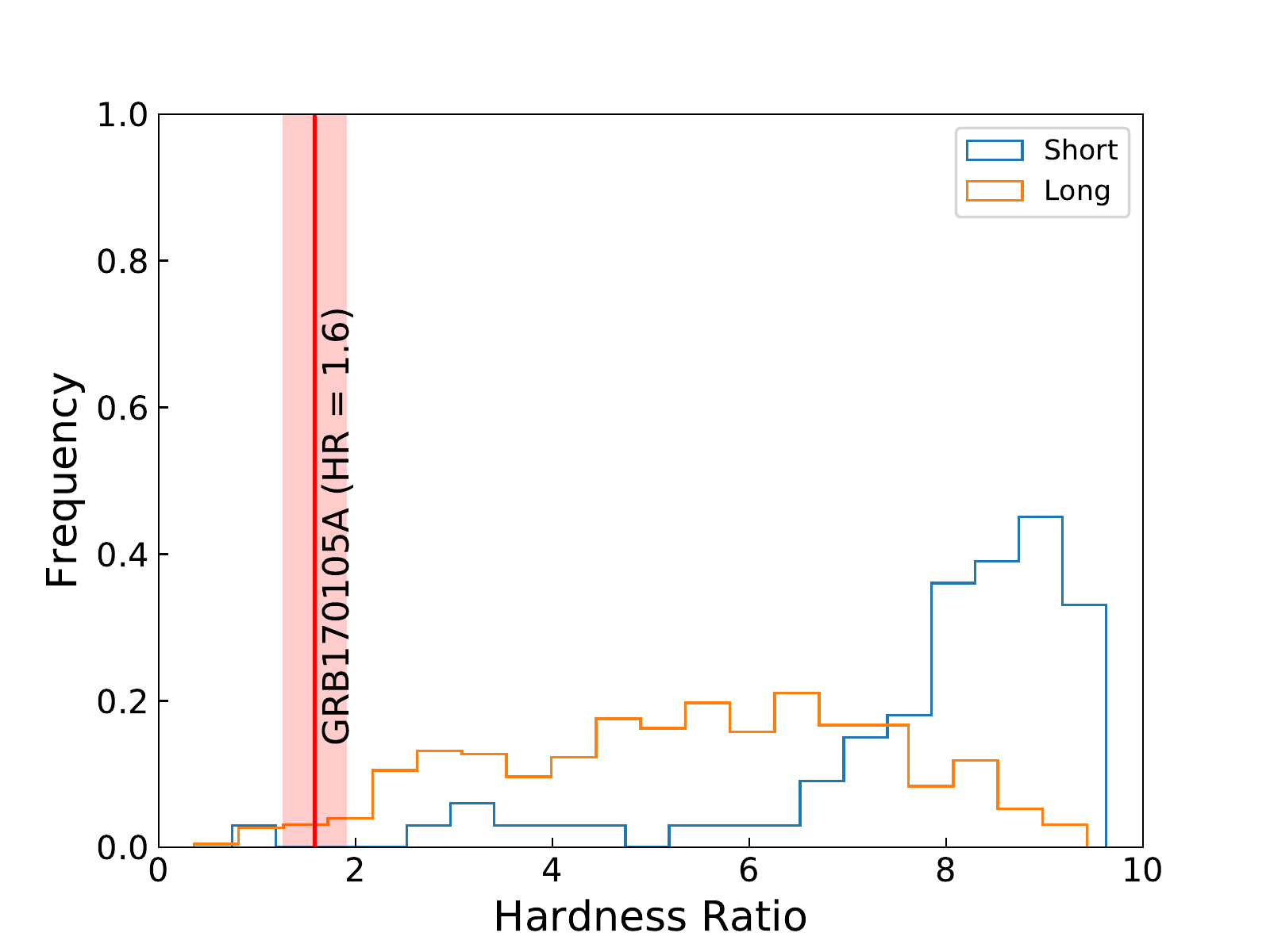}
\caption{Simulated (70--200~keV)/(20--50~keV) hardness ratio distributions for our GRB sample. We use GRB spectral parameters from the IceCube GRB Web, and simulate CZTI hardness ratios assuming that these GRBs were at the same position as ATLAS17aeu in the CZTI instrument reference frame. The red line and the shaded red region mark HR$= 1.6\pm0.3$ for GRB~170105A. We see that it is softer than most long and short GRBs. In particular, only one simulated short hard burst has a smaller hardness ratio than GRB~170105A.}
\label{fig:hardhist}
\end{figure}

Figure~\ref{fig:xspec} shows the observed spectrum of this GRB. The $\sim$1070 photons are not enough for a detailed spectral analysis. Instead, we define a hardness ratio (HR) as the ratio of counts in 70--200~keV to the counts in 20--50~keV. 
Using these bands, the HR of GRB~170105A is $1.6\pm0.3$. To put this in the context of other GRBs, we have to take into account the direction-dependent response of CZTI. The $T_{90}$ and HR analysis of all GRBs detected by CZTI will be reported elsewhere. Here, we take the more straightforward route of comparing it to the simulated HR of other GRBs, assuming they were in the same direction as GRB~170105A in the CZTI reference frame. Our sample comprises GRBs from the GRB Web service\footnote{\url{http://grbweb.icecube.wisc.edu/}} \citep{grbweb} in the time period from 1 January 2010 to 9 March 2017. This service conveniently tabulates the $T_{90}$ values and Band function spectral parameters for all GRBs. On examining the spectral properties, we find a large number of GRBs with peak energy exactly 200~keV, 205~keV, or 1000~keV, likely default values in the fit. We also find GRBs with negative values for $T_{90}$. We omit all these from further consideration, to get a final reference sample of 578 GRBs. We divide the remaining GRBs into short and long based on a cutoff value $T_{90}$=2.0~s. We then simulate the Band model spectra of these GRBs, fold them through our GEANT4 response, and calculate the HR for each. The resultant distribution of HRs shows that GRB~170105A is softer than most GRBs (Figure~\ref{fig:hardhist}), as noted in the raw spectrum itself. Such a soft spectrum is consistent with the expectations from a long GRB.

\subsection{Radio observations}\label{subsec:radio}
We observed the position of ATLAS17aeu with the VLA in its most extended configuration (A configuration) on three epochs~\citep{GW170104_VLA}, under our approved target of opportunity program\footnote{VLA/16A-206; PI: A. Corsi}. Our first two observations of ATLAS17aeu were carried out in C-band (nominal central frequency of $\approx 6$\,GHz). Our third and last observation spanned three bands (S-X-K$_{\rm u}$-bands) covering the frequency range 2.8-14\,GHz. We used J0921+6215 as the phase calibrator, and 3C286 or 3C48 as flux density and bandpass calibrators. VLA data were calibrated using the automated VLA calibration pipeline available in the Common Astronomy Software Applications package \citep[CASA;][]{McMullin2007}, and imaged using the CLEAN algorithm. Additional flagging was performed where needed after visual inspection of the calibrated data. Flux errors were calculated as the quadrature sum of the map R.M.S. noise, plus a $\approx 5\%$ fractional error on the measured flux which accounts for systematics in the absolute flux calibration. Combining all observations, we obtain the source position as RA = 09:13:13.91, Dec = +61:05:33.6 --- consistent with the optical position \citep[RA = 09:13:13.89, Dec = +61:05:33.6;][]{GW170104_ATLAS}. The source is point-like even at the highest resolution of 0.23$^{\prime\prime}$ full width at half power.  

ATLAS17aeu was also observed with AMI-LA~\citep{zbb+08} radio telescope between 08 Jan and 24 Jan 2017. Observations were made with the new digital correlator having 4096 channels across a 5 GHz bandwidth between 13--18~GHz. The nearby bright source J0921+6215 was observed every $\sim$10 minutes for complex gain calibration. The AMI-LA data were binned to 8$\times$0.625~GHz channels, and subsequently flagged for RFI excision and calibrated with a fully-automated pipeline, AMI-REDUCE~\citep[cf.][]{dtd+09,psg+13}. Daily measurements of 3C48 and 3C286 were used for the absolute flux calibration, which is accurate to about 10\%. The calibrated and RFI-flagged data were then imported into CASA and imaged on a 512$\times$512 square pixel grid (4 arcsec pix$^{-1}$), and  flux densities were measured using the {\it pyfits} module in Python.

\begin{figure}[bht]
\includegraphics[width=\columnwidth]{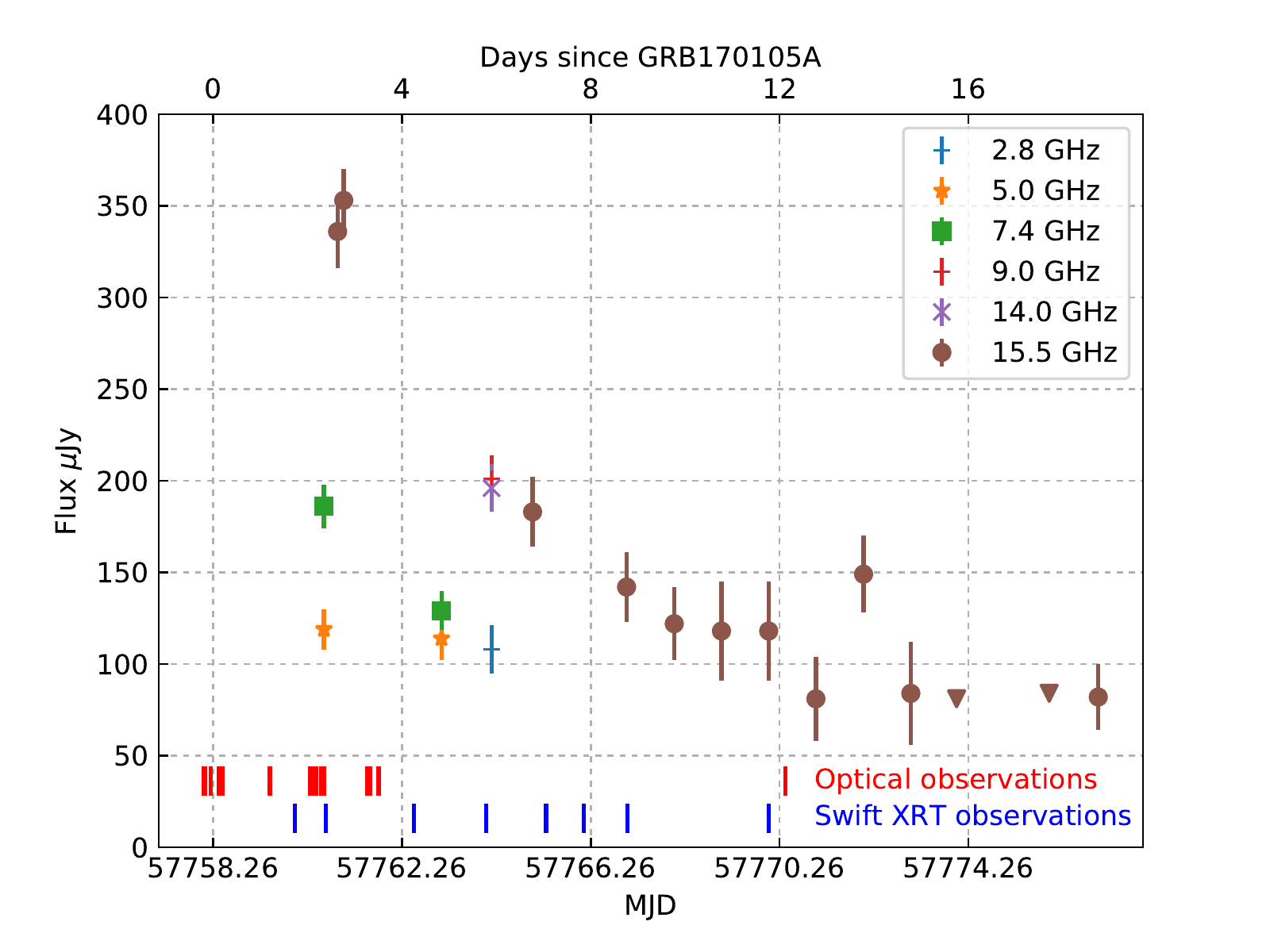}
\caption{Radio evolution of the afterglow of GRB~170105A. Data are given in Table~\ref{Tab:radio}. The bottom axis is in MJD, while the top axis shows time elapsed since GRB~170105A (MJD 57758.25980). Short red lines at the bottom indicate the epochs of optical observations (Table~\ref{tab:obs_opt}). Short blue lines indicate epochs of {\em Swift} XRT observations.}
\label{fig:radiolight}
\end{figure}

The results of our VLA and AMI follow-up of ATLAS17aeu are given in Table~\ref{Tab:radio} and Figure~\ref{fig:radiolight}.

\floattable
\begin{deluxetable}{ccrlc}
\tablewidth{\columnwidth}
\tablecaption{VLA and AMI observations of ATLAS17aeu.\label{Tab:radio}}
\tablehead{\colhead{Time} & \colhead{Instru-} &\colhead{Freq} &\colhead{Flux} &  \colhead{Reference}\\
\colhead{(MJD)} & \colhead{ment} & \colhead{(GHz)} & \colhead{$(\mu$Jy)}
}
\startdata
57760.61 & VLA &5.0 & $119\pm11$ & \citet{GW170104_VLA}\tablenotemark{a}\\
"         & VLA & 7.4 & $186\pm12$ & \citet{GW170104_VLA}\tablenotemark{a}\\
57760.90  & AMI   & 15.5 & $336\pm20$& \citet{GW170104_AMI}\\
57761.03 & AMI & 15.5 & 353 $\pm$ 17 & This work \\
57763.10 & VLA &5.0 & $114\pm12$ & ''\\
"         & VLA & 7.4 & $129\pm11$& ''\\
57764.16 & VLA &2.8 & $108\pm13$ & ''\\
"         & VLA &9.0 & $201\pm13$ & ''\\
"         & VLA &14  & $196\pm13$ & ''\\
57765.03 & AMI & 15.5 & 183 $\pm$ 19 & '' \\
57767.02 & AMI & 15.5 & 142 $\pm$ 19 & '' \\
57768.03 & AMI & 15.5 & 122 $\pm$ 20 & '' \\
57769.03 & AMI & 15.5 & 118 $\pm$ 27 & '' \\
57770.03 & AMI & 15.5 & 118 $\pm$ 27 & '' \\
57771.03 & AMI & 15.5 &  81 $\pm$ 23 & '' \\
57772.04 & AMI & 15.5 & 149 $\pm$ 21 & '' \\
57773.04 & AMI & 15.5 &  84 $\pm$ 28 & '' \\
57774.01 & AMI & 15.5 &  $<$ 81      & '' \\
57775.97 & AMI & 15.5 &  $<$ 84      & '' \\
57777.01 & AMI & 15.5 &  82 $\pm$ 18 & '' \\
\enddata
\tablenotetext{a}{These are updated values compared to the 6~GHz flux reported by \citet{GW170104_VLA}.}
\end{deluxetable} 

\subsection{Host Galaxy}
On UT 2017-01-07.5, two days after explosion, we obtained a low-resolution 
spectrum of ATLAS17aeu with the Palomar 200-inch Double Beam 
Spectrograph~\citep{og82} covering the wavelength range from 3300--10000~\AA. 
The spectrum, with an integration time of 60~minutes, shows no
significant absorption or emission features. Specifically, there is no 
Galactic H$\alpha$ emission around 6563~\AA\ to a 3$\sigma$ flux limit of 
${3.3\times10^{-17}}$~ergs~s$^{-1}$~cm$^{-2}$~\AA\,$^{-1}$, disfavoring a 
cataclysmic variable outburst as the origin.  
Continuum is detected at least as far blue as
3800\AA, placing an upper limit on the redshift of $z<3.2$ from the
absence of a Lyman break. This is consistent with the 
Gemini/GMOS spectrum reported by~\citet{GW170104_Pan}.

In late-time imaging LMI on the DCT, we do not detect any emission
\textit{at the location of ATLAS17aeu} --- limits directly underlying
the transient location measured with a 1\arcsec\ aperture are provided
in Table~\ref{tab:obs_opt}.  However, we identify a faint source offset from the location of
ATLAS17aeu by $\approx 1.5$\arcsec to the East, and measure magnitudes of
$r = 24.16 \pm 0.12$ and $i = 24.45 \pm 0.20$.  Using the formalism described
in \citet{pmm+12} and galaxy count rates from \citet{hpm+97}, we estimate an \textit{a posteriori} probability of chance alignment for this source $\approx 0.04$.  Thus it is a reasonable host candidate for ATLAS17aeu (see also \citealt{GW170104_GRAWITA}).  However, the measured offset of 1.5$^{\prime\prime}$ is much larger than is typical for GRB hosts at cosmological distances \citep{Bloom+2002,Blanchard+2015}, as GRB host galaxies tend to be quite compact.  
\citet{sts+17} detect a fainter source (``Galaxy~A'', $r = 25.59 \pm 0.16$) at the exact location of ATLAS17aeu, which may be the host galaxy. Our late time DCT imaging provides a marginal detection of this source, with magnitudes $r^\prime = 25.54 \pm 0.39, i^\prime = 25.15 \pm 0.38$, and $z^\prime = 24.40 \pm 0.41$. However, the measurements are low confidence (2.5$\sigma$), and we do not consider it a firm detection. Instead, we provide upper limits in Table~\ref{tab:obs_opt}.

Deep imaging of the afterglow location was obtained with the J-band filter 
on the NIRC2 camera on Keck~II, utilizing laser guide star adaptive optics 
corrections and obtaining 12 dithered 300\,s exposures. We do not detect the 
galaxy and measure a 5$\sigma$ point source limiting magnitude of J $>$22.8 (Vega mag) or J $>$ 23.7 (AB mag).
We caution, however, that this $J$-band limit is calculated for an unresolved point spread function --- brighter extended sources could be resolved out and remain undetected in our data.

\subsection{Afterglow properties}

\begin{figure}[!bth]
\includegraphics[width=\columnwidth]{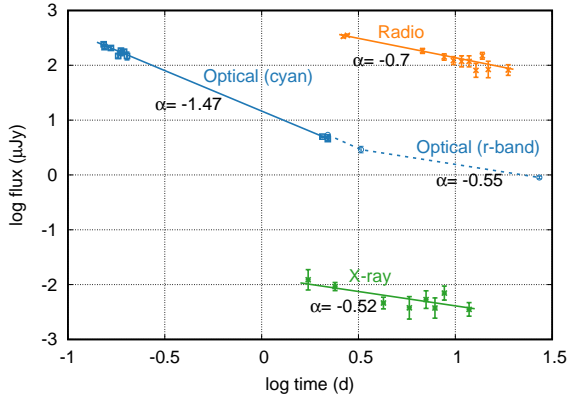}
\caption{Flux evolution of ATLAS17aeu, the afterglow of GRB 170105A. Radio (AMI) data are shown in orange. In green are shown X-ray fluxes at 1~keV, calculated from publicly available Swift-XRT data, adopting a count-rate to flux scaling from \citet{GW170104_SwiftXRT2}. Optical data are displayed in blue, with cyan band shown as squares and r-band as circles. The solid line through each data set represents the best-fit power-law decay with T0 fixed to the time of GRB 170105A (MJD 57758.259803).  The slopes of the fitted lines are quoted next to them.  The dashed line is not a fit, but just joins the r-band data points.}
\label{fig:allflux}
\end{figure}

\begin{figure}[hbt]
\includegraphics[width=\columnwidth]{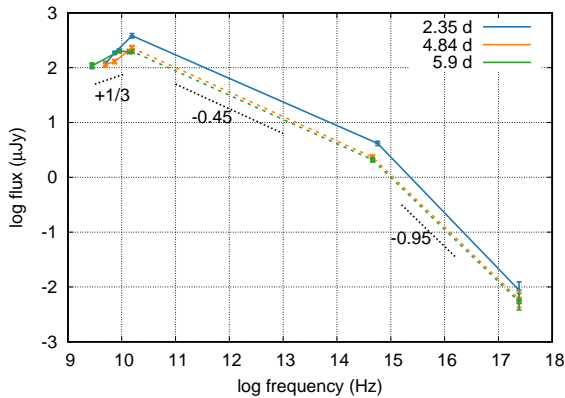}
\caption{Broadband spectral energy distributions (SED) of ATLAS17aeu, the afterglow of GRB~170105A. Different colors show the SED evaluated at the three epochs of VLA observations (Table~\ref{Tab:radio}). The 15.5~GHz, optical and X-ray points in this figure are obtained from the light curve fits shown in Figure~\ref{fig:allflux}. The optical flux at the first epoch is evaluated from the cyan band fit, while those as the latter two epochs are from the line joining the r-band points.  To guide the eye, dotted black lines with indicated spectral slopes are shown alongside the SED.}
\label{fig:sed}
\end{figure}

Light curves of ATLAS17aeu derived from radio (AMI), optical and Swift 
XRT data are shown in Figure~\ref{fig:allflux}. The broadband properties of the afterglow appear similar to those of other long GRBs (See \citealt{kkz+11,ebp+09,cf12} respectively for optical, X-ray and radio afterglow studies). The optical flux decays 
rapidly in the beginning, but appears to slow down significantly after 
$\sim 3$ days.  Radio and X-ray lightcurves, covering mostly this later 
period, also show similar, shallow decay slopes.  There could be many 
possible reasons for the flattening of the decay slope, including 
multi-component jets~\citep{berger+03}, spine-sheath emission~\citep{ramirez-ruiz+02}, late energy injection~\citep{zhang+02}, 
enhancement in ambient density~\citep{geng+14} etc, but the available data do not 
provide enough information to distinguish between them. There is no obvious 
signature of a jet break \citep{rhoads99} caused by the 
sideways expansion of the beamed afterglow.

Spectral energy distributions constructed from the VLA data, along with 
contemporaneous fluxes at higher frequencies evaluated from the light 
curves in Figure~\ref{fig:allflux} are shown in Figure~\ref{fig:sed}.  The overall spectral shape resembles that of synchrotron emission from a GRB afterglow.  However, the relatively sparse 
spectral and temporal coverage of our data cannot rule out the 
possibility of physically distinct emission regions (for example, a
forward and a reverse shock) being involved at 
different bands.

\section{Summary}
AstroSat CZTI covered 50.3\% of the GW170104 probability region on sky, but did not detect any temporally coincident excess hard X-ray emission. We calculate a flux upper limit of $4.6 \times 10^{-7}~{\rm erg~cm}^{-2}{\rm~s}^{-1}$ for any emission from this event at 1~s timescales.

We collected data from various optical telescopes worldwide and fit a power law to the optical lightcurve of ATLAS17aeu to discover that the explosion time of this transient is offset from the gravitational wave trigger by $21.1 \pm 1.1$ hours, but temporally consistent with GRB~170105A. Combining this information with AstroSat CZTI and IPN localisations of the GRB, we conclude that ATLAS17aeu is the afterglow of GRB~170105A. We examine the multi-wavelength observations of ATLAS17aeu in the context of standard afterglow models, and conclude that the observations are broadly consistent with typical long GRB afterglows.

Our effort demonstrates the advantage of having a wide network of instruments for electromagnetic followup of gravitational wave candidates. GRB~170105A was not detected by the sensitive {\em Swift}, Integral, or {\em Fermi} satellites --- but this keystone of our inference was obtained from the relatively new AstroSat CZTI and POLAR instruments. 
This underscores the importance of developing more broadband, truly all-sky monitors, lest we risk missing any interesting transients or electromagnetic counterparts to gravitational wave sources. 

The extensive multi-wavelength data obtained from this source demonstrates the efficacy of the GROWTH network for transient followup. 
In future observing runs of advanced gravitational wave detectors, such active collaboration among a geographically well-distributed network of telescopes with diverse capabilities should play a key role in the detection and characterisation of electromagnetic counterparts to GW sources. 

\section*{Acknowledgements}

CZT--Imager is built by a consortium of Institutes across India. The Tata Institute of Fundamental Research, Mumbai, led the effort with instrument design and development. Vikram Sarabhai Space Centre, Thiruvananthapuram provided the electronic design, assembly and testing. ISRO Satellite Centre (ISAC), Bengaluru provided the mechanical design, quality consultation and project management. The Inter University Centre for Astronomy and Astrophysics (IUCAA), Pune did the Coded Mask design, instrument calibration, and Payload Operation Centre. Space Application Centre (SAC) at Ahmedabad provided the analysis software. Physical Research Laboratory (PRL) Ahmedabad, provided the polarisation detection algorithm and ground calibration. A vast number of industries participated in the fabrication and the University sector pitched in by participating in the test and evaluation of the payload.
The Indian Space Research Organisation funded, managed and facilitated the project.

This work was supported by the GROWTH project funded by the National Science Foundation under Grant No 1545949. GROWTH is a collaborative project between California Institute of Technology (USA), Pomona College (USA), San Diego State University (USA), Los Alamos National Laboratory (USA), University of Maryland College Park (USA), University of Wisconsin Milwaukee (USA), Tokyo Institute of Technology (Japan), National Central University (Taiwan), Indian Institute of Astrophysics (India), Inter-University Center for Astronomy and Astrophysics (India), Weizmann Institute of Science (Israel), The Oskar Klein Centre at Stockholm University (Sweden), Humboldt University (Germany). The authors acknowledge the support of the Science and Engineering Research Board, Department of Science and Technology, India. 

These results made use of the Discovery Channel Telescope at Lowell Observatory. Lowell is a private, non-profit institution dedicated to astrophysical research and public appreciation of astronomy and operates the DCT in partnership with Boston University, the University of Maryland, the University of Toledo, Northern Arizona University and Yale University.  The Large Monolithic Imager was built by Lowell Observatory using funds provided by the National Science Foundation (AST-1005313).The author(s) also acknowledge the support of Japan Society for the Promotion of Science. We acknowledge support for MITSuME Telescope at Akeno by the Inter-University Research Program of the Institute for Cosmic Ray Research, the University of Tokyo. We thank Yoichi Yatsu, Yutaro Tachibana, Yoshihiko Saito, Kotaro Morita for help with Akeno observations. The AMI telescope gratefully acknowledges support from the European Research Council under grant ERC-2012-StG-307215 LODESTONE, the UK Science and Technology Facilities Council (STFC) and the University of Cambridge

A.C. acknowledges support from the National Science Foundation under CAREER Grant No. 1455090. V.B. acknowledges support from the INSPIRE program of the Department of Science and Technology, Government of India.	J. Mao is supported by the National Natural Science Foundation of China 11673062, 11661161010, the Hundred Talent Program of the Chinese Academy of Sciences, the Major Program of the Chinese Academy of Sciences (KJZD-EW-M06), and the Oversea Talent Program of Yunnan Province.

This research has made use of the NASA/ IPAC Infrared Science Archive, which is operated by the Jet Propulsion Laboratory, California Institute of Technology, under contract with the National Aeronautics and Space Administration. PyRAF is a product of the Space Telescope Science Institute, which is operated by AURA for NASA.

\software{Python \citep{python}, Astropy \citep{astropy}, Numpy \citep{numpy}, Scipy \citep{scipy},  DAOPHOT ~\citep{daophot}, FTOOLS \citep{ftools}, GEANT4 \citep{geant4}, CASA \citep{McMullin2007}, AMI-REDUCE \citep{dtd+09,psg+13}, IRAF \citep{iraf1,iraf2}.}

\bibliographystyle{apj}
\bibliography{grb170105.bib}

\begin{thebibliography}{102}
\expandafter\ifx\csname natexlab\endcsname\relax\def\natexlab#1{#1}\fi

\bibitem[{{Aab} {et~al.}(2016){Aab}, {Abreu}, {Aglietta}, {Al Samarai},
  {Albuquerque}, {Allekotte}, {Almela}, {Alvarez Castillo},
  {Alvarez-Mu{\~n}iz}, {Ambrosio}, \& et~al.}]{2016PhRvD..94l2007A}
{Aab}, A., {Abreu}, P., {Aglietta}, M., {Al Samarai}, I., {Albuquerque},
  I.~F.~M., {et~al.} 2016, \prd, 94, 122007

\bibitem[{Abbott {et~al.}(2016{\natexlab{a}})Abbott, Abbott, Abbott, Abernathy,
  Acernese, Ackley, Adams, Adams, Addesso, Adhikari, Adya, Affeldt, Agathos,
  Agatsuma, Aggarwal, Aguiar, Aiello, Ain, Ajith, Allen, Allocca, Altin,
  Anderson, Anderson, Arai, Araya, Arceneaux, Areeda, Arnaud, Arun, Ascenzi,
  Ashton, Ast, Aston, Astone, Aufmuth, Aulbert, Babak, Bacon, Bader, Baker,
  Baldaccini, Ballardin, Ballmer, Barayoga, Barclay, Barish, Barker, Barone,
  Barr, Barsotti, Barsuglia, Barta, Bartlett, Bartos, Bassiri, Basti, Batch,
  Baune, Bavigadda, Bazzan, Bejger, Bell, Berger, Bergmann, Berry, Bersanetti,
  Bertolini, Betzwieser, Bhagwat, Bhandare, Bilenko, Billingsley, Birch,
  Birney, Birnholtz, Biscans, Bisht, Bitossi, Biwer, Bizouard, Blackburn,
  Blair, Blair, Blair, Bloemen, Bock, Boer, Bogaert, Bogan, Bohe, Bond, Bondu,
  Bonnand, Boom, Bork, Boschi, Bose, Bouffanais, Bozzi, Bradaschia, Brady,
  Braginsky, Branchesi, Brau, Briant, Brillet, Brinkmann, Brisson, Brockill,
  Broida, Brooks, Brown, Brown, Brown, Brunett, Buchanan, Buikema, Bulik,
  Bulten, Buonanno, Buskulic, Buy, Byer, Cabero, Cadonati, Cagnoli, Cahillane,
  Bustillo, Callister, Calloni, Camp, Cannon, Cao, Capano, Capocasa,
  Carbognani, Caride, Diaz, Casentini, Caudill, Cavagli{\`{a}}, Cavalier,
  Cavalieri, Cella, Cepeda, Baiardi, Cerretani, Cesarini, Chamberlin, Chan,
  Chao, Charlton, Chassande-Mottin, Cheeseboro, Chen, Chen, Cheng, Chincarini,
  Chiummo, Cho, Cho, Chow, Christensen, Chu, Chua, Chung, Ciani, Clara, Clark,
  Cleva, Coccia, Cohadon, Colla, Collette, Cominsky, Constancio, Conte, Conti,
  Cook, Corbitt, Cornish, Corsi, Cortese, Costa, Coughlin, Coughlin, Coulon,
  Countryman, Couvares, Cowan, Coward, Cowart, Coyne, Coyne, Craig, Creighton,
  Cripe, Crowder, Cumming, Cunningham, Cuoco, Canton, Danilishin, D'Antonio,
  Danzmann, Darman, Dasgupta, Costa, Dattilo, Dave, Davier, Davies, Daw, Day,
  De, DeBra, Debreczeni, Degallaix, {De Laurentis}, Del'eglise, {Del Pozzo},
  Denker, Dent, Dergachev, {De Rosa}, DeRosa, DeSalvo, Devine, Dhurandhar,
  D{\'{i}}az, {Di Fiore}, {Di Giovanni}, {Di Girolamo}, {Di Lieto}, {Di Pace},
  {Di Palma}, {Di Virgilio}, Dolique, Donovan, Dooley, Doravari, Douglas,
  Downes, Drago, Drever, Driggers, Ducrot, Dwyer, Edo, Edwards, Effler,
  Eggenstein, Ehrens, Eichholz, Eikenberry, Engels, Essick, Etzel, Evans,
  Evans, Everett, Factourovich, Fafone, Fair, Fairhurst, Fan, Fang, Farinon,
  Farr, Farr, Favata, Fays, Fehrmann, Fejer, Fenyvesi, Ferrante, Ferreira,
  Ferrini, Fidecaro, Fiori, Fiorucci, Fisher, Flaminio, Fletcher, Fong,
  Fournier, Frasca, Frasconi, Frei, Freise, Frey, Frey, Fritschel, Frolov,
  Fulda, Fyffe, Gabbard, Gaebel, Gair, Gammaitoni, Gaonkar, Garufi, Gaur,
  Gehrels, Gemme, Geng, Genin, Gennai, George, Gergely, Germain, Ghosh, Ghosh,
  Ghosh, Giaime, Giardina, Giazotto, Gill, Glaefke, Goetz, Goetz, Gondan,
  Gonz{\'{a}}lez, Castro, Gopakumar, Gordon, Gorodetsky, Gossan, Gosselin,
  Gouaty, Grado, Graef, Graff, Granata, Grant, Gras, Gray, Greco, Green, Groot,
  Grote, Grunewald, Guidi, Guo, Gupta, Gupta, Gushwa, Gustafson, Gustafson,
  Hacker, Hall, Hall, Hamilton, Hammond, Haney, Hanke, Hanks, Hanna, Hannam,
  Hanson, Hardwick, Harms, Harry, Harry, Hart, Hartman, Haster, Haughian,
  Healy, Heidmann, Heintze, Heitmann, Hello, Hemming, Hendry, Heng, Hennig,
  Henry, Heptonstall, Heurs, Hild, Hoak, Hofman, Holt, Holz, Hopkins, Hough,
  Houston, Howell, Hu, Huang, Huerta, Huet, Hughey, Husa, Huttner, Huynh-Dinh,
  Indik, Ingram, Inta, Isa, Isac, Isi, Isogai, Iyer, Izumi, Jacqmin, Jang,
  Jani, Jaranowski, Jawahar, Jian, Jim{\'{e}}nez-Forteza, Johnson,
  Johnson-McDaniel, Jones, Jones, Jonker, Ju, K, Kalaghatgi, Kalogera,
  Kandhasamy, Kang, Kanner, Kapadia, Karki, Karvinen, Kasprzack, Katsavounidis,
  Katzman, Kaufer, Kaur, Kawabe, K{\'{e}}f{\'{e}}lian, Kehl, Keitel, Kelley,
  Kells, Kennedy, Key, Khalili, Khan, Khan, Khan, Khazanov, Kijbunchoo, Kim,
  Kim, Kim, Kim, Kim, Kim, Kim, Kimbrell, King, King, Kissel, Klein, Kleybolte,
  Klimenko, Koehlenbeck, Koley, Kondrashov, Kontos, Korobko, Korth, Kowalska,
  Kozak, Kringel, Krishnan, Kr{\'{o}}lak, Krueger, Kuehn, Kumar, Kumar, Kuo,
  Kutynia, Lackey, Landry, Lange, Lantz, Lasky, Laxen, Lazzarini, Lazzaro,
  Leaci, Leavey, Lebigot, Lee, Lee, Lee, Lee, Lenon, Leonardi, Leong, Leroy,
  Letendre, Levin, Lewis, Li, Libson, Littenberg, Lockerbie, Lombardi, London,
  Lord, Lorenzini, Loriette, Lormand, Losurdo, Lough, Lousto, L{\"{u}}ck,
  Lundgren, Lynch, Ma, Machenschalk, MacInnis, Macleod, Magana-Sandoval,
  Zertuche, Magee, Majorana, Maksimovic, Malvezzi, Man, Mandel, Mandic,
  Mangano, Mansell, Manske, Mantovani, Marchesoni, Marion, M{\'{a}}rka,
  M{\'{a}}rka, Markosyan, Maros, Martelli, Martellini, Martin, Martynov, Marx,
  Mason, Masserot, Massinger, Masso-Reid, Mastrogiovanni, Matichard, Matone,
  Mavalvala, Mazumder, McCarthy, McClelland, McCormick, McGuire, McIntyre,
  McIver, McManus, McRae, McWilliams, Meacher, Meadors, Meidam, Melatos,
  Mendell, Mercer, Merilh, Merzougui, Meshkov, Messenger, Messick, Metzdorff,
  Meyers, Mezzani, Miao, Michel, Middleton, Mikhailov, Milano, Miller, Miller,
  Miller, Miller, Millhouse, Minenkov, Ming, Mirshekari, Mishra, Mitra,
  Mitrofanov, Mitselmakher, Mittleman, Moggi, Mohan, Mohapatra, Montani, Moore,
  Moore, Moraru, Moreno, Morriss, Mossavi, Mours, Mow-Lowry, Mueller, Muir,
  Mukherjee, Mukherjee, Mukherjee, Mukund, Mullavey, Munch, Murphy, Murray,
  Mytidis, Nardecchia, Naticchioni, Nayak, Nedkova, Nelemans, Nelson, Neri,
  Neunzert, Newton, Nguyen, Nielsen, Nissanke, Nitz, Nocera, Nolting,
  Normandin, Nuttall, Oberling, Ochsner, O'Dell, Oelker, Ogin, Oh, Oh, Ohme,
  Oliver, Oppermann, Oram, O'Reilly, O'Shaughnessy, Ottaway, Overmier, Owen,
  Pai, Pai, Palamos, Palashov, Palomba, Pal-Singh, Pan, Pan, Pankow, Pannarale,
  Pant, Paoletti, Paoli, Papa, Paris, Parker, Pascucci, Pasqualetti,
  Passaquieti, Passuello, Patricelli, Patrick, Pearlstone, Pedraza, Pedurand,
  Pekowsky, Pele, Penn, Perreca, Perri, Pfeiffer, Phelps, Piccinni, Pichot,
  Piergiovanni, Pierro, Pillant, Pinard, Pinto, Pitkin, Poe, Poggiani,
  Popolizio, Porter, Post, Powell, Prasad, Predoi, Prestegard, Price,
  Prijatelj, Principe, Privitera, Prix, Prodi, Prokhorov, Puncken, Punturo,
  Puppo, P{\"{u}}rrer, Qi, Qin, Qiu, Quetschke, Quintero, Quitzow-James, Raab,
  Rabeling, Radkins, Raffai, Raja, Rajan, Rakhmanov, Rapagnani, Raymond,
  Razzano, Re, Read, Reed, Regimbau, Rei, Reid, Reitze, Rew, Reyes, Ricci,
  Riles, Rizzo, Robertson, Robie, Robinet, Rocchi, Rolland, Rollins, Roma,
  Romano, Romano, Romanov, Romie, Rosi{\'{n}}ska, Rowan, R{\"{u}}diger, Ruggi,
  Ryan, Sachdev, Sadecki, Sadeghian, Sakellariadou, Salconi, Saleem, Salemi,
  Samajdar, Sammut, Sanchez, Sandberg, Sandeen, Sanders, Sassolas,
  Sathyaprakash, Saulson, Sauter, Savage, Sawadsky, Schale, Schilling, Schmidt,
  Schmidt, Schnabel, Schofield, Sch{\"{o}}nbeck, Schreiber, Schuette, Schutz,
  Scott, Scott, Sellers, Sengupta, Sentenac, Sequino, Sergeev, Setyawati,
  Shaddock, Shaffer, Shahriar, Shaltev, Shapiro, Shawhan, Sheperd, Shoemaker,
  Shoemaker, Siellez, Siemens, Sieniawska, Sigg, Silva, Singer, Singer, Singh,
  Singh, Singhal, Sintes, Slagmolen, Smith, Smith, Smith, Son, Sorazu,
  Sorrentino, Souradeep, Srivastava, Staley, Steinke, Steinlechner,
  Steinlechner, Steinmeyer, Stephens, Stevenson, Stone, Strain, Straniero,
  Stratta, Strauss, Strigin, Sturani, Stuver, Summerscales, Sun, Sunil, Sutton,
  Swinkels, Szczepa{\'{n}}czyk, Tacca, Talukder, Tanner, T{\'{a}}pai, Tarabrin,
  Taracchini, Taylor, Theeg, Thirugnanasambandam, Thomas, Thomas, Thomas,
  Thorne, Thrane, Tiwari, Tiwari, Tokmakov, Toland, Tomlinson, Tonelli,
  Tornasi, Torres, Torrie, T{\"{o}}yr{\"{a}}, Travasso, Traylor, Trifir{\`{o}},
  Tringali, Trozzo, Tse, Turconi, Tuyenbayev, Ugolini, Unnikrishnan, Urban,
  Usman, Vahlbruch, Vajente, Valdes, Vallisneri, van Bakel, van Beuzekom,
  van~den Brand, Broeck, Vander-Hyde, van~der Schaaf, van Heijningen, van
  Veggel, Vardaro, Vass, Vas{\'{u}}th, Vaulin, Vecchio, Vedovato, Veitch,
  Veitch, Venkateswara, Verkindt, Vetrano, Vicer{\'{e}}, Vinciguerra, Vine,
  Vinet, Vitale, Vo, Vocca, Vorvick, Voss, Vousden, Vyatchanin, Wade, Wade,
  Wade, Walker, Wallace, Walsh, Wang, Wang, Wang, Wang, Wang, Ward, Warner,
  Was, Weaver, Wei, Weinert, Weinstein, Weiss, Wen, Wessels, Westphal, Wette,
  Whelan, Whitcomb, Whiting, Williams, Williamson, Willis, Willke, Wimmer,
  Winkler, Wipf, Wittel, Woan, Woehler, Worden, Wright, Wu, Wu, Yablon, Yam,
  Yamamoto, Yancey, Yu, Yvert, Zadro{\.{z}}ny, Zangrando, Zanolin, Zendri,
  Zevin, Zhang, Zhang, Zhang, Zhao, Zhou, Zhou, Zhu, Zucker, Zuraw, \&
  Zweizig}]{lsc16}
Abbott, B.~P., Abbott, R., Abbott, T.~D., Abernathy, M.~R., Acernese, F.,
  {et~al.} 2016{\natexlab{a}}, Physical Review X, 6, 041015

\bibitem[{Abbott {et~al.}(2016{\natexlab{b}})Abbott, Abbott, Abbott, Abernathy,
  Acernese, Ackley, Adams, Adams, Addesso, Adhikari, Adya, Affeldt, Agathos,
  Agatsuma, Aggarwal, Aguiar, Aiello, Ain, Ajith, Allen, Allocca, Altin,
  Anderson, Anderson, Arai, Araya, Arceneaux, Areeda, Arnaud, Arun, Ascenzi,
  Ashton, Ast, Aston, Astone, Aufmuth, Aulbert, Babak, Bacon, Bader, Baker,
  Baldaccini, Ballardin, Ballmer, Barayoga, Barclay, Barish, Barker, Barone,
  Barr, Barsotti, Barsuglia, Barta, Barthelmy, Bartlett, Bartos, Bassiri,
  Basti, Batch, Baune, Bavigadda, Bazzan, Behnke, Bejger, Bell, Bell, Berger,
  Bergman, Bergmann, Berry, Bersanetti, Bertolini, Betzwieser, Bhagwat,
  Bhandare, Bilenko, Billingsley, Birch, Birney, Biscans, Bisht, Bitossi,
  Biwer, Bizouard, Blackburn, Blair, Blair, Blair, Bloemen, Bock, Bodiya, Boer,
  Bogaert, Bogan, Bohe, Bojtos, Bond, Bondu, Bonnand, Boom, Bork, Boschi, Bose,
  Bouffanais, Bozzi, Bradaschia, Brady, Braginsky, Branchesi, Brau, Briant,
  Brillet, Brinkmann, Brisson, Brockill, Brooks, Brown, Brown, Brown, Buchanan,
  Buikema, Bulik, Bulten, Buonanno, Buskulic, Buy, Byer, Cadonati, Cagnoli,
  Cahillane, Bustillo, Callister, Calloni, Camp, Cannon, Cao, Capano, Capocasa,
  Carbognani, Caride, Diaz, Casentini, Caudill, Cavagli{\'{a}}, Cavalier,
  Cavalieri, Cella, Cepeda, Baiardi, Cerretani, Cesarini, Chakraborty,
  Chalermsongsak, Chamberlin, Chan, Chao, Charlton, Chassande-Mottin, Chen,
  Chen, Cheng, Chincarini, Chiummo, Cho, Cho, Chow, Christensen, Chu, Chua,
  Chung, Ciani, Clara, Clark, Cleva, Coccia, Cohadon, Colla, Collette,
  Cominsky, {Constancio, M.}, Conte, Conti, Cook, Corbitt, Cornish, Corsi,
  Cortese, Costa, Coughlin, Coughlin, Coulon, Countryman, Couvares, Cowan,
  Coward, Cowart, Coyne, Coyne, Craig, Creighton, Cripe, Crowder, Cumming,
  Cunningham, Cuoco, {Dal Canton}, Danilishin, D'Antonio, Danzmann, Darman,
  Dattilo, Dave, Daveloza, Davier, Davies, Daw, Day, DeBra, Debreczeni,
  Degallaix, {De Laurentis}, Del{\'{e}}glise, {Del Pozzo}, Denker, Dent,
  Dereli, Dergachev, DeRosa, {De Rosa}, DeSalvo, Dhurandhar, D{\'{i}}az, {Di
  Fiore}, {Di Giovanni}, {Di Lieto}, {Di Pace}, {Di Palma}, {Di Virgilio},
  Dojcinoski, Dolique, Donovan, Dooley, Doravari, Douglas, Downes, Drago,
  Drever, Driggers, Du, Ducrot, Dwyer, Edo, Edwards, Effler, Eggenstein,
  Ehrens, Eichholz, Eikenberry, Engels, Essick, Etzel, Evans, Evans, Everett,
  Factourovich, Fafone, Fair, Fairhurst, Fan, Fang, Farinon, Farr, Farr,
  Favata, Fays, Fehrmann, Fejer, Ferrante, Ferreira, Ferrini, Fidecaro, Fiori,
  Fiorucci, Fisher, Flaminio, Fletcher, Fournier, Franco, Frasca, Frasconi,
  Frei, Freise, Frey, Frey, Fricke, Fritschel, Frolov, Fulda, Fyffe, Gabbard,
  Gair, Gammaitoni, Gaonkar, Garufi, Gatto, Gaur, Gehrels, Gemme, Gendre,
  Genin, Gennai, George, Gergely, Germain, Ghosh, Ghosh, Giaime, Giardina,
  Giazotto, Gill, Glaefke, Goetz, Goetz, Gondan, Gonz{\'{a}}lez, Castro,
  Gopakumar, Gordon, Gorodetsky, Gossan, Gosselin, Gouaty, Graef, Graff,
  Granata, Grant, Gras, Gray, Greco, Green, Groot, Grote, Grunewald, Guidi,
  Guo, Gupta, Gupta, Gushwa, Gustafson, Gustafson, Hacker, Hall, Hall, Hammond,
  Haney, Hanke, Hanks, Hanna, Hannam, Hanson, Hardwick, Haris, Harms, Harry,
  Harry, Hart, Hartman, Haster, Haughian, Heidmann, Heintze, Heitmann, Hello,
  Hemming, Hendry, Heng, Hennig, Heptonstall, Heurs, Hild, Hoak, Hodge, Hofman,
  Hollitt, Holt, Holz, Hopkins, Hosken, Hough, Houston, Howell, Hu, Huang,
  Huerta, Huet, Hughey, Husa, Huttner, Huynh-Dinh, Idrisy, Indik, Ingram, Inta,
  Isa, Isac, Isi, Islas, Isogai, Iyer, Izumi, Jacqmin, Jang, Jani, Jaranowski,
  Jawahar, Jim{\'{e}}nez-Forteza, Johnson, Jones, Jones, Jonker, Ju,
  Kalaghatgi, Kalogera, Kandhasamy, Kang, Kanner, Karki, Kasprzack,
  Katsavounidis, Katzman, Kaufer, Kaur, Kawabe, Kawazoe, K{\'{e}}f{\'{e}}lian,
  Kehl, Keitel, Kelley, Kells, Kennedy, Key, Khalaidovski, Khalili, Khan, Khan,
  Khan, Khazanov, Kijbunchoo, Kim, Kim, Kim, Kim, Kim, Kim, King, King, Kinzel,
  Kissel, Kleybolte, Klimenko, Koehlenbeck, Kokeyama, Koley, Kondrashov,
  Kontos, Korobko, Korth, Kowalska, Kozak, Kringel, Kr{\'{o}}lak, Krueger,
  Kuehn, Kumar, Kuo, Kutynia, Lackey, Landry, Lange, Lantz, Lasky, Lazzarini,
  Lazzaro, Leaci, Leavey, Lebigot, Lee, Lee, Lee, Lee, Lenon, Leonardi, Leong,
  Leroy, Letendre, Levin, Levine, Li, Libson, Littenberg, Lockerbie, Logue,
  Lombardi, Lord, Lorenzini, Loriette, Lormand, Losurdo, Lough, L{\"{u}}ck,
  Lundgren, Luo, Lynch, Ma, MacDonald, Machenschalk, MacInnis, Macleod,
  Maga{\~{n}}a-Sandoval, Magee, Mageswaran, Majorana, Maksimovic, Malvezzi,
  Man, Mandel, Mandic, Mangano, Mansell, Manske, Mantovani, Marchesoni, Marion,
  M{\'{a}}rka, M{\'{a}}rka, Markosyan, Maros, Martelli, Martellini, Martin,
  Martin, Martynov, Marx, Mason, Masserot, Massinger, Masso-Reid, Matichard,
  Matone, Mavalvala, Mazumder, Mazzolo, McCarthy, McClelland, McCormick,
  McGuire, McIntyre, McIver, McManus, McWilliams, Meacher, Meadors, Meidam,
  Melatos, Mendell, Mendoza-Gandara, Mercer, Merilh, Merzougui, Meshkov,
  Messenger, Messick, Meyers, Mezzani, Miao, Michel, Middleton, Mikhailov,
  Milano, Miller, Millhouse, Minenkov, Ming, Mirshekari, Mishra, Mitra,
  Mitrofanov, Mitselmakher, Mittleman, Moggi, Mohan, Mohapatra, Montani, Moore,
  Moore, Moraru, Moreno, Morriss, Mossavi, Mours, Mow-Lowry, Mueller, Mueller,
  Muir, Mukherjee, Mukherjee, Mukherjee, Mukund, Mullavey, Munch, Murphy,
  Murray, Mytidis, Nardecchia, Naticchioni, Nayak, Necula, Nedkova, Nelemans,
  Neri, Neunzert, Newton, Nguyen, Nielsen, Nissanke, Nitz, Nocera, Nolting,
  Normandin, Nuttall, Oberling, Ochsner, O'Dell, Oelker, Ogin, Oh, Oh, Ohme,
  Oliver, Oppermann, Oram, O'Reilly, O'Shaughnessy, Ottaway, Ottens, Overmier,
  Owen, Pai, Pai, Palamos, Palashov, Palliyaguru, Palomba, Pal-Singh, Pan,
  Pankow, Pannarale, Pant, Paoletti, Paoli, Papa, Paris, Parker, Pascucci,
  Pasqualetti, Passaquieti, Passuello, Patricelli, Patrick, Pearlstone,
  Pedraza, Pedurand, Pekowsky, Pele, Penn, Perreca, Phelps, Piccinni, Pichot,
  Piergiovanni, Pierro, Pillant, Pinard, Pinto, Pitkin, Poggiani, Popolizio,
  Post, Powell, Prasad, Predoi, Premachandra, Prestegard, Price, Prijatelj,
  Principe, Privitera, Prodi, Prokhorov, Puncken, Punturo, Puppo, P{\"{u}}rrer,
  Qi, Qin, Quetschke, Quintero, Quitzow-James, Raab, Rabeling, Radkins, Raffai,
  Raja, Rakhmanov, Rapagnani, Raymond, Razzano, Re, Read, Reed, Regimbau, Rei,
  Reid, Reitze, Rew, Reyes, Ricci, Riles, Robertson, Robie, Robinet, Rocchi,
  Rolland, Rollins, Roma, Romano, Romanov, Romie, Rosi{\'{n}}ska, Rowan,
  R{\"{u}}diger, Ruggi, Ryan, Sachdev, Sadecki, Sadeghian, Salconi, Saleem,
  Salemi, Samajdar, Sammut, Sanchez, Sandberg, Sandeen, Sanders, Sassolas,
  Sathyaprakash, Saulson, Sauter, Savage, Sawadsky, Schale, Schilling, Schmidt,
  Schmidt, Schnabel, Schofield, Sch{\"{o}}nbeck, Schreiber, Schuette, Schutz,
  Scott, Scott, Sellers, Sentenac, Sequino, Sergeev, Serna, Setyawati, Sevigny,
  Shaddock, Shah, Shahriar, Shaltev, Shao, Shapiro, Shawhan, Sheperd,
  Shoemaker, Shoemaker, Siellez, Siemens, Sigg, Silva, Simakov, Singer, Singh,
  Singh, Singhal, Sintes, Slagmolen, Smith, Smith, Smith, Son, Sorazu,
  Sorrentino, Souradeep, Srivastava, Staley, Steinke, Steinlechner,
  Steinlechner, Steinmeyer, Stephens, Stone, Strain, Straniero, Stratta,
  Strauss, Strigin, Sturani, Stuver, Summerscales, Sun, Sutton, Swinkels,
  Szczepa{\'{n}}czyk, Tacca, Talukder, Tanner, T{\'{a}}pai, Tarabrin,
  Taracchini, Taylor, Theeg, Thirugnanasambandam, Thomas, Thomas, Thomas,
  Thorne, Thorne, Thrane, Tiwari, Tiwari, Tokmakov, Tomlinson, Tonelli, Torres,
  Torrie, T{\"{o}}yr{\"{a}}, Travasso, Traylor, Trifir{\`{o}}, Tringali,
  Trozzo, Tse, Turconi, Tuyenbayev, Ugolini, Unnikrishnan, Urban, Usman,
  Vahlbruch, Vajente, Valdes, van Bakel, van Beuzekom, van~den Brand, {Van Den
  Broeck}, Vander-Hyde, van~der Schaaf, van Heijningen, van Veggel, Vardaro,
  Vass, Vas{\'{u}}th, Vaulin, Vecchio, Vedovato, Veitch, Veitch, Venkateswara,
  Verkindt, Vetrano, Vicer{\'{e}}, Vinciguerra, Vine, Vinet, Vitale, Vo, Vocca,
  Vorvick, Voss, Vousden, Vyatchanin, Wade, Wade, Wade, Walker, Wallace, Walsh,
  Wang, Wang, Wang, Wang, Wang, Ward, Warner, Was, Weaver, Wei, Weinert,
  Weinstein, Weiss, Welborn, Wen, We{\ss}els, Westphal, Wette, Whelan, White,
  Whiting, Williams, Williamson, Willis, Willke, Wimmer, Winkler, Wipf, Wittel,
  Woan, Worden, Wright, Wu, Yablon, Yam, Yamamoto, Yancey, Yap, Yu, Yvert,
  Zadro{\.{z}}ny, Zangrando, Zanolin, Zendri, Zevin, Zhang, Zhang, Zhang,
  Zhang, Zhao, Zhou, Zhou, Zhu, Zucker, Zuraw, Zweizig, Collaboration,
  Collaboration, Allison, Bannister, Bell, Chatterjee, Chippendale, Edwards,
  Harvey-Smith, Heywood, Hotan, Indermuehle, Marvil, McConnell, Murphy,
  Popping, Reynolds, Sault, Voronkov, Whiting, Collaboration, Castro-Tirado,
  Cunniffe, Jel{\'{i}}nek, Tello, Oates, Hu, Kub{\'{a}}nek, Guziy,
  Castell{\'{o}}n, Garc{\'{i}}a-Cerezo, Mu{\~{n}}oz, {P{\'{e}}rez del Pulgar},
  Castillo-Carri{\'{o}}n, {Castro Cer{\'{o}}n}, Hudec, Caballero-Garc{\'{i}}a,
  P{\'{a}}ta, Vitek, Adame, Konig, Rend{\'{o}}n, {Mateo Sanguino},
  Fern{\'{a}}ndez-Mu{\~{n}}oz, Yock, Rattenbury, Allen, Querel, Jeong, Park,
  Bai, Cui, Fan, Wang, Hiriart, Lee, Claret, S{\'{a}}nchez-Ram{\'{i}}rez,
  Pandey, Mediavilla, Sabau-Graziati, Collaboration, Abbott, Abdalla, Allam,
  Annis, Armstrong, Benoit-L{\'{e}}vy, Berger, Bernstein, Bertin, Brout,
  Buckley-Geer, Burke, Capozzi, Carretero, Castander, Chornock, Cowperthwaite,
  Crocce, Cunha, D'Andrea, da~Costa, Desai, Diehl, Dietrich, Doctor,
  Drlica-Wagner, Drout, Eifler, Estrada, Evrard, Fernandez, Finley, Flaugher,
  Foley, Fong, Fosalba, Fox, Frieman, Fryer, Gaztanaga, Gerdes, Goldstein,
  Gruen, Gruendl, Gutierrez, Herner, Honscheid, James, Johnson, Johnson,
  Karliner, Kasen, Kent, Kessler, Kim, Kind, Kuehn, Kuropatkin, Lahav, Li,
  Lima, Lin, Maia, Margutti, Marriner, Martini, Matheson, Melchior, Metzger,
  Miller, Miquel, Neilsen, Nichol, Nord, Nugent, Ogando, Petravick, Plazas,
  Quataert, Roe, Romer, Roodman, Rosell, Rykoff, Sako, Sanchez, Scarpine,
  Schindler, Schubnell, Scolnic, Sevilla-Noarbe, Sheldon, Smith, Smith,
  Soares-Santos, Sobreira, Stebbins, Suchyta, Swanson, Tarle, Thaler, Thomas,
  Thomas, Tucker, Vikram, Walker, Wechsler, Wester, Yanny, Zhang, Zuntz,
  Survey, Collaborations, Connaughton, Burns, Goldstein, Briggs, Zhang, Hui,
  Jenke, Wilson-Hodge, Bhat, Bissaldi, Cleveland, Fitzpatrick, Giles, Gibby,
  Greiner, von Kienlin, Kippen, McBreen, Mailyan, Meegan, Paciesas, Preece,
  Roberts, Sparke, Stanbro, Toelge, Veres, Yu, Blackburn, Collaboration,
  Ackermann, Ajello, Albert, Anderson, Atwood, Axelsson, Baldini, Barbiellini,
  Bastieri, Bellazzini, Bissaldi, Blandford, Bloom, Bonino, Bottacini, Brandt,
  Bruel, Buson, Caliandro, Cameron, Caragiulo, Caraveo, Cavazzuti, Charles,
  Chekhtman, Chiang, Chiaro, Ciprini, Cohen-Tanugi, Cominsky, Costanza, Cuoco,
  D'Ammando, de~Palma, Desiante, Digel, {Di Lalla}, {Di Mauro}, {Di Venere},
  Dom{\'{i}}nguez, Drell, Dubois, Favuzzi, Ferrara, Franckowiak, Fukazawa,
  Funk, Fusco, Gargano, Gasparrini, Giglietto, Giommi, Giordano, Giroletti,
  Glanzman, Godfrey, Gomez-Vargas, Green, Grenier, Grove, Guiriec, Hadasch,
  Harding, Hays, Hewitt, Hill, Horan, Jogler, J{\'{o}}hannesson, Johnson,
  Kensei, Kocevski, Kuss, {La Mura}, Larsson, Latronico, Li, Li, Longo,
  Loparco, Lovellette, Lubrano, Magill, Maldera, Manfreda, Marelli, Mayer,
  Mazziotta, McEnery, Meyer, Michelson, Mirabal, Mizuno, Moiseev, Monzani,
  Moretti, Morselli, Moskalenko, Negro, Nuss, Ohsugi, Omodei, Orienti, Orlando,
  Ormes, Paneque, Perkins, Pesce-Rollins, Piron, Pivato, Porter, Racusin,
  Rain{\`{o}}, Rando, Razzaque, Reimer, Reimer, Salvetti, {Saz Parkinson},
  Sgr{\`{o}}, Simone, Siskind, Spada, Spandre, Spinelli, Suson, Tajima, Thayer,
  Thompson, Tibaldo, Torres, Troja, Uchiyama, Venters, Vianello, Wood, Wood,
  Zhu, Zimmer, Collaboration, Brocato, Cappellaro, Covino, Grado, Nicastro,
  Palazzi, Pian, Amati, Antonelli, Capaccioli, D'Avanzo, D'Elia, Getman,
  Giuffrida, Iannicola, Limatola, Lisi, Marinoni, Marrese, Melandri,
  Piranomonte, Possenti, Pulone, Rossi, Stamerra, Stella, Testa, Tomasella,
  Yang, (GRAWITA), Bazzano, Bozzo, Brandt, Courvoisier, Ferrigno, Hanlon,
  Kuulkers, Laurent, Mereghetti, Roques, Savchenko, Ubertini, Collaboration,
  Kasliwal, Singer, Cao, Duggan, Kulkarni, Bhalerao, Miller, Barlow, Bellm,
  Manulis, Rana, Laher, Masci, Surace, Rebbapragada, Cook, {Van Sistine},
  Sesar, Perley, Ferreti, Prince, Kendrick, Horesh, Collaboration, Hurley,
  Golenetskii, Aptekar, Frederiks, Svinkin, Rau, von Kienlin, Zhang, Smith,
  Cline, Krimm, Network, Abe, Doi, Fujisawa, Kawabata, Morokuma, Motohara,
  Tanaka, Ohta, Yanagisawa, Yoshida, Collaboration, Baltay, Rabinowitz, Ellman,
  Rostami, Survey, Bersier, Bode, Collins, Copperwheat, Darnley, Galloway,
  Gomboc, Kobayashi, Mazzali, Mundell, Piascik, Pollacco, Steele, Ulaczyk,
  Collaboration, Broderick, Fender, Jonker, Rowlinson, Stappers, Wijers,
  Collaboration, Lipunov, Gorbovskoy, Tyurina, Kornilov, Balanutsa, Kuznetsov,
  Buckley, Rebolo, Serra-Ricart, Israelian, Budnev, Gress, Ivanov, Poleshuk,
  Tlatov, Yurkov, Collaboration, Kawai, Serino, Negoro, Nakahira, Mihara,
  Tomida, Ueno, Tsunemi, Matsuoka, Collaboration, Croft, Feng, Franzen,
  Gaensler, Johnston-Hollitt, Kaplan, Morales, Tingay, Wayth, Williams,
  Collaboration, Smartt, Chambers, Smith, Huber, Young, Wright, Schultz,
  Denneau, Flewelling, Magnier, Primak, Rest, Sherstyuk, Stalder, Stubbs,
  Tonry, Waters, Willman, Collaboration, {Olivares E.}, Campbell, Kotak,
  Sollerman, Smith, Dennefeld, Anderson, Botticella, Chen, Valle, Elias-Rosa,
  Fraser, Inserra, Kankare, Kupfer, Harmanen, Galbany, {Le Guillou}, Lyman,
  Maguire, Mitra, Nicholl, Razza, Terreran, Valenti, Gal-Yam, Collaboration,
  {\'{C}}wiek, {\'{C}}wiok, Mankiewicz, Opiela, Zaremba, {\.{Z}}arnecki,
  Collaboration, Onken, Scalzo, Schmidt, Wolf, Yuan, Collaboration, Evans,
  Kennea, Burrows, Campana, Cenko, Giommi, Marshall, Nousek, O'Brien, Osborne,
  Palmer, Perri, Siegel, Tagliaferri, Collaboration, Klotz, Turpin, Laugier,
  Collaboration, Collaboration, Collaboration, Collaboration, Beroiz,
  Pe{\~{n}}uela, Macri, Oelkers, Lambas, Vrech, Cabral, Colazo, Dominguez,
  Sanchez, Gurovich, Lares, Marshall, DePoy, Padilla, Pereyra, Benacquista,
  Collaboration, Tanvir, Wiersema, Levan, Steeghs, Hjorth, Fynbo, Malesani,
  Milvang-Jensen, Watson, Irwin, Fernandez, McMahon, Banerji, Gonzalez-Solares,
  Schulze, {de Ugarte Postigo}, Thoene, Cano, Rosswog, \&
  Collaboration}]{aaa+16}
---. 2016{\natexlab{b}}, ApJL, 826, L13

\bibitem[{Abbott {et~al.}(2016{\natexlab{c}})Abbott, Abbott, Abbott, Abernathy,
  Acernese, Ackley, Adams, Adams, Addesso, Adhikari, Adya, Affeldt, Agathos,
  Agatsuma, Aggarwal, Aguiar, Aiello, Ain, Ajith, Allen, Allocca, Altin,
  Anderson, Anderson, Arai, Araya, Arceneaux, Areeda, Arnaud, Arun, Ascenzi,
  Ashton, Ast, Aston, Astone, Aufmuth, Aulbert, Babak, Bacon, Bader, Baker,
  Baldaccini, Ballardin, Ballmer, Barayoga, Barclay, Barish, Barker, Barone,
  Barr, Barsotti, Barsuglia, Barta, Barthelmy, Bartlett, Bartos, Bassiri,
  Basti, Batch, Baune, Bavigadda, Bazzan, Behnke, Bejger, Bell, Bell, Berger,
  Bergman, Bergmann, Berry, Bersanetti, Bertolini, Betzwieser, Bhagwat,
  Bhandare, Bilenko, Billingsley, Birch, Birney, Biscans, Bisht, Bitossi,
  Biwer, Bizouard, Blackburn, Blair, Blair, Blair, Bloemen, Bock, Bodiya, Boer,
  Bogaert, Bogan, Bohe, Bojtos, Bond, Bondu, Bonnand, Boom, Bork, Boschi, Bose,
  Bouffanais, Bozzi, Bradaschia, Brady, Braginsky, Branchesi, Brau, Briant,
  Brillet, Brinkmann, Brisson, Brockill, Brooks, Brown, Brown, Brown, Buchanan,
  Buikema, Bulik, Bulten, Buonanno, Buskulic, Buy, Byer, Cadonati, Cagnoli,
  Cahillane, Bustillo, Callister, Calloni, Camp, Cannon, Cao, Capano, Capocasa,
  Carbognani, Caride, Diaz, Casentini, Caudill, Cavagli{\`{a}}, Cavalier,
  Cavalieri, Cella, Cepeda, Baiardi, Cerretani, Cesarini, Chakraborty,
  Chalermsongsak, Chamberlin, Chan, Chao, Charlton, Chassande-Mottin, Chen,
  Chen, Cheng, Chincarini, Chiummo, Cho, Cho, Chow, Christensen, Chu, Chua,
  Chung, Ciani, Clara, Clark, Cleva, Coccia, Cohadon, Colla, Collette,
  Cominsky, Constancio, Conte, Conti, Cook, Corbitt, Cornish, Corsi, Cortese,
  Costa, Coughlin, Coughlin, Coulon, Countryman, Couvares, Cowan, Coward,
  Cowart, Coyne, Coyne, Craig, Creighton, Cripe, Crowder, Cumming, Cunningham,
  Cuoco, Canton, Danilishin, D'Antonio, Danzmann, Darman, Dattilo, Dave,
  Daveloza, Davier, Davies, Daw, Day, DeBra, Debreczeni, Degallaix, {De
  Laurentis}, Del{\'{e}}glise, {Del Pozzo}, Denker, Dent, Dereli, Dergachev,
  DeRosa, {De Rosa}, DeSalvo, Dhurandhar, D{\'{i}}az, {Di Fiore}, {Di
  Giovanni}, {Di Lieto}, {Di Pace}, {Di Palma}, {Di Virgilio}, Dojcinoski,
  Dolique, Donovan, Dooley, Doravari, Douglas, Downes, Drago, Drever, Driggers,
  Du, Ducrot, Dwyer, Edo, Edwards, Effler, Eggenstein, Ehrens, Eichholz,
  Eikenberry, Engels, Essick, Etzel, Evans, Evans, Everett, Factourovich,
  Fafone, Fair, Fairhurst, Fan, Fang, Farinon, Farr, Farr, Favata, Fays,
  Fehrmann, Fejer, Ferrante, Ferreira, Ferrini, Fidecaro, Fiori, Fiorucci,
  Fisher, Flaminio, Fletcher, Fournier, Franco, Frasca, Frasconi, Frei, Freise,
  Frey, Frey, Fricke, Fritschel, Frolov, Fulda, Fyffe, Gabbard, Gair,
  Gammaitoni, Gaonkar, Garufi, Gatto, Gaur, Gehrels, Gemme, Gendre, Genin,
  Gennai, George, Gergely, Germain, Ghosh, Ghosh, Giaime, Giardina, Giazotto,
  Gill, Glaefke, Goetz, Goetz, Gondan, Gonz{\'{a}}lez, Castro, Gopakumar,
  Gordon, Gorodetsky, Gossan, Gosselin, Gouaty, Graef, Graff, Granata, Grant,
  Gras, Gray, Greco, Green, Groot, Grote, Grunewald, Guidi, Guo, Gupta, Gupta,
  Gushwa, Gustafson, Gustafson, Hacker, Hall, Hall, Hammond, Haney, Hanke,
  Hanks, Hanna, Hannam, Hanson, Hardwick, Haris, Harms, Harry, Harry, Hart,
  Hartman, Haster, Haughian, Heidmann, Heintze, Heitmann, Hello, Hemming,
  Hendry, Heng, Hennig, Heptonstall, Heurs, Hild, Hoak, Hodge, Hofman, Hollitt,
  Holt, Holz, Hopkins, Hosken, Hough, Houston, Howell, Hu, Huang, Huerta, Huet,
  Hughey, Husa, Huttner, Huynh-Dinh, Idrisy, Indik, Ingram, Inta, Isa, Isac,
  Isi, Islas, Isogai, Iyer, Izumi, Jacqmin, Jang, Jani, Jaranowski, Jawahar,
  Jim{\'{e}}nez-Forteza, Johnson, Jones, Jones, Jonker, Ju, Kalaghatgi,
  Kalogera, Kandhasamy, Kang, Kanner, Karki, Kasprzack, Katsavounidis, Katzman,
  Kaufer, Kaur, Kawabe, Kawazoe, K{\'{e}}f{\'{e}}lian, Kehl, Keitel, Kelley,
  Kells, Kennedy, Key, Khalaidovski, Khalili, Khan, Khan, Khan, Khazanov,
  Kijbunchoo, Kim, Kim, Kim, Kim, Kim, Kim, King, King, Kinzel, Kissel,
  Kleybolte, Klimenko, Koehlenbeck, Kokeyama, Koley, Kondrashov, Kontos,
  Korobko, Korth, Kowalska, Kozak, Kringel, Kr{\'{o}}lak, Krueger, Kuehn,
  Kumar, Kuo, Kutynia, Lackey, Landry, Lange, Lantz, Lasky, Lazzarini, Lazzaro,
  Leaci, Leavey, Lebigot, Lee, Lee, Lee, Lee, Lenon, Leonardi, Leong, Leroy,
  Letendre, Levin, Levine, Li, Libson, Littenberg, Lockerbie, Logue, Lombardi,
  Lord, Lorenzini, Loriette, Lormand, Losurdo, Lough, L{\"{u}}ck, Lundgren,
  Luo, Lynch, Ma, MacDonald, Machenschalk, MacInnis, Macleod,
  Maga{\~{n}}a-Sandoval, Magee, Mageswaran, Majorana, Maksimovic, Malvezzi,
  Man, Mandel, Mandic, Mangano, Mansell, Manske, Mantovani, Marchesoni, Marion,
  M{\'{a}}rka, M{\'{a}}rka, Markosyan, Maros, Martelli, Martellini, Martin,
  Martin, Martynov, Marx, Mason, Masserot, Massinger, Masso-Reid, Matichard,
  Matone, Mavalvala, Mazumder, Mazzolo, McCarthy, McClelland, McCormick,
  McGuire, McIntyre, McIver, McManus, McWilliams, Meacher, Meadors, Meidam,
  Melatos, Mendell, Mendoza-Gandara, Mercer, Merilh, Merzougui, Meshkov,
  Messenger, Messick, Meyers, Mezzani, Miao, Michel, Middleton, Mikhailov,
  Milano, Miller, Millhouse, Minenkov, Ming, Mirshekari, Mishra, Mitra,
  Mitrofanov, Mitselmakher, Mittleman, Moggi, Mohan, Mohapatra, Montani, Moore,
  Moore, Moraru, Moreno, Morriss, Mossavi, Mours, Mow-Lowry, Mueller, Mueller,
  Muir, Mukherjee, Mukherjee, Mukherjee, Mukund, Mullavey, Munch, Murphy,
  Murray, Mytidis, Nardecchia, Naticchioni, Nayak, Necula, Nedkova, Nelemans,
  Neri, Neunzert, Newton, Nguyen, Nielsen, Nissanke, Nitz, Nocera, Nolting,
  Normandin, Nuttall, Oberling, Ochsner, O'Dell, Oelker, Ogin, Oh, Oh, Ohme,
  Oliver, Oppermann, Oram, O'Reilly, O'Shaughnessy, Ottaway, Ottens, Overmier,
  Owen, Pai, Pai, Palamos, Palashov, Palliyaguru, Palomba, Pal-Singh, Pan,
  Pankow, Pannarale, Pant, Paoletti, Paoli, Papa, Paris, Parker, Pascucci,
  Pasqualetti, Passaquieti, Passuello, Patricelli, Patrick, Pearlstone,
  Pedraza, Pedurand, Pekowsky, Pele, Penn, Perreca, Phelps, Piccinni, Pichot,
  Piergiovanni, Pierro, Pillant, Pinard, Pinto, Pitkin, Poggiani, Popolizio,
  Post, Powell, Prasad, Predoi, Premachandra, Prestegard, Price, Prijatelj,
  Principe, Privitera, Prodi, Prokhorov, Puncken, Punturo, Puppo, P{\"{u}}rrer,
  Qi, Qin, Quetschke, Quintero, Quitzow-James, Raab, Rabeling, Radkins, Raffai,
  Raja, Rakhmanov, Rapagnani, Raymond, Razzano, Re, Read, Reed, Regimbau, Rei,
  Reid, Reitze, Rew, Reyes, Ricci, Riles, Robertson, Robie, Robinet, Rocchi,
  Rolland, Rollins, Roma, Romano, Romanov, Romie, Rosi{\'{n}}ska, Rowan,
  R{\"{u}}diger, Ruggi, Ryan, Sachdev, Sadecki, Sadeghian, Salconi, Saleem,
  Salemi, Samajdar, Sammut, Sanchez, Sandberg, Sandeen, Sanders, Sassolas,
  Sathyaprakash, Saulson, Sauter, Savage, Sawadsky, Schale, Schilling, Schmidt,
  Schmidt, Schnabel, Schofield, Sch{\"{o}}nbeck, Schreiber, Schuette, Schutz,
  Scott, Scott, Sellers, Sentenac, Sequino, Sergeev, Serna, Setyawati, Sevigny,
  Shaddock, Shah, Shahriar, Shaltev, Shao, Shapiro, Shawhan, Sheperd,
  Shoemaker, Shoemaker, Siellez, Siemens, Sigg, Silva, Simakov, Singer, Singh,
  Singh, Singhal, Sintes, Slagmolen, Smith, Smith, Smith, Son, Sorazu,
  Sorrentino, Souradeep, Srivastava, Staley, Steinke, Steinlechner,
  Steinlechner, Steinmeyer, Stephens, Stone, Strain, Straniero, Stratta,
  Strauss, Strigin, Sturani, Stuver, Summerscales, Sun, Sutton, Swinkels,
  Szczepa{\'{n}}czyk, Tacca, Talukder, Tanner, T{\'{a}}pai, Tarabrin,
  Taracchini, Taylor, Theeg, Thirugnanasambandam, Thomas, Thomas, Thomas,
  Thorne, Thorne, Thrane, Tiwari, Tiwari, Tokmakov, Tomlinson, Tonelli, Torres,
  Torrie, T{\"{o}}yr{\"{a}}, Travasso, Traylor, Trifir{\`{o}}, Tringali,
  Trozzo, Tse, Turconi, Tuyenbayev, Ugolini, Unnikrishnan, Urban, Usman,
  Vahlbruch, Vajente, Valdes, van Bakel, van Beuzekom, van~den Brand, Broeck,
  Vander-Hyde, van~der Schaaf, van Heijningen, van Veggel, Vardaro, Vass,
  Vas{\'{u}}th, Vaulin, Vecchio, Vedovato, Veitch, Veitch, Venkateswara,
  Verkindt, Vetrano, Vicer{\'{e}}, Vinciguerra, Vine, Vinet, Vitale, Vo, Vocca,
  Vorvick, Voss, Vousden, Vyatchanin, Wade, Wade, Wade, Walker, Wallace, Walsh,
  Wang, Wang, Wang, Wang, Wang, Ward, Warner, Was, Weaver, Wei, Weinert,
  Weinstein, Weiss, Welborn, Wen, We{\ss}els, Westphal, Wette, Whelan, White,
  Whiting, Williams, Williamson, Willis, Willke, Wimmer, Winkler, Wipf, Wittel,
  Woan, Worden, Wright, Wu, Yablon, Yam, Yamamoto, Yancey, Yap, Yu, Yvert,
  Zadro{\.{z}}ny, Zangrando, Zanolin, Zendri, Zevin, Zhang, Zhang, Zhang,
  Zhang, Zhao, Zhou, Zhou, Zhu, Zucker, Zuraw, Zweizig, Allison, Bannister,
  Bell, Chatterjee, Chippendale, Edwards, Harvey-Smith, Heywood, Hotan,
  Indermuehle, Marvil, McConnell, Murphy, Popping, Reynolds, Sault, Voronkov,
  Whiting, Castro-Tirado, Cunniffe, Jel{\'{i}}nek, Tello, Oates, Hu,
  Kub{\'{a}}nek, Guziy, Castell{\'{o}}n, Garc{\'{i}}a-Cerezo, Mu{\~{n}}oz, del
  Pulgar, Castillo-Carri{\'{o}}n, Cer{\'{o}}n, Hudec, Caballero-Garc{\'{i}}a,
  P{\'{a}}ta, Vitek, Adame, Konig, Rend{\'{o}}n, Sanguino,
  Fern{\'{a}}ndez-Mu{\~{n}}oz, Yock, Rattenbury, Allen, Querel, Jeong, Park,
  Bai, Cui, Fan, Wang, Hiriart, Lee, Claret, S{\'{a}}nchez-Ram{\'{i}}rez,
  Pandey, Mediavilla, Sabau-Graziati, Abbott, Abdalla, Allam, Annis, Armstrong,
  Benoit-L{\'{e}}vy, Berger, Bernstein, Bertin, Brout, Buckley-Geer, Burke,
  Capozzi, Carretero, Castander, Chornock, Cowperthwaite, Crocce, Cunha,
  D'Andrea, da~Costa, Desai, Diehl, Dietrich, Doctor, Drlica-Wagner, Drout,
  Eifler, Estrada, Evrard, Fernandez, Finley, Flaugher, Foley, Fong, Fosalba,
  Fox, Frieman, Fryer, Gaztanaga, Gerdes, Goldstein, Gruen, Gruendl, Gutierrez,
  Herner, Honscheid, James, Johnson, Johnson, Karliner, Kasen, Kent, Kessler,
  Kim, Kind, Kuehn, Kuropatkin, Lahav, Li, Lima, Lin, Maia, Margutti, Marriner,
  Martini, Matheson, Melchior, Metzger, Miller, Miquel, Neilsen, Nichol, Nord,
  Nugent, Ogando, Petravick, Plazas, Quataert, Roe, Romer, Roodman, Rosell,
  Rykoff, Sako, Sanchez, Scarpine, Schindler, Schubnell, Scolnic,
  Sevilla-Noarbe, Sheldon, Smith, Smith, Soares-Santos, Sobreira, Stebbins,
  Suchyta, Swanson, Tarle, Thaler, Thomas, Thomas, Tucker, Vikram, Walker,
  Wechsler, Wester, Yanny, Zhang, Zuntz, Connaughton, Burns, Goldstein, Briggs,
  Zhang, Hui, Jenke, Wilson-Hodge, Bhat, Bissaldi, Cleveland, Fitzpatrick,
  Giles, Gibby, Greiner, von Kienlin, Kippen, McBreen, Mailyan, Meegan,
  Paciesas, Preece, Roberts, Sparke, Stanbro, Toelge, Veres, Yu, Blackburn,
  Ackermann, Ajello, Albert, Anderson, Atwood, Axelsson, Baldini, Barbiellini,
  Bastieri, Bellazzini, Bissaldi, Blandford, Bloom, Bonino, Bottacini, Brandt,
  Bruel, Buson, Caliandro, Cameron, Caragiulo, Caraveo, Cavazzuti, Charles,
  Chekhtman, Chiang, Chiaro, Ciprini, Cohen-Tanugi, Cominsky, Costanza, Cuoco,
  D'Ammando, de~Palma, Desiante, Digel, {Di Lalla}, {Di Mauro}, {Di Venere},
  Dom{\'{i}}nguez, Drell, Dubois, Favuzzi, Ferrara, Franckowiak, Fukazawa,
  Funk, Fusco, Gargano, Gasparrini, Giglietto, Giommi, Giordano, Giroletti,
  Glanzman, Godfrey, Gomez-Vargas, Green, Grenier, Grove, Guiriec, Hadasch,
  Harding, Hays, Hewitt, Hill, Horan, Jogler, J{\'{o}}hannesson, Johnson,
  Kensei, Kocevski, Kuss, {La Mura}, Larsson, Latronico, Li, Li, Longo,
  Loparco, Lovellette, Lubrano, Magill, Maldera, Manfreda, Marelli, Mayer,
  Mazziotta, McEnery, Meyer, Michelson, Mirabal, Mizuno, Moiseev, Monzani,
  Moretti, Morselli, Moskalenko, Negro, Nuss, Ohsugi, Omodei, Orienti, Orlando,
  Ormes, Paneque, Perkins, Pesce-Rollins, Piron, Pivato, Porter, Racusin,
  Rain{\`{o}}, Rando, Razzaque, Reimer, Reimer, Salvetti, Parkinson,
  Sgr{\`{o}}, Simone, Siskind, Spada, Spandre, Spinelli, Suson, Tajima, Thayer,
  Thompson, Tibaldo, Torres, Troja, Uchiyama, Venters, Vianello, Wood, Wood,
  Zhu, Zimmer, Brocato, Cappellaro, Covino, Grado, Nicastro, Palazzi, Pian,
  Amati, Antonelli, Capaccioli, D'Avanzo, D'Elia, Getman, Giuffrida, Iannicola,
  Limatola, Lisi, Marinoni, Marrese, Melandri, Piranomonte, Possenti, Pulone,
  Rossi, Stamerra, Stella, Testa, Tomasella, Yang, Bazzano, Bozzo, Brandt,
  Courvoisier, Ferrigno, Hanlon, Kuulkers, Laurent, Mereghetti, Roques,
  Savchenko, Ubertini, Kasliwal, Singer, Cao, Duggan, Kulkarni, Bhalerao,
  Miller, Barlow, Bellm, Manulis, Rana, Laher, Masci, Surace, Rebbapragada,
  Cook, {Van Sistine}, Sesar, Perley, Ferreti, Prince, Kendrick, Horesh,
  Hurley, Golenetskii, Aptekar, Frederiks, Svinkin, Rau, von Kienlin, Zhang,
  Smith, Cline, Krimm, Abe, Doi, Fujisawa, Kawabata, Morokuma, Motohara,
  Tanaka, Ohta, Yanagisawa, Yoshida, Baltay, Rabinowitz, Ellman, Rostami,
  Bersier, Bode, Collins, Copperwheat, Darnley, Galloway, Gomboc, Kobayashi,
  Mazzali, Mundell, Piascik, Pollacco, Steele, Ulaczyk, Broderick, Fender,
  Jonker, Rowlinson, Stappers, Wijers, Lipunov, Gorbovskoy, Tyurina, Kornilov,
  Balanutsa, Kuznetsov, Buckley, Rebolo, Serra-Ricart, Israelian, Budnev,
  Gress, Ivanov, Poleshuk, Tlatov, Yurkov, Kawai, Serino, Negoro, Nakahira,
  Mihara, Tomida, Ueno, Tsunemi, Matsuoka, Croft, Feng, Franzen, Gaensler,
  Johnston-Hollitt, Kaplan, Morales, Tingay, Wayth, Williams, Smartt, Chambers,
  Smith, Huber, Young, Wright, Schultz, Denneau, Flewelling, Magnier, Primak,
  Rest, Sherstyuk, Stalder, Stubbs, Tonry, Waters, Willman, E., Campbell,
  Kotak, Sollerman, Smith, Dennefeld, Anderson, Botticella, Chen, Valle,
  Elias-Rosa, Fraser, Inserra, Kankare, Kupfer, Harmanen, Galbany, Guillou,
  Lyman, Maguire, Mitra, Nicholl, Razza, Terreran, Valenti, Gal-Yam,
  {\'{C}}wiek, {\'{C}}wiok, Mankiewicz, Opiela, Zaremba, {\.{Z}}arnecki, Onken,
  Scalzo, Schmidt, Wolf, Yuan, Evans, Kennea, Burrows, Campana, Cenko, Giommi,
  Marshall, Nousek, O'Brien, Osborne, Palmer, Perri, Siegel, Tagliaferri,
  Klotz, Turpin, Laugier, Beroiz, Pe{\~{n}}uela, Macri, Oelkers, Lambas, Vrech,
  Cabral, Colazo, Dominguez, Sanchez, Gurovich, Lares, Marshall, DePoy,
  Padilla, Pereyra, Benacquista, Tanvir, Wiersema, Levan, Steeghs, Hjorth,
  Fynbo, Malesani, Milvang-Jensen, Watson, Irwin, Fernandez, McMahon, Banerji,
  Gonzalez-Solares, Schulze, Postigo, Thoene, Cano, Rosswog, Dominguez,
  Sanchez, Gurovich, Lares, Marshall, DePoy, Padilla, Pereyra, Benacquista,
  Collaboration, Tanvir, Wiersema, Levan, Steeghs, Hjorth, Fynbo, Malesani,
  Milvang-Jensen, Watson, Irwin, Fernandez, McMahon, Banerji, Gonzalez-Solares,
  Schulze, Postigo, Thoene, Cano, Rosswog, \& Collaboration}]{aaa+16b}
---. 2016{\natexlab{c}}, ApJSS, 225, 8

\bibitem[{Abbott {et~al.}(2017)Abbott, Abbott, Abbott, Acernese, Ackley, Adams,
  Adams, Addesso, Adhikari, Adya, \& {LCV}}]{GW170104_main}
Abbott, B.~P., Abbott, R., Abbott, T.~D., Acernese, F., Ackley, K., {et~al.}
  2017, Physical Review Letters, 118, 221101

\bibitem[{{Abe} {et~al.}(2016){Abe}, {Haga}, {Hayato}, {Ikeda}, {Iyogi},
  {Kameda}, {Kishimoto}, {Miura}, {Moriyama}, {Nakahata}, {Nakajima}, {Nakano},
  {Nakayama}, {Orii}, {Sekiya}, {Shiozawa}, {Takeda}, {Tanaka}, {Tasaka},
  {Tomura}, {Akutsu}, {Kajita}, {Kaneyuki}, {Nishimura}, {Richard}, {Okumura},
  {Labarga}, {Fernandez}, {Blaszczyk}, {Gustafson}, {Kachulis}, {Kearns},
  {Raaf}, {Stone}, {Sulak}, {Berkman}, {Nantais}, {Tobayama}, {Goldhaber},
  {Kropp}, {Mine}, {Weatherly}, {Smy}, {Sobel}, {Takhistov}, {Ganezer},
  {Hartfiel}, {Hill}, {Hong}, {Kim}, {Lim}, {Park}, {Himmel}, {Li},
  {O'Sullivan}, {Scholberg}, {Walter}, {Ishizuka}, {Nakamura}, {Jang}, {Choi},
  {Learned}, {Matsuno}, {Smith}, {Friend}, {Hasegawa}, {Ishida}, {Ishii},
  {Kobayashi}, {Nakadaira}, {Nakamura}, {Oyama}, {Sakashita}, {Sekiguchi},
  {Tsukamoto}, {Suzuki}, {Takeuchi}, {Yano}, {Cao}, {Hiraki}, {Hirota},
  {Huang}, {Jiang}, {Minamino}, {Nakaya}, {Patel}, {Wendell}, {Suzuki},
  {Fukuda}, {Itow}, {Suzuki}, {Mijakowski}, {Frankiewicz}, {Hignight}, {Imber},
  {Jung}, {Li}, {Palomino}, {Santucci}, {Wilking}, {Yanagisawa}, {Fukuda},
  {Ishino}, {Kayano}, {Kibayashi}, {Koshio}, {Mori}, {Sakuda}, {Xu}, {Kuno},
  {Tacik}, {Kim}, {Okazawa}, {Choi}, {Nishijima}, {Koshiba}, {Totsuka}, {Suda},
  {Yokoyama}, {Bronner}, {Calland}, {Hartz}, {Martens}, {Marti}, {Suzuki},
  {Vagins}, {Martin}, {Tanaka}, {Konaka}, {Chen}, {Wan}, {Zhang}, {Wilkes}, \&
  {Super-Kamiokande Collaboration}}]{2016ApJ...830L..11A}
{Abe}, K., {Haga}, K., {Hayato}, Y., {Ikeda}, M., {Iyogi}, K., {et~al.} 2016,
  \apjl, 830, L11

\bibitem[{{Ackermann} {et~al.}(2016){Ackermann}, {Ajello}, {Albert},
  {Anderson}, {Arimoto}, {Atwood}, {Axelsson}, {Baldini}, {Ballet},
  {Barbiellini}, {Baring}, {Bastieri}, {Becerra Gonzalez}, {Bellazzini},
  {Bissaldi}, {Blandford}, {Bloom}, {Bonino}, {Bottacini}, {Brandt}, {Bregeon},
  {Britto}, {Bruel}, {Buehler}, {Burnett}, {Buson}, {Caliandro}, {Cameron},
  {Caputo}, {Caragiulo}, {Caraveo}, {Casandjian}, {Cavazzuti}, {Charles},
  {Chekhtman}, {Chiang}, {Chiaro}, {Ciprini}, {Cohen-Tanugi}, {Cominsky},
  {Condon}, {Costanza}, {Cuoco}, {Cutini}, {D'Ammando}, {de Palma}, {Desiante},
  {Digel}, {Di Lalla}, {Di Mauro}, {Di Venere}, {Dom{\'{\i}}nguez}, {Drell},
  {Dubois}, {Dumora}, {Favuzzi}, {Fegan}, {Ferrara}, {Franckowiak}, {Fukazawa},
  {Funk}, {Fusco}, {Gargano}, {Gasparrini}, {Gehrels}, {Giglietto}, {Giomi},
  {Giommi}, {Giordano}, {Giroletti}, {Glanzman}, {Godfrey}, {Gomez-Vargas},
  {Granot}, {Green}, {Grenier}, {Grondin}, {Grove}, {Guillemot}, {Guiriec},
  {Hadasch}, {Harding}, {Hays}, {Hewitt}, {Hill}, {Horan}, {Jogler},
  {J{\'o}hannesson}, {Kamae}, {Kensei}, {Kocevski}, {Kuss}, {La Mura},
  {Larsson}, {Latronico}, {Lemoine-Goumard}, {Li}, {Li}, {Longo}, {Loparco},
  {Lovellette}, {Lubrano}, {Madejski}, {Magill}, {Maldera}, {Manfreda},
  {Marelli}, {Mayer}, {Mazziotta}, {McEnery}, {Meyer}, {Michelson}, {Mirabal},
  {Mizuno}, {Moiseev}, {Monzani}, {Moretti}, {Morselli}, {Moskalenko},
  {Murgia}, {Negro}, {Nuss}, {Ohsugi}, {Omodei}, {Orienti}, {Orlando}, {Ormes},
  {Paneque}, {Perkins}, {Pesce-Rollins}, {Piron}, {Pivato}, {Porter},
  {Racusin}, {Rain{\`o}}, {Rando}, {Razzaque}, {Reimer}, {Reimer}, {Reposeur},
  {Ritz}, {Rochester}, {Romani}, {Saz Parkinson}, {Sgr{\`o}}, {Simone},
  {Siskind}, {Smith}, {Spada}, {Spandre}, {Spinelli}, {Suson}, {Tajima},
  {Thayer}, {Thayer}, {Thompson}, {Tibaldo}, {Torres}, {Troja}, {Uchiyama},
  {Venters}, {Vianello}, {Wood}, {Wood}, {Zaharijas}, {Zhu}, \&
  {Zimmer}}]{2016ApJ...823L...2A}
{Ackermann}, M., {Ajello}, M., {Albert}, A., {Anderson}, B., {Arimoto}, M.,
  {et~al.} 2016, \apjl, 823, L2

\bibitem[{{Adriani} {et~al.}(2016){Adriani}, {Akaike}, {Asano}, {Asaoka},
  {Bagliesi}, {Bigongiari}, {Binns}, {Bonechi}, {Bongi}, {Brogi}, {Buckley},
  {Cannady}, {Castellini}, {Checchia}, {Cherry}, {Collazuol}, {Di Felice},
  {Ebisawa}, {Fuke}, {Guzik}, {Hams}, {Hareyama}, {Hasebe}, {Hibino},
  {Ichimura}, {Ioka}, {Ishizaki}, {Israel}, {Javaid}, {Kasahara}, {Kataoka},
  {Kataoka}, {Katayose}, {Kato}, {Kawanaka}, {Kawakubo}, {Kitamura},
  {Krawczynski}, {Krizmanic}, {Kuramata}, {Lomtadze}, {Maestro}, {Marrocchesi},
  {Messineo}, {Mitchell}, {Miyake}, {Mizutani}, {Moiseev}, {Mori}, {Mori},
  {Mori}, {Motz}, {Munakata}, {Murakami}, {Nakagawa}, {Nakahira}, {Nishimura},
  {Okuno}, {Ormes}, {Ozawa}, {Pacini}, {Palma}, {Papini}, {Penacchioni},
  {Rauch}, {Ricciarini}, {Sakai}, {Sakamoto}, {Sasaki}, {Shimizu}, {Shiomi},
  {Sparvoli}, {Spillantini}, {Stolzi}, {Takahashi}, {Takayanagi}, {Takita},
  {Tamura}, {Tateyama}, {Terasawa}, {Tomida}, {Torii}, {Tsunesada}, {Uchihori},
  {Ueno}, {Vannuccini}, {Wefel}, {Yamaoka}, {Yanagita}, {Yoshida}, {Yoshida},
  \& {Yuda}}]{2016ApJ...829L..20A}
{Adriani}, O., {Akaike}, Y., {Asano}, K., {Asaoka}, Y., {Bagliesi}, M.~G.,
  {et~al.} 2016, \apjl, 829, L20

\bibitem[{Agostinelli {et~al.}(2003)Agostinelli, Allison, Amako, Apostolakis,
  Araujo, Arce, Asai, Axen, Banerjee, Barrand, Behner, Bellagamba, Boudreau,
  Broglia, Brunengo, Burkhardt, Chauvie, Chuma, Chytracek, Cooperman, Cosmo,
  Degtyarenko, Dell'Acqua, Depaola, Dietrich, Enami, Feliciello, Ferguson,
  Fesefeldt, Folger, Foppiano, Forti, Garelli, Giani, Giannitrapani, Gibin,
  Cadenas, González, Abril, Greeniaus, Greiner, Grichine, Grossheim, Guatelli,
  Gumplinger, Hamatsu, Hashimoto, Hasui, Heikkinen, Howard, Ivanchenko,
  Johnson, Jones, Kallenbach, Kanaya, Kawabata, Kawabata, Kawaguti, Kelner,
  Kent, Kimura, Kodama, Kokoulin, Kossov, Kurashige, Lamanna, Lampén, Lara,
  Lefebure, Lei, Liendl, Lockman, Longo, Magni, Maire, Medernach, Minamimoto,
  de~Freitas, Morita, Murakami, Nagamatu, Nartallo, Nieminen, Nishimura,
  Ohtsubo, Okamura, O'Neale, Oohata, Paech, Perl, Pfeiffer, Pia, Ranjard,
  Rybin, Sadilov, Salvo, Santin, Sasaki, Savvas, Sawada, Scherer, Sei,
  Sirotenko, Smith, Starkov, Stoecker, Sulkimo, Takahata, Tanaka, Tcherniaev,
  Tehrani, Tropeano, Truscott, Uno, Urban, Urban, Verderi, Walkden, Wander,
  Weber, Wellisch, Wenaus, Williams, Wright, Yamada, Yoshida, \&
  Zschiesche}]{geant4}
Agostinelli, S., Allison, J., Amako, K., Apostolakis, J., Araujo, H., {et~al.}
  2003, Nuclear Instruments and Methods in Physics Research Section A:
  Accelerators, Spectrometers, Detectors and Associated Equipment, 506, 250

\bibitem[{Aguilar(2011)}]{grbweb}
Aguilar, J.-A. 2011, in Proceedings of the 32nd International Cosmic Ray
  Conference (ICRC2011) OG2.3-2.4: Cosmic Ray Origin and Galactic Phenomena,
  Vol.~8, Beijing, China, 235

\bibitem[{{Annis} {et~al.}(2016){Annis}, {Soares-Santos}, {Berger}, {Brout},
  {Chen}, {Chornock}, {Cowperthwaite}, {Diehl}, {Doctor}, {Drlica-Wagner},
  {Drout}, {Farr}, {Finley}, {Flaugher}, {Foley}, {Frieman}, {Gruendl},
  {Herner}, {Holz}, {Kessler}, {Lin}, {Marriner}, {Neilsen}, {Rest}, {Sako},
  {Smith}, {Smith}, {Sobreira}, {Walker}, {Yanny}, {Abbott}, {Abdalla},
  {Allam}, {Benoit-L{\'e}vy}, {Bernstein}, {Bertin}, {Buckley-Geer}, {Burke},
  {Capozzi}, {Carnero Rosell}, {Carrasco Kind}, {Carretero}, {Castander},
  {Cenko}, {Crocce}, {Cunha}, {D'Andrea}, {da Costa}, {Desai}, {Dietrich},
  {Eifler}, {Evrard}, {Fernandez}, {Fischer}, {Fong}, {Fosalba}, {Fox},
  {Fryer}, {Garcia-Bellido}, {Gaztanaga}, {Gerdes}, {Goldstein}, {Gruen},
  {Gutierrez}, {Honscheid}, {James}, {Karliner}, {Kasen}, {Kent}, {Kuehn},
  {Kuropatkin}, {Lahav}, {Li}, {Lima}, {Maia}, {Martini}, {Metzger}, {Miller},
  {Miquel}, {Mohr}, {Nichol}, {Nord}, {Ogando}, {Peoples}, {Petravic},
  {Plazas}, {Quataert}, {Romer}, {Roodman}, {Rykoff}, {Sanchez}, {Santiago},
  {Scarpine}, {Schindler}, {Schubnell}, {Sevilla-Noarbe}, {Sheldon}, {Smith},
  {Stebbins}, {Swanson}, {Tarle}, {Thaler}, {Thomas}, {Tucker}, {Vikram},
  {Wechsler}, {Weller}, {Wester}, \& {DES Collaboration}}]{2016ApJ...823L..34A}
{Annis}, J., {Soares-Santos}, M., {Berger}, E., {Brout}, D., {Chen}, H.,
  {et~al.} 2016, \apjl, 823, L34

\bibitem[{{ANTARES Collaboration} {et~al.}(2017){ANTARES Collaboration},
  {IceCube Collaboration}, {LIGO Scientific Collaboration}, \& {Virgo
  Collaboration}}]{2017arXiv170306298A}
{ANTARES Collaboration}, {IceCube Collaboration}, {LIGO Scientific
  Collaboration}, \& {Virgo Collaboration}. 2017, ArXiv e-prints, 1703.06298

\bibitem[{{Berger} {et~al.}(2003){Berger}, {Kulkarni}, {Pooley}, {Frail},
  {McIntyre}, {Wark}, {Sari}, {Soderberg}, {Fox}, {Yost}, \&
  {Price}}]{berger+03}
{Berger}, E., {Kulkarni}, S.~R., {Pooley}, G., {Frail}, D.~A., {McIntyre}, V.,
  {et~al.} 2003, \nat, 426, 154

\bibitem[{Bertin {et~al.}(2002)Bertin, Mellier, Radovich, Missonnier, Didelon,
  \& Morin}]{bmr+02}
Bertin, E., Mellier, Y., Radovich, M., Missonnier, G., Didelon, P., \& Morin,
  B. 2002, in Astronomical Society of the Pacific Conference Series, Vol. 281,
  Astronomical Data Analysis Software and Systems XI, ed. D.~Bohlender,
  D.~Durand, \& T.~Handley, 228

\bibitem[{Berton {et~al.}(2017)Berton, La~Mura, Chen, Tomasella, Cappellaro,
  Melandri, Piranomonte, Greco, D'Avanzo, Stratta, Ascenzi, Botticella, DElia,
  Palazzi, Rossi, Covino, Amati, Antonelli, Branchesi, Campana, Getman, Grado,
  Limatola, Lisi, Nicastro, Pian, Pulone, Tagliaferri, Testa, Yang, Brocato, \&
  {GRavitational Wave Inaf TeAm}}]{GW170104_asiago}
Berton, M., La~Mura, G., Chen, S., Tomasella, L., Cappellaro, E., {et~al.}
  2017, GRB Coordinates Network, 20386, 1

\bibitem[{Bhalerao {et~al.}(2016{\natexlab{a}})Bhalerao, Bhattacharya, Vibhute,
  Pawar, Rao, Hingar, Khanna, Kutty, Malkar, Patil, Arora, Sinha, Priya,
  Samuel, Sreekumar, Vinod, Mithun, Vadawale, Vagshette, Navalgund, Sarma,
  Pandiyan, Seetha, \& Subbarao}]{bbv+16}
Bhalerao, V., Bhattacharya, D., Vibhute, A., Pawar, P., Rao, A.~R., {et~al.}
  2016{\natexlab{a}}, ArXiv e-prints, 1608.03408

\bibitem[{Bhalerao {et~al.}(2016{\natexlab{b}})Bhalerao, Bhattacharya, Vibhute,
  Bose, Dewangan, Misra, Mitra, Rao, Souradeep, \& Vadawale}]{bbv+16b}
Bhalerao, V.~B., Bhattacharya, D., Vibhute, A., Bose, S., Dewangan, G.~C.,
  {et~al.} 2016{\natexlab{b}}, GRB Coordinates Network, 19401, 1

\bibitem[{{Blackburn}(1995)}]{ftools}
{Blackburn}, J.~K. 1995, in Astronomical Society of the Pacific Conference
  Series, Vol.~77, Astronomical Data Analysis Software and Systems IV, ed.
  R.~A. {Shaw}, H.~E. {Payne}, \& J.~J.~E. {Hayes}, 367

\bibitem[{{Blanchard} {et~al.}(2016){Blanchard}, {Berger}, \&
  {Fong}}]{Blanchard+2015}
{Blanchard}, P.~K., {Berger}, E., \& {Fong}, W.-f. 2016, \apj, 817, 144

\bibitem[{{Blanton} \& {Roweis}(2007)}]{br07}
{Blanton}, M.~R., \& {Roweis}, S. 2007, \aj, 133, 734

\bibitem[{{Bloom} {et~al.}(2002){Bloom}, {Kulkarni}, \&
  {Djorgovski}}]{Bloom+2002}
{Bloom}, J.~S., {Kulkarni}, S.~R., \& {Djorgovski}, S.~G. 2002, \aj, 123, 1111

\bibitem[{Bogomolov {et~al.}(2017)Bogomolov, Svertilov, Amelushkin,
  V.O.Barinova, M.I.Panasyuk, A.V.Bogomolov, A.F.Iyudin, V.V.Kalegaev, D.Nguen,
  Petrov, I.V.Yashin, P.S.Kazarian, N.L.Dzhioeva, Lipunov, Gorbovskoy,
  N.Tyurina, D.Kuvshinov, M.Panchenko, Park, Lee, \& Jeong}]{GW170104_lom}
Bogomolov, V.~V., Svertilov, S., Amelushkin, A., V.O.Barinova, M.I.Panasyuk,
  {et~al.} 2017, GRB Coordinates Network, 20402, 1

\bibitem[{Burns {et~al.}(2017)Burns, Blackburn, Briggs, Broida, Camp, Canton,
  Christensen, Connaughton, Goldstein, Hamburg, Hui, Jenke, Kocevski, Leroy,
  Littenberg, McEnery, Preece, Racusin, Shawhan, Siellez, Singer, Veitch,
  Veres, \& Wilson-Hodge}]{GW170104_FermiGBM}
Burns, E., Blackburn, L., Briggs, S., Broida, J., Camp, J., {et~al.} 2017, GRB
  Coordinates Network, 20365, 1

\bibitem[{Cenko \& Troja(2017)}]{GW170104_channel}
Cenko, S., \& Troja, E. 2017, GRB Coordinates Network, 20397, 1

\bibitem[{Cenko {et~al.}(2013)Cenko, Kulkarni, Horesh, Corsi, Fox, Carpenter,
  Frail, Nugent, Perley, Gruber, Gal-Yam, Groot, Hallinan, Ofek, Rau, MacLeod,
  Miller, Bloom, Filippenko, Kasliwal, Law, Morgan, Polishook, Poznanski,
  Quimby, Sesar, Shen, Silverman, \& Sternberg}]{ckh+13}
Cenko, S.~B., Kulkarni, S.~R., Horesh, A., Corsi, A., Fox, D.~B., {et~al.}
  2013, ApJ, 769, 130

\bibitem[{Cenko {et~al.}(2015)Cenko, Urban, Perley, Horesh, Corsi, Fox, Cao,
  Kasliwal, Lien, Arcavi, Bloom, Butler, Cucchiara, de~Diego, Filippenko,
  Gal-Yam, Gehrels, Georgiev, Gonzalez, Graham, Greiner, Kann, Klein, Knust,
  Kulkarni, Kutyrev, Laher, Lee, Nugent, Prochaska, Ramirez-Ruiz, Richer,
  Rubin, Urata, Varela, Watson, \& Wozniak}]{cup+15}
Cenko, S.~B., Urban, A.~L., Perley, D.~A., Horesh, A., Corsi, A., {et~al.}
  2015, ApJL, 803, L24

\bibitem[{{Chambers} {et~al.}(2016){Chambers}, {Magnier}, {Metcalfe},
  {Flewelling}, {Huber}, {Waters}, {Denneau}, {Draper}, {Farrow}, {Finkbeiner},
  {Holmberg}, {Koppenhoefer}, {Price}, {Saglia}, {Schlafly}, {Smartt},
  {Sweeney}, {Wainscoat}, {Burgett}, {Grav}, {Heasley}, {Hodapp}, {Jedicke},
  {Kaiser}, {Kudritzki}, {Luppino}, {Lupton}, {Monet}, {Morgan}, {Onaka},
  {Stubbs}, {Tonry}, {Banados}, {Bell}, {Bender}, {Bernard}, {Botticella},
  {Casertano}, {Chastel}, {Chen}, {Chen}, {Cole}, {Deacon}, {Frenk},
  {Fitzsimmons}, {Gezari}, {Goessl}, {Goggia}, {Goldman}, {Grebel}, {Hambly},
  {Hasinger}, {Heavens}, {Heckman}, {Henderson}, {Henning}, {Holman}, {Hopp},
  {Ip}, {Isani}, {Keyes}, {Koekemoer}, {Kotak}, {Long}, {Lucey}, {Liu},
  {Martin}, {McLean}, {Morganson}, {Murphy}, {Nieto-Santisteban}, {Norberg},
  {Peacock}, {Pier}, {Postman}, {Primak}, {Rae}, {Rest}, {Riess}, {Riffeser},
  {Rix}, {Roser}, {Schilbach}, {Schultz}, {Scolnic}, {Szalay}, {Seitz},
  {Shiao}, {Small}, {Smith}, {Soderblom}, {Taylor}, {Thakar}, {Thiel},
  {Thilker}, {Urata}, {Valenti}, {Walter}, {Watters}, {Werner}, {White},
  {Wood-Vasey}, \& {Wyse}}]{cmm+16}
{Chambers}, K.~C., {Magnier}, E.~A., {Metcalfe}, N., {Flewelling}, H.~A.,
  {Huber}, M.~E., {et~al.} 2016, ArXiv e-prints, 1612.05560

\bibitem[{Chambers {et~al.}(2017)Chambers, Smartt, Tonry, Denneau, A.~Heinze,
  talder, Weiland, Stubbs, Smith, Chen, Kruehler, Young, Rest, Coughlin, Huber,
  Wright, Flewelling, Magnier, Schultz, Waters, \& Wainscoat}]{GW170104_Pan}
Chambers, K.~C., Smartt, S., Tonry, J., Denneau, L., A.~Heinze, B., {et~al.}
  2017, GRB Coordinates Network, 20407, 1

\bibitem[{{Chandra} \& {Frail}(2012)}]{cf12}
{Chandra}, P., \& {Frail}, D.~A. 2012, \apj, 746, 156

\bibitem[{{Connaughton} {et~al.}(2016){Connaughton}, {Burns}, {Goldstein},
  {Blackburn}, {Briggs}, {Zhang}, {Camp}, {Christensen}, {Hui}, {Jenke},
  {Littenberg}, {McEnery}, {Racusin}, {Shawhan}, {Singer}, {Veitch},
  {Wilson-Hodge}, {Bhat}, {Bissaldi}, {Cleveland}, {Fitzpatrick}, {Giles},
  {Gibby}, {von Kienlin}, {Kippen}, {McBreen}, {Mailyan}, {Meegan}, {Paciesas},
  {Preece}, {Roberts}, {Sparke}, {Stanbro}, {Toelge}, \&
  {Veres}}]{2016ApJ...826L...6C}
{Connaughton}, V., {Burns}, E., {Goldstein}, A., {Blackburn}, L., {Briggs},
  M.~S., {et~al.} 2016, \apjl, 826, L6

\bibitem[{Corsi {et~al.}(2017)Corsi, Kasliwal, Frail, \&
  Palliyaguru}]{GW170104_VLA}
Corsi, A., Kasliwal, M.~M., Frail, D.~A., \& Palliyaguru, N.~T. 2017, GRB
  Coordinates Network, 20396, 1

\bibitem[{{Cowperthwaite} {et~al.}(2016){Cowperthwaite}, {Berger},
  {Soares-Santos}, {Annis}, {Brout}, {Brown}, {Buckley-Geer}, {Cenko}, {Chen},
  {Chornock}, {Diehl}, {Doctor}, {Drlica-Wagner}, {Drout}, {Farr}, {Finley},
  {Foley}, {Fong}, {Fox}, {Frieman}, {Garcia-Bellido}, {Gill}, {Gruendl},
  {Herner}, {Holz}, {Kasen}, {Kessler}, {Lin}, {Margutti}, {Marriner},
  {Matheson}, {Metzger}, {Neilsen}, {Quataert}, {Rest}, {Sako}, {Scolnic},
  {Smith}, {Sobreira}, {Strampelli}, {Villar}, {Walker}, {Wester}, {Williams},
  {Yanny}, {Abbott}, {Abdalla}, {Allam}, {Armstrong}, {Bechtol},
  {Benoit-L{\'e}vy}, {Bertin}, {Brooks}, {Burke}, {Carnero Rosell}, {Carrasco
  Kind}, {Carretero}, {Castander}, {Cunha}, {D'Andrea}, {da Costa}, {Desai},
  {Dietrich}, {Evrard}, {Fausti Neto}, {Fosalba}, {Gerdes}, {Giannantonio},
  {Goldstein}, {Gruen}, {Gutierrez}, {Honscheid}, {James}, {Johnson},
  {Johnson}, {Krause}, {Kuehn}, {Kuropatkin}, {Lima}, {Maia}, {Marshall},
  {Menanteau}, {Miquel}, {Mohr}, {Nichol}, {Nord}, {Ogando}, {Plazas}, {Reil},
  {Romer}, {Sanchez}, {Scarpine}, {Sevilla-Noarbe}, {Smith}, {Suchyta},
  {Tarle}, {Thomas}, {Thomas}, {Tucker}, {Weller}, \& {DES
  Collaboration}}]{2016ApJ...826L..29C}
{Cowperthwaite}, P.~S., {Berger}, E., {Soares-Santos}, M., {Annis}, J.,
  {Brout}, D., {et~al.} 2016, \apjl, 826, L29

\bibitem[{Davies {et~al.}(2009)Davies, Franzen, Davies, Davis, Feroz,
  Genova-Santos, Grainge, Green, Hobson, Hurley-Walker, Lasenby, Lopez-Caniego,
  Olamaie, Padilla-Torres, Pooley, Rebolo, Rodriguez-Gonzalvez, Saunders,
  Scaife, Scott, Shimwell, Titterington, Waldram, Watson, \& Zwart}]{dtd+09}
Davies, M.~L., Franzen, T. M.~O., Davies, R.~D., Davis, R.~J., Feroz, F.,
  {et~al.} 2009, Monthly Notices of the Royal Astronomical Society, 400, 984

\bibitem[{Evans {et~al.}(2017)Evans, Kennea, Barthelmy, Beardmore, Breeveld,
  Burrows, Campana, Cenko, G.Cusumano, D'Ai, D'Avanzo, V.D'Elia, Gehrels,
  Giommi, Gronwall, Krimm, Kuin, Lien, Malesani, Marshall, Melandri, Mingo,
  Nousek, Oates, O'Brien, Osborne, Pagani, Page, Palmer, Perri, Racusin,
  Sbarufatti, Siegel, Tagliaferri, \& Troja}]{GW170104_SwiftXRT2}
Evans, P., Kennea, J., Barthelmy, S., Beardmore, A., Breeveld, A., {et~al.}
  2017, GRB Coordinates Network, 20415, 1

\bibitem[{{Evans} {et~al.}(2009){Evans}, {Beardmore}, {Page}, {Osborne},
  {O'Brien}, {Willingale}, {Starling}, {Burrows}, {Godet}, {Vetere}, {Racusin},
  {Goad}, {Wiersema}, {Angelini}, {Capalbi}, {Chincarini}, {Gehrels}, {Kennea},
  {Margutti}, {Morris}, {Mountford}, {Pagani}, {Perri}, {Romano}, \&
  {Tanvir}}]{ebp+09}
{Evans}, P.~A., {Beardmore}, A.~P., {Page}, K.~L., {Osborne}, J.~P., {O'Brien},
  P.~T., {et~al.} 2009, \mnras, 397, 1177

\bibitem[{{Evans} {et~al.}(2016){Evans}, {Kennea}, {Barthelmy}, {Beardmore},
  {Burrows}, {Campana}, {Cenko}, {Gehrels}, {Giommi}, {Gronwall}, {Marshall},
  {Malesani}, {Markwardt}, {Mingo}, {Nousek}, {O'Brien}, {Osborne}, {Pagani},
  {Page}, {Palmer}, {Perri}, {Racusin}, {Siegel}, {Sbarufatti}, \&
  {Tagliaferri}}]{2016MNRAS.460L..40E}
{Evans}, P.~A., {Kennea}, J.~A., {Barthelmy}, S.~D., {Beardmore}, A.~P.,
  {Burrows}, D.~N., {et~al.} 2016, \mnras, 460, L40

\bibitem[{{Flewelling} {et~al.}(2016){Flewelling}, {Magnier}, {Chambers},
  {Heasley}, {Holmberg}, {Huber}, {Sweeney}, {Waters}, {Chen}, {Farrow},
  {Hasinger}, {Henderson}, {Long}, {Metcalfe}, {Nieto-Santisteban}, {Norberg},
  {Saglia}, {Szalay}, {Rest}, {Thakar}, {Tonry}, {Valenti}, {Werner}, {White},
  {Denneau}, {Draper}, {Hodapp}, {Jedicke}, {Kaiser}, {Kudritzki}, {Price},
  {Wainscoat}, {Chastel}, {McClean}, {Postman}, \& {Shiao}}]{fmc+16}
{Flewelling}, H.~A., {Magnier}, E.~A., {Chambers}, K.~C., {Heasley}, J.~N.,
  {Holmberg}, C., {et~al.} 2016, ArXiv e-prints, 1612.05243

\bibitem[{{Geng} {et~al.}(2014){Geng}, {Wu}, {Li}, {Huang}, \& {Dai}}]{geng+14}
{Geng}, J.~J., {Wu}, X.~F., {Li}, L., {Huang}, Y.~F., \& {Dai}, Z.~G. 2014,
  \apj, 792, 31

\bibitem[{{Hogg} {et~al.}(1997){Hogg}, {Pahre}, {McCarthy}, {Cohen},
  {Blandford}, {Smail}, \& {Soifer}}]{hpm+97}
{Hogg}, D.~W., {Pahre}, M.~A., {McCarthy}, J.~K., {Cohen}, J.~G., {Blandford},
  R., {Smail}, I., \& {Soifer}, B.~T. 1997, \mnras, 288, 404

\bibitem[{Jinzhong {et~al.}(2017)Jinzhong, Dong, Zhang, Zhang, Niu, guangxin
  (XAO), shuguo (XAO), taozhi (XAO), \& fangfang(XAO)}]{GW_EWE}
Jinzhong, L., Dong, X., Zhang, Y., Zhang, X., Niu, H.~X., {et~al.} 2017, GRB
  Coordinates Network, 20394, 1

\bibitem[{Jones {et~al.}(2001--)Jones, Oliphant, Peterson, {et~al.}}]{scipy}
Jones, E., Oliphant, T., Peterson, P., {et~al.} 2001--, {SciPy}: Open source
  scientific tools for {Python}, [Online; accessed 2017-07-20]

\bibitem[{{Kann} {et~al.}(2011){Kann}, {Klose}, {Zhang}, {Covino}, {Butler},
  {Malesani}, {Nakar}, {Wilson}, {Antonelli}, {Chincarini}, {Cobb}, {D'Avanzo},
  {D'Elia}, {Della Valle}, {Ferrero}, {Fugazza}, {Gorosabel}, {Israel},
  {Mannucci}, {Piranomonte}, {Schulze}, {Stella}, {Tagliaferri}, \&
  {Wiersema}}]{kkz+11}
{Kann}, D.~A., {Klose}, S., {Zhang}, B., {Covino}, S., {Butler}, N.~R.,
  {et~al.} 2011, \apj, 734, 96

\bibitem[{Kasliwal {et~al.}(2017{\natexlab{a}})Kasliwal, Adams, Vedantham,
  Bhalerao, Cenko, \& Quimby}]{GW170104_Palo}
Kasliwal, M., Adams, S., Vedantham, H., Bhalerao, V., Cenko, S.~B., \& Quimby,
  R. 2017{\natexlab{a}}, GRB Coordinates Network, 20393, 1

\bibitem[{Kasliwal {et~al.}(2017{\natexlab{b}})Kasliwal, Singer,
  Karamehmetoglu, Cenko, Quimby, Cook, (Caltech), (IUCAA), (IUCAA), (OKC),
  (OKC), (OKC), Seattle), \& (Northwestern/Adler)}]{GW_iPTF}
Kasliwal, M., Singer, L., Karamehmetoglu, E., Cenko, S., Quimby, R., {et~al.}
  2017{\natexlab{b}}, GRB Coordinates Network, 20398, 1

\bibitem[{{Kasliwal} {et~al.}(2016){Kasliwal}, {Cenko}, {Singer}, {Corsi},
  {Cao}, {Barlow}, {Bhalerao}, {Bellm}, {Cook}, {Duggan}, {Ferretti}, {Frail},
  {Horesh}, {Kendrick}, {Kulkarni}, {Lunnan}, {Palliyaguru}, {Laher}, {Masci},
  {Manulis}, {Miller}, {Nugent}, {Perley}, {Prince}, {Quimby}, {Rana},
  {Rebbapragada}, {Sesar}, {Singhal}, {Surace}, \& {Van Sistine}}]{kcs+16}
{Kasliwal}, M.~M., {Cenko}, S.~B., {Singer}, L.~P., {Corsi}, A., {Cao}, Y.,
  {et~al.} 2016, \apjl, 824, L24

\bibitem[{Kong {et~al.}(2017)Kong, Mao, Hou, Wang, \& Bai}]{GW170104_GMG}
Kong, A., Mao, J., Hou, X., Wang, J., \& Bai, J.-M. 2017, GRB Coordinates
  Network, 20421, 1

\bibitem[{{Koshut} {et~al.}(1995){Koshut}, {Paciesas}, {Kouveliotou}, {van
  Paradijs}, {Pendleton}, {Fishman}, \& {Meegan}}]{kpk+95}
{Koshut}, T.~M., {Paciesas}, W.~S., {Kouveliotou}, C., {van Paradijs}, J.,
  {Pendleton}, G.~N., {Fishman}, G.~J., \& {Meegan}, C.~A. 1995, in Bulletin of
  the American Astronomical Society, Vol.~27, American Astronomical Society
  Meeting Abstracts \#186, 886

\bibitem[{{Kotani} {et~al.}(2005){Kotani}, {Kawai}, {Yanagisawa}, {Watanabe},
  {Arimoto}, {Fukushima}, {Hattori}, {Inata}, {Izumiura}, {Kataoka}, {Koyano},
  {Kubota}, {Kuroda}, {Mori}, {Nagayama}, {Ohta}, {Okada}, {Okita}, {Sato},
  {Serino}, {Shimizu}, {Shimokawabe}, {Suzuki}, {Toda}, {Ushiyama}, {Yatsu},
  {Yoshida}, \& {Yoshida}}]{mitsume}
{Kotani}, T., {Kawai}, N., {Yanagisawa}, K., {Watanabe}, J., {Arimoto}, M.,
  {et~al.} 2005, Nuovo Cimento C Geophysics Space Physics C, 28, 755

\bibitem[{{Kouveliotou} {et~al.}(1993){Kouveliotou}, {Meegan}, {Fishman},
  {Bhat}, {Briggs}, {Koshut}, {Paciesas}, \& {Pendleton}}]{kmf+93}
{Kouveliotou}, C., {Meegan}, C.~A., {Fishman}, G.~J., {Bhat}, N.~P., {Briggs},
  M.~S., {Koshut}, T.~M., {Paciesas}, W.~S., \& {Pendleton}, G.~N. 1993, \apjl,
  413, L101

\bibitem[{Lien {et~al.}(2017)Lien, Barthelmy, Palmer, Sakamoto, Breeveld,
  Beardmore, Burrows, Campana, Cenko, Cusumano, D'Ai, D'Avanzo, V.D'Elia,
  Evans, Gehrels, Giommi, Gronwall, Kennea, Krimm, Kuin, Malesani, Marshall,
  Melandri, Mingo, Nousek, Oates, O'Brien, Osborne, Pagani, Page, Perri,
  Racusin, Sbarufatti, Siegel, Tagliaferri, \& Troja}]{GW170104_bat}
Lien, A., Barthelmy, S., Palmer, D., Sakamoto, T., Breeveld, A.~A., {et~al.}
  2017, GRB Coordinates Network, 20422, 1

\bibitem[{{LIGO Scientific Collaboration} \&
  {Virgo}(2017{\natexlab{a}})}]{GW170104_detect}
{LIGO Scientific Collaboration}, \& {Virgo}. 2017{\natexlab{a}}, GRB
  Coordinates Network, 20364, 1

\bibitem[{{LIGO Scientific Collaboration} \&
  {Virgo}(2017{\natexlab{b}})}]{GW170104_finalskymap}
---. 2017{\natexlab{b}}, GRB Coordinates Network, 21056, 1

\bibitem[{Lipunov {et~al.}(2017{\natexlab{a}})Lipunov, Tyurina, Gorbovskoy,
  Kornilov, P.Balanutsa, A.Kuznetsov, V.Shumkov, M.I.Panchenko (Lomonosov
  Moscow State~University, D.Buckley, Potter, Observatory), R.~Rebolo,
  Israelian, de~Astrofisica~de Canarias), Tlatov, of~the Pulkovo~Observatory),
  Budnev, Gress, University), V.Yurkov, Yu.Sergienko, \& A.Gabovich
  (Blagoveschensk Educational State~University}]{GW_Global}
Lipunov, V., Tyurina, N., Gorbovskoy, E., Kornilov, V., P.Balanutsa, {et~al.}
  2017{\natexlab{a}}, GRB Coordinates Network, 20381, 1

\bibitem[{Lipunov {et~al.}(2017{\natexlab{b}})Lipunov, Gress, Tyurina,
  Gorbovskoy, Kornilov, P.Balanutsa, A.Kuznetsov, V.Shumkov, M.I.Panchenko
  Lomonosov Moscow State~University, D.Buckley, Potter, Observatory, Rebolo,
  Serra-Ricart, Israelian, de~Astrofisica~de Canarias, Tlatov, of~the
  Pulkovo~Observatory, Budnev, Gress, University, V.Yurkov, Yu.Sergienko, \&
  University}]{GW_Global2nd}
Lipunov, V., Gress, O., Tyurina, N., Gorbovskoy, E., Kornilov, V., {et~al.}
  2017{\natexlab{b}}, GRB Coordinates Network, 20392, 1

\bibitem[{Marcinkowski {et~al.}(2017)Marcinkowski, Xiao, \& Hajdas}]{mxh17}
Marcinkowski, R., Xiao, H., \& Hajdas, W. 2017, GRB Coordinates Network, 20387,
  1

\bibitem[{{McMullin} {et~al.}(2007){McMullin}, {Waters}, {Schiebel}, {Young},
  \& {Golap}}]{McMullin2007}
{McMullin}, J.~P., {Waters}, B., {Schiebel}, D., {Young}, W., \& {Golap}, K.
  2007, in Astronomical Society of the Pacific Conference Series, Vol. 376,
  Astronomical Data Analysis Software and Systems XVI, ed. R.~A. {Shaw},
  F.~{Hill}, \& D.~J. {Bell}, 127

\bibitem[{Melandri {et~al.}(2017{\natexlab{a}})Melandri, Piranomonte, D~Avanzo,
  Greco, Stratta, Ascenzi, Botticella, Palazzi, Rossi, Cappellaro, Covino,
  Elia, Amati, Antonelli, , Branchesi, Campana, Getman, Grado, Limatola, Lisi,
  Nicastro, Pian, Pulone, Tagliaferri, Testa, Tomasella, Yang, Brocato,
  Fabrizio, \& Mainella}]{GW170104_TNG}
Melandri, A., Piranomonte, S., D~Avanzo, P., Greco, G., Stratta, G., {et~al.}
  2017{\natexlab{a}}, GRB Coordinates Network, 20416, 1

\bibitem[{Melandri {et~al.}(2017{\natexlab{b}})Melandri, Piranomonte,
  Branchesi, Covino, D'Avanzo, D'Elia, Rossi, Amati, Antonelli, Ascenzi,
  Botticella, Campana, Cappellaro, Getman, Grado, Greco, Limatola, Lisi,
  Nicastro, Palazzi, Pian, Pulone, Stratta, Tagliaferri, Testa, Tomasella,
  Yang, Brocato, \& {GRavitational Wave Inaf TeAm
  (GRAWITA)}}]{GW170104_GRAWITA}
Melandri, A., Piranomonte, S., Branchesi, M., Covino, S., D'Avanzo, P.,
  {et~al.} 2017{\natexlab{b}}, GRB Coordinates Network, 20735, 1

\bibitem[{Mooley {et~al.}(2017)Mooley, Fender, \& Horesh}]{GW170104_AMI}
Mooley, K.~P., Fender, R.~P., \& Horesh, A. 2017, GRB Coordinates Network,
  20425, 1

\bibitem[{{Morokuma} {et~al.}(2016){Morokuma}, {Tanaka}, {Asakura}, {Abe},
  {Tristram}, {Utsumi}, {Doi}, {Fujisawa}, {Itoh}, {Itoh}, {Kawabata}, {Kawai},
  {Kuroda}, {Matsubayashi}, {Motohara}, {Murata}, {Nagayama}, {Ohta}, {Saito},
  {Tamura}, {Tominaga}, {Uemura}, {Yanagisawa}, {Yatsu}, \&
  {Yoshida}}]{2016PASJ...68L...9M}
{Morokuma}, T., {Tanaka}, M., {Asakura}, Y., {Abe}, F., {Tristram}, P.~J.,
  {et~al.} 2016, \pasj, 68, L9

\bibitem[{Oke \& Gunn(1982)}]{og82}
Oke, J.~B., \& Gunn, J.~E. 1982, Publications of the Astronomical Society of
  the Pacific, 94, 586

\bibitem[{{Palliyaguru} {et~al.}(2016){Palliyaguru}, {Corsi}, {Kasliwal},
  {Cenko}, {Frail}, {Perley}, {Mishra}, {Singer}, {Gal-Yam}, {Nugent}, \&
  {Surace}}]{pck+16}
{Palliyaguru}, N.~T., {Corsi}, A., {Kasliwal}, M.~M., {Cenko}, S.~B., {Frail},
  D.~A., {et~al.} 2016, \apjl, 829, L28

\bibitem[{{Perley} {et~al.}(2012){Perley}, {Modjaz}, {Morgan}, {Cenko},
  {Bloom}, {Butler}, {Filippenko}, \& {Miller}}]{pmm+12}
{Perley}, D.~A., {Modjaz}, M., {Morgan}, A.~N., {Cenko}, S.~B., {Bloom}, J.~S.,
  {Butler}, N.~R., {Filippenko}, A.~V., \& {Miller}, A.~A. 2012, \apj, 758, 122

\bibitem[{Perrott {et~al.}(2013)Perrott, Scaife, Green, Davies, Franzen,
  Grainge, Hobson, Hurley-Walker, Lasenby, Olamaie, Pooley,
  Rodr{\'{i}}guez-Gonz{\'{a}}lvez, Rumsey, Saunders, Schammel, Scott, Shimwell,
  Titterington, Waldram, \& {AMI Consortium}}]{psg+13}
Perrott, Y.~C., Scaife, A. M.~M., Green, D.~A., Davies, M.~L., Franzen, T.
  M.~O., {et~al.} 2013, Monthly Notices of the Royal Astronomical Society, 429,
  3330

\bibitem[{{Racusin} {et~al.}(2017){Racusin}, {Burns}, {Goldstein},
  {Connaughton}, {Wilson-Hodge}, {Jenke}, {Blackburn}, {Briggs}, {Broida},
  {Camp}, {Christensen}, {Hui}, {Littenberg}, {Shawhan}, {Singer}, {Veitch},
  {Bhat}, {Cleveland}, {Fitzpatrick}, {Gibby}, {von Kienlin}, {McBreen},
  {Mailyan}, {Meegan}, {Paciesas}, {Preece}, {Roberts}, {Stanbro}, {Veres},
  {Zhang}, {Fermi LAT Collaboration}, {Ackermann}, {Albert}, {Atwood},
  {Axelsson}, {Baldini}, {Ballet}, {Barbiellini}, {Baring}, {Bastieri},
  {Bellazzini}, {Bissaldi}, {Blandford}, {Bloom}, {Bonino}, {Bregeon}, {Bruel},
  {Buson}, {Caliandro}, {Cameron}, {Caputo}, {Caragiulo}, {Caraveo},
  {Cavazzuti}, {Charles}, {Chiang}, {Ciprini}, {Costanza}, {Cuoco}, {Cutini},
  {D'Ammando}, {de Palma}, {Desiante}, {Digel}, {Di Lalla}, {Di Mauro}, {Di
  Venere}, {Drell}, {Favuzzi}, {Ferrara}, {Focke}, {Fukazawa}, {Funk}, {Fusco},
  {Gargano}, {Gasparrini}, {Giglietto}, {Gill}, {Giroletti}, {Glanzman},
  {Granot}, {Green}, {Grove}, {Guillemot}, {Guiriec}, {Harding}, {Jogler},
  {J{\'o}hannesson}, {Kamae}, {Kensei}, {Kocevski}, {Kuss}, {Larsson},
  {Latronico}, {Li}, {Longo}, {Loparco}, {Lubrano}, {Magill}, {Maldera},
  {Malyshev}, {Mazziotta}, {McEnery}, {Michelson}, {Mizuno}, {Monzani},
  {Morselli}, {Moskalenko}, {Negro}, {Nuss}, {Omodei}, {Orienti}, {Orlando},
  {Ormes}, {Paneque}, {Perkins}, {Pesce-Rollins}, {Piron}, {Pivato}, {Porter},
  {Principe}, {Rain{\`o}}, {Rando}, {Razzano}, {Razzaque}, {Reimer}, {Reimer},
  {Saz Parkinson}, {Scargle}, {Sgr{\`o}}, {Simone}, {Siskind}, {Smith},
  {Spada}, {Spinelli}, {Suson}, {Tajima}, {Thayer}, {Torres}, {Troja},
  {Uchiyama}, {Vianello}, {Wood}, \& {Wood}}]{2017ApJ...835...82R}
{Racusin}, J.~L., {Burns}, E., {Goldstein}, A., {Connaughton}, V.,
  {Wilson-Hodge}, C.~A., {et~al.} 2017, \apj, 835, 82

\bibitem[{{Ramirez-Ruiz} {et~al.}(2002){Ramirez-Ruiz}, {Celotti}, \&
  {Rees}}]{ramirez-ruiz+02}
{Ramirez-Ruiz}, E., {Celotti}, A., \& {Rees}, M.~J. 2002, \mnras, 337, 1349

\bibitem[{{Rao} {et~al.}(2016){Rao}, {Chand}, {Hingar}, {Iyyani}, {Khanna},
  {Kutty}, {Malkar}, {Paul}, {Bhalerao}, {Bhattacharya}, {Dewangan}, {Pawar},
  {Vibhute}, {Chattopadhyay}, {Mithun}, {Vadawale}, {Vagshette}, {Basak},
  {Pradeep}, {Samuel}, {Sreekumar}, {Vinod}, {Navalgund}, {Pandiyan}, {Sarma},
  {Seetha}, \& {Subbarao}}]{rch+16}
{Rao}, A.~R., {Chand}, V., {Hingar}, M.~K., {Iyyani}, S., {Khanna}, R.,
  {et~al.} 2016, \apj, 833, 86

\bibitem[{{Rhoads}(1999)}]{rhoads99}
{Rhoads}, J.~E. 1999, \apj, 525, 737

\bibitem[{Rossum(1995)}]{python}
Rossum, G. 1995, Python Reference Manual, Tech. rep., Amsterdam, The
  Netherlands, The Netherlands

\bibitem[{Sakamoto {et~al.}(2017)Sakamoto, Yoshida, Kawakubo, Moriyama, Yamada,
  Yamaoka, Nakahira, Takahashi, Asaoka, Ozawa, Torii, Shimizu, Tamura,
  Ishizaki, Cherry, Ricciarini, \& Marrocchesi}]{GW170104_Calet}
Sakamoto, T., Yoshida, A., Kawakubo, Y., Moriyama, M., Yamada, Y., {et~al.}
  2017, GRB Coordinates Network, 20399, 1

\bibitem[{{Savchenko} {et~al.}(2016){Savchenko}, {Ferrigno}, {Mereghetti},
  {Natalucci}, {Bazzano}, {Bozzo}, {Brandt}, {Courvoisier}, {Diehl}, {Hanlon},
  {von Kienlin}, {Kuulkers}, {Laurent}, {Lebrun}, {Roques}, {Ubertini}, \&
  {Weidenspointner}}]{2016ApJ...820L..36S}
{Savchenko}, V., {Ferrigno}, C., {Mereghetti}, S., {Natalucci}, L., {Bazzano},
  A., {et~al.} 2016, \apjl, 820, L36

\bibitem[{Savchenko {et~al.}(2017)Savchenko, Ferrigno, Mereghetti, Kuulkers,
  Bazzano, E.~Bozzo, Brandt, Diehl, Hanlon, Laurent, Lutovinov, Roques,
  Sunyaev, \& Ubertini}]{GW170104_Integral}
Savchenko, V., Ferrigno, C., Mereghetti, S., Kuulkers, E., Bazzano, A.,
  {et~al.} 2017, GRB Coordinates Network, 20366, 1

\bibitem[{{Schlegel} {et~al.}(1998){Schlegel}, {Finkbeiner}, \&
  {Davis}}]{sfd98}
{Schlegel}, D.~J., {Finkbeiner}, D.~P., \& {Davis}, M. 1998, \apj, 500, 525

\bibitem[{{SDSS Collaboration} {et~al.}(2016){SDSS Collaboration}, {Albareti},
  {Allende Prieto}, {Almeida}, {Anders}, {Anderson}, {Andrews},
  {Aragon-Salamanca}, {Argudo-Fernandez}, {Armengaud}, \& et~al.}]{sdss16}
{SDSS Collaboration}, {Albareti}, F.~D., {Allende Prieto}, C., {Almeida}, A.,
  {Anders}, F., {et~al.} 2016, ArXiv e-prints, 1608.02013

\bibitem[{Serino {et~al.}(2017)Serino, Kawai, Sugita, Negoro, Ueno, Tomida,
  Nakahira, Ishikawa, Sugawara, Nakagawa, Mihara, Sugizaki, Iwakiri, Shidatsu,
  Sugimoto, Takagi, Matsuoka, N.Isobe, Yoshii, Tachibana, Ono, Fujiwara,
  Harita, Muraki, Yoshida, Sakamoto, Kawakubo, Kitaoka, Tsunemi, Shomura,
  Nakajima, Tanaka, Masumitsu, Kawase, Ueda, Kawamuro, Hori, Tanimoto, Oda,
  Tsuboi, Nakamura, Sasaki, Yamauchi, Furuya, \& Yamaoka}]{GW170104_maxi}
Serino, M., Kawai, N., Sugita, S., Negoro, H., Ueno, S., {et~al.} 2017, GRB
  Coordinates Network, 20507, 1

\bibitem[{Sharma {et~al.}(2017)Sharma, Bhalerao, Bhattacharya, Rao, \&
  Vadawale}]{czti_grb170105A}
Sharma, V., Bhalerao, V., Bhattacharya, D., Rao, A.~R., \& Vadawale, S. 2017,
  GRB Coordinates Network, 20389, 1

\bibitem[{{Simcoe} {et~al.}(2000){Simcoe}, {Metzger}, {Small}, \&
  {Araya}}]{sms+00}
{Simcoe}, R.~A., {Metzger}, M.~R., {Small}, T.~A., \& {Araya}, G. 2000, in
  Bulletin of the American Astronomical Society, Vol.~32, American Astronomical
  Society Meeting Abstracts \#196, 758

\bibitem[{Singer {et~al.}(2017)Singer, Kupfer, Roy, Kasliwal, Cenko, \&
  Barlow}]{GW_CiPTF}
Singer, L., Kupfer, T., Roy, R., Kasliwal, M., Cenko, S.~B., \& Barlow, T.
  2017, GRB Coordinates Network, 20401, 1

\bibitem[{Smartt {et~al.}(2017)Smartt, Smith, Huber, Chambers, Young, D.R.,
  (QUB), (Harvard), Denneau, Flewelling, Heinze, (IfA), (STScI), (IfA),
  Schultz, Tonry, Waters, Wainscoat, \& (IfA)}]{GW_PANSTAR}
Smartt, S., Smith, K., Huber, M., Chambers, K., Young, {et~al.} 2017, GRB
  Coordinates Network, 20410, 1

\bibitem[{{Smartt} {et~al.}(2016{\natexlab{a}}){Smartt}, {Chambers}, {Smith},
  {Huber}, {Young}, {Chen}, {Inserra}, {Wright}, {Coughlin}, {Denneau},
  {Flewelling}, {Heinze}, {Jerkstrand}, {Magnier}, {Maguire}, {Mueller},
  {Rest}, {Sherstyuk}, {Stalder}, {Schultz}, {Stubbs}, {Tonry}, {Waters},
  {Wainscoat}, {Della Valle}, {Dennefeld}, {Dimitriadis}, {Firth}, {Fraser},
  {Frohmaier}, {Gal-Yam}, {Harmanen}, {Kankare}, {Kotak}, {Kromer}, {Mandel},
  {Sollerman}, {Gibson}, {Primak}, \& {Willman}}]{2016ApJ...827L..40S}
{Smartt}, S.~J., {Chambers}, K.~C., {Smith}, K.~W., {Huber}, M.~E., {Young},
  D.~R., {et~al.} 2016{\natexlab{a}}, \apjl, 827, L40

\bibitem[{{Smartt} {et~al.}(2016{\natexlab{b}}){Smartt}, {Chambers}, {Smith},
  {Huber}, {Young}, {Cappellaro}, {Wright}, {Coughlin}, {Schultz}, {Denneau},
  {Flewelling}, {Heinze}, {Magnier}, {Primak}, {Rest}, {Sherstyuk}, {Stalder},
  {Stubbs}, {Tonry}, {Waters}, {Willman}, {Anderson}, {Baltay}, {Botticella},
  {Campbell}, {Dennefeld}, {Chen}, {Della Valle}, {Elias-Rosa}, {Fraser},
  {Inserra}, {Kankare}, {Kotak}, {Kupfer}, {Harmanen}, {Galbany}, {Gal-Yam},
  {Le Guillou}, {Lyman}, {Maguire}, {Mitra}, {Nicholl}, {Olivares E},
  {Rabinowitz}, {Razza}, {Sollerman}, {Smith}, {Terreran}, {Valenti}, {Gibson},
  \& {Goggia}}]{2016MNRAS.462.4094S}
---. 2016{\natexlab{b}}, \mnras, 462, 4094

\bibitem[{{Soares-Santos} {et~al.}(2016){Soares-Santos}, {Kessler}, {Berger},
  {Annis}, {Brout}, {Buckley-Geer}, {Chen}, {Cowperthwaite}, {Diehl}, {Doctor},
  {Drlica-Wagner}, {Farr}, {Finley}, {Flaugher}, {Foley}, {Frieman}, {Gruendl},
  {Herner}, {Holz}, {Lin}, {Marriner}, {Neilsen}, {Rest}, {Sako}, {Scolnic},
  {Sobreira}, {Walker}, {Wester}, {Yanny}, {Abbott}, {Abdalla}, {Allam},
  {Armstrong}, {Banerji}, {Benoit-L{\'e}vy}, {Bernstein}, {Bertin}, {Brown},
  {Burke}, {Capozzi}, {Carnero Rosell}, {Carrasco Kind}, {Carretero},
  {Castander}, {Cenko}, {Chornock}, {Crocce}, {D'Andrea}, {da Costa}, {Desai},
  {Dietrich}, {Drout}, {Eifler}, {Estrada}, {Evrard}, {Fairhurst}, {Fernandez},
  {Fischer}, {Fong}, {Fosalba}, {Fox}, {Fryer}, {Garcia-Bellido}, {Gaztanaga},
  {Gerdes}, {Goldstein}, {Gruen}, {Gutierrez}, {Honscheid}, {James},
  {Karliner}, {Kasen}, {Kent}, {Kuropatkin}, {Kuehn}, {Lahav}, {Li}, {Lima},
  {Maia}, {Margutti}, {Martini}, {Matheson}, {McMahon}, {Metzger}, {Miller},
  {Miquel}, {Mohr}, {Nichol}, {Nord}, {Ogando}, {Peoples}, {Plazas},
  {Quataert}, {Romer}, {Roodman}, {Rykoff}, {Sanchez}, {Scarpine}, {Schindler},
  {Schubnell}, {Sevilla-Noarbe}, {Sheldon}, {Smith}, {Smith}, {Smith},
  {Stebbins}, {Sutton}, {Swanson}, {Tarle}, {Thaler}, {Thomas}, {Tucker},
  {Vikram}, {Wechsler}, {Weller}, \& {DES Collaboration}}]{2016ApJ...823L..33S}
{Soares-Santos}, M., {Kessler}, R., {Berger}, E., {Annis}, J., {Brout}, D.,
  {et~al.} 2016, \apjl, 823, L33

\bibitem[{{Stalder} {et~al.}(2017){Stalder}, {Tonry}, {Smartt}, {Coughlin},
  {Chambers}, {Stubbs}, {Chen}, {Kankare}, {Smith}, {Denneau}, {Sherstyuk},
  {Heinze}, {Weiland}, {Rest}, {Young}, {Huber}, {Flewelling}, {Lowe},
  {Magnier}, {Schultz}, {Waters}, {Wainscoat}, {Willman}, {Wright}, {Chu},
  {Sanders}, {Inserra}, {Maguire}, \& {Kotak}}]{sts+17}
{Stalder}, B., {Tonry}, J., {Smartt}, S.~J., {Coughlin}, M., {Chambers}, K.~C.,
  {et~al.} 2017, ArXiv e-prints

\bibitem[{Steeghs {et~al.}(2017)Steeghs, Pollacco, Ulaczyk, Cutter, West,
  Levan, Galloway, Rol, Thrane, Dhillon, Dyer, Littlefair, Daw, Maund,
  Mullaney, Ramsay, Brien, \& Starling}]{GW170104_SWASP}
Steeghs, D., Pollacco, D., Ulaczyk, K., Cutter, R., West, R., {et~al.} 2017,
  GRB Coordinates Network, 20434, 1

\bibitem[{{Stetson}(1987)}]{daophot}
{Stetson}, P.~B. 1987, \pasp, 99, 191

\bibitem[{Svinkin {et~al.}(2017{\natexlab{a}})Svinkin, Golenetskii, Aptekar,
  Frederiks, Oleynik, Ulanov, Tsvetkova, Lysenko, Kozlova, \&
  Cline}]{GW170104_konus}
Svinkin, D., Golenetskii, S., Aptekar, R., Frederiks, D., Oleynik, P., {et~al.}
  2017{\natexlab{a}}, GRB Coordinates Network, 20794, 1

\bibitem[{Svinkin {et~al.}(2017{\natexlab{b}})Svinkin, Golenetskii, Aptekar,
  Frederiks, Tsvetkova, Kozlova, Cline, Hurley, von Kienlin, Zhang, Rau,
  Savchenko, Bozzo, \& Ferrigno}]{GW170104_IPN}
Svinkin, D., Golenetskii, S., Aptekar, R., Frederiks, D., Tsvetkova, A.,
  {et~al.} 2017{\natexlab{b}}, GRB Coordinates Network, 20406, 1

\bibitem[{{Tavani} {et~al.}(2016){Tavani}, {Pittori}, {Verrecchia},
  {Bulgarelli}, {Giuliani}, {Donnarumma}, {Argan}, {Trois}, {Lucarelli},
  {Marisaldi}, {Del Monte}, {Evangelista}, {Fioretti}, {Zoli}, {Piano},
  {Munar-Adrover}, {Antonelli}, {Barbiellini}, {Caraveo}, {Cattaneo}, {Costa},
  {Feroci}, {Ferrari}, {Longo}, {Mereghetti}, {Minervini}, {Morselli},
  {Pacciani}, {Pellizzoni}, {Picozza}, {Pilia}, {Rappoldi}, {Sabatini},
  {Vercellone}, {Vittorini}, {Giommi}, {Colafrancesco}, {Cardillo}, {Galli}, \&
  {Fuschino}}]{2016ApJ...825L...4T}
{Tavani}, M., {Pittori}, C., {Verrecchia}, F., {Bulgarelli}, A., {Giuliani},
  A., {et~al.} 2016, \apjl, 825, L4

\bibitem[{Tavani {et~al.}(2017{\natexlab{a}})Tavani, Verrecchia, Minervini,
  Giuliani, Bulgarelli, Zoli, Pittori, Donnarumma, Munar-Adrover, Piano,
  Lucarelli, Cardillo, Longo, Ursi, Fuschino, Evangelista, Marisaldi, Argan,
  Pilia, Trois, \& Fioretti}]{GW170104_AGILEGRID}
Tavani, M., Verrecchia, F., Minervini, G., Giuliani, A., Bulgarelli, A.,
  {et~al.} 2017{\natexlab{a}}, GRB Coordinates Network, 20395, 1

\bibitem[{Tavani {et~al.}(2017{\natexlab{b}})Tavani, Ursi, Fuschino,
  Evangelista, Donnarumma, Verrecchia, Minervini, Marisaldi, A.~Bulgarelli,
  Pittori, Lucarelli, Piano, Munar-Adrover, Argan, Cardillo, Giuliani, Pilia,
  Trois, \& Longo}]{GW170104_AGILE}
Tavani, M., Ursi, A., Fuschino, F., Evangelista, Y., Donnarumma, I., {et~al.}
  2017{\natexlab{b}}, GRB Coordinates Network, 20375, 1

\bibitem[{{The Astropy Collaboration} {et~al.}(2013){The Astropy
  Collaboration}, Robitaille, Tollerud, Greenfield, Droettboom, Bray, Aldcroft,
  Davis, Ginsburg, Price-Whelan, Kerzendorf, Conley, Crighton, Barbary, Muna,
  Ferguson, Grollier, Parikh, Nair, G{\"{u}}nther, Deil, Woillez, Conseil,
  Kramer, Turner, Singer, Fox, Weaver, Zabalza, Edwards, {Azalee Bostroem},
  Burke, Casey, Crawford, Dencheva, Ely, Jenness, Labrie, Lim, Pierfederici,
  Pontzen, Ptak, Refsdal, Servillat, \& Streicher}]{astropy}
{The Astropy Collaboration}, Robitaille, T.~P., Tollerud, E.~J., Greenfield,
  P., Droettboom, M., {et~al.} 2013, Astronomy {\&} Astrophysics, 558, A33

\bibitem[{{Tody}(1986)}]{iraf1}
{Tody}, D. 1986, in \procspie, Vol. 627, Instrumentation in astronomy VI, ed.
  D.~L. {Crawford}, 733

\bibitem[{{Tody}(1993)}]{iraf2}
{Tody}, D. 1993, in Astronomical Society of the Pacific Conference Series,
  Vol.~52, Astronomical Data Analysis Software and Systems II, ed. R.~J.
  {Hanisch}, R.~J.~V. {Brissenden}, \& J.~{Barnes}, 173

\bibitem[{Tonry {et~al.}(2017)Tonry, Denneau, Heinze, Stalder, Weiland, Stubbs,
  Smith, Smartt, , Young, Rest, Chambers, Coughlin, Huber, Wright, Flewelling,
  Magnier, Schultz, Waters, \& Wainscoat}]{GW170104_ATLAS}
Tonry, J., Denneau, L., Heinze, A., Stalder, B., Weiland, H., {et~al.} 2017,
  GRB Coordinates Network, 20382, 1

\bibitem[{{Tonry}(2011)}]{ATLAS}
{Tonry}, J.~L. 2011, \pasp, 123, 58

\bibitem[{Tyurina {et~al.}(2017)Tyurina, Lipunov, Gress, Gorbovskoy, Kornilov,
  P.Balanutsa, A.Kuznetsov, V.Shumkov, M.I.Panchenko, A.V.Krylov, I.Gorbunov
  Lomonosov Moscow State~University, D.Buckley, Potter, Observatory, Rebolo,
  Serra-Ricart, Israelian, de~Astrofisica~de Canarias, Tlatov, of~the
  Pulkovo~Observatory, Budnev, Gress, University, V.Yurkov, Yu.Sergienko, \&
  Blagoveschensk}]{GW_GLOBALOT}
Tyurina, N., Lipunov, V., Gress, O., Gorbovskoy, E., Kornilov, V., {et~al.}
  2017, GRB Coordinates Network, 20493, 1

\bibitem[{van~der Walt {et~al.}(2011)van~der Walt, Colbert, \&
  Varoquaux}]{numpy}
van~der Walt, S., Colbert, S.~C., \& Varoquaux, G. 2011, Computing in Science
  {\&} Engineering, 13, 22

\bibitem[{Vianello {et~al.}(2017)Vianello, Kocevski, Longo, McEnery, Perkins,
  \& Racusin}]{GW170104_LAT}
Vianello, G., Kocevski, D., Longo, F., McEnery, J., Perkins, J.~S., \& Racusin,
  J. 2017, GRB Coordinates Network, 20374, 1

\bibitem[{{Wilson} {et~al.}(2003){Wilson}, {Eikenberry}, {Henderson},
  {Hayward}, {Carson}, {Pirger}, {Barry}, {Brandl}, {Houck}, {Fitzgerald}, \&
  {Stolberg}}]{weh+03}
{Wilson}, J.~C., {Eikenberry}, S.~S., {Henderson}, C.~P., {Hayward}, T.~L.,
  {Carson}, J.~C., {et~al.} 2003, in \procspie, Vol. 4841, Instrument Design
  and Performance for Optical/Infrared Ground-based Telescopes, ed. M.~{Iye} \&
  A.~F.~M. {Moorwood}, 451--458

\bibitem[{Xu {et~al.}(2017)Xu, Liu, Niu, Zhang, Zhang, Pu, Ma, Yang, Song,
  Zhou, Zhang, Zhao, Li, Zhaori, \& andJ.R. Mao}]{GW170104_Xu}
Xu, D., Liu, J., Niu, H., Zhang, Y., Zhang, X., {et~al.} 2017, GRB Coordinates
  Network, 20417, 1

\bibitem[{{Zhang} \& {M{\'e}sz{\'a}ros}(2002)}]{zhang+02}
{Zhang}, B., \& {M{\'e}sz{\'a}ros}, P. 2002, \apj, 566, 712

\bibitem[{Zwart {et~al.}(2008)Zwart, Barker, Biddulph, Bly, Boysen, Brown,
  Clementson, Crofts, Culverhouse, Czeres, Dace, Davies, D'Alessandro, Doherty,
  Duggan, Ely, Felvus, Feroz, Flynn, Franzen, Geisb{\"{u}}sch,
  G{\'{e}}nova-Santos, Grainge, Grainger, Hammett, Hills, Hobson, Holler,
  Hurley-Walker, Jilley, Jones, Kaneko, Kneissl, Lancaster, Lasenby, Marshall,
  Newton, Norris, Northrop, Odell, Pober, Pooley, Quy,
  Rodr{\'{i}}guez-Gonz{\'{a}}lvez, Saunders, Scaife, Schofield, Scott, Shaw,
  Shimwell, Smith, Taylor, Titterington, Veli{\'{c}}, Waldram, West, Wood,
  Yassin, \& Consortium}]{zbb+08}
Zwart, J. T.~L., Barker, R.~W., Biddulph, P., Bly, D., Boysen, R.~C., {et~al.}
  2008, Monthly Notices of the Royal Astronomical Society, 391, 1545

\end{thebibliography}

\end{document}